\documentstyle[prl,aps,psfig]{revtex}
\def\Bf#1{\mbox{\boldmath{$#1$}}}
\def\bF#1{\mbox{\scriptsize\boldmath{$#1$}}}

\def\Ieq#1#2{{\mathstrut_{#1}^{#2}}}

\def\wt#1{\widetilde{#1}}
\def\wh#1{\widehat{#1}}
\def\ol#1{\overline{#1}}

\def\tJ{$t$-$J$ }
\begin{document}
%.\draft
%.\twocolumn[\hsize\textwidth\columnwidth\hsize\csname
%.@twocolumnfalse\endcsname

\title{
On the break in the single-particle energy dispersions and
the `universal' nodal Fermi velocity in the
high-temperature copper oxide superconductors }

\author{\sc Behnam Farid}

\address{Spinoza Institute, Department of Physics and Astronomy,
University of Utrecht,\\
Leuvenlaan 4, 3584 CE Utrecht, The Netherlands
\footnote{Electronic address: B.Farid@phys.uu.nl.} }

\date{Received 6 August 2003 and accepted 3 September 2003}
%\date{6 August 2003}
%\date{\today}

\maketitle

\begin{abstract}
\leftskip 54.8pt
\rightskip 54.8pt
Recent data from  angle-resolved photoemission experiments published 
by Zhou {\it et al.} [{\sl Nature}, {\bf 423}, 398 (2003)] concerning 
a number of hole-doped copper-oxide based high-temperature 
superconductors reveal that in the nodal directions of the underlying 
square Brillouin zones (i.e. the directions along which the 
${\rm d}$-wave superconducting gap is vanishing) the Fermi velocities 
for some finite range of ${\Bf k}$ inside the Fermi sea and away from 
the nodal Fermi wave vector ${\Bf k}_{\sc f}$ are to within an 
experimental uncertainty of approximately $20$\% the same both in 
all the compounds investigated and over a wide range of doping 
concentrations {\sl and} that, in line with earlier experimental 
observations, at some characteristic wave vector ${\Bf k}_{\star}$ 
away from ${\Bf k}_{\sc f}$ ($\|{\Bf k}_{\star}-{\Bf k}_{\sc f}\|$ 
typically amounts to approximately $5$\% of $\|{\Bf k}_{\sc f}\|$) 
the Fermi velocities undergo a sudden change, with this change (roughly 
speaking, an {\sl increase} for $\|{\Bf k}\| < \|{\Bf k}_{\star}\|$) 
being the greatest (smallest) in the case of underdoped (overdoped) 
compounds. We demonstrate that these observations establish 
{\sl four} essential facts: firstly, the ground-state momentum 
distribution function ${\sf n}({\Bf k})$ must be discontinuous at 
${\Bf k} ={\Bf k}_{\star}$; with ${\Bf v}_{{\bF k}_{\star}^-}^{<}$ 
and ${\Bf v}_{{\bF k}_{\star}^+}^{<}$ denoting the measured velocities 
close to ${\Bf k}={\Bf k}_{\star}$, $\|{\Bf k}_{\star}^-\| < 
\|{\Bf k}_{\star}\| < \|{\Bf k}_{\star}^+\|$, and $Z_{{\bF k}_{\star}} 
\equiv {\sf n}({\Bf k}_{\star}^-)-{\sf n}({\Bf k}_{\star}^+) > 0$, 
we obtain
\begin{eqnarray}
{\Bf v}_{{\bF k}_{\star}^+}^{<} = 
{\Bf v}_{{\bF k}_{\star}^-}^{<} -
\frac{Z_{{\bF k}_{\star}}}{{\sf n}({\Bf k}_{\star}^-) }\,
{\Bf v}_{{\bF k}_{\star}}, 
\nonumber
\end{eqnarray}
in which ${\Bf v}_{{\bF k}_{\star}}$ is the `Fermi' velocity
corresponding to the case in which $Z_{{\bF k}_{\star}} =0$ (for
two-body interaction potentials of shorter range than the Coulomb
potential, ${\Bf v}_{{\bF k}_{\star}}\approx 
{\Bf v}_{{\bF k}_{\star}^+}^{<}$ whereby 
${\Bf v}_{{\bF k}_{\star}^+}^{<} <
{\Bf v}_{{\bF k}_{\star}^-}^{<}\,{\small\Ieq\sim<}\, 
2\, {\Bf v}_{{\bF k}_{\star}^+}^{<}$);
secondly, the single-particle spectral function $A({\Bf k};\varepsilon)$ 
must at ${\Bf k}={\Bf k}_{\star}$ possess a coherent contribution 
corresponding to a well-defined quasiparticle excitation at an energy 
of approximately $70$~meV below the Fermi energy; thirdly, the amount 
of discontinuity $Z_{{\bF k}_{\sc f}}$ in ${\sf n}({\Bf k})$ at the 
nodal Fermi points must be small and ideally vanishing; 
fourthly, the long range of the two-body Coulomb potential is of 
vital importance for realization of a certain aspect of the 
observed behaviour in the Fermi velocities (specifically) in the 
underdoped regime. The condition $Z_{{\bF k}_{\sc f}}=0$ conforms 
with the observation through an earlier angle-resolved photoemission 
experiment by Valla {\it et al.} [{\sl Science}, {\bf 285}, 2110 
(1999)], on the optimally doped compound 
Bi$_2$Sr$_2$CaCu$_2$O$_{8+\delta}$, which shows that the imaginary part 
of the self-energy along the nodal directions of the Brillouin zone
and in the vicinity of the nodal Fermi points satisfies the scaling 
behaviour characteristic of marginal Fermi-liquid metallic states, for 
which $Z_{{\bF k}_{\sc f}}$ is indeed vanishing. We present arguments 
advocating the viewpoint that the observed `kink' in the measured 
energy dispersions {\sl cannot} be a direct consequence of 
electron-phonon interaction, although a finite $Z_{{\bF k}_{\star}}$ 
may possibly arise from this interaction. In other words, even though 
possibly vital, the role played by phonons in bringing about the latter 
`kink' must be indirect. Our approach further provides a consistent 
interpretation of the observed sudden decrease in the width of the 
so-called `momentum distribution curve' on $\|{\Bf k}\|$ increasing 
above $\|{\Bf k}_{\star}\|$. 
\end{abstract}

\vspace{1.5cm}
\noindent
\underline{\sf Preprint number: SPIN-2003/15 }
\vspace{1.0cm}

\noindent\underline{\sf arXiv:cond-mat/0308090} \\
\underline{\sc Philosophical Magazine, Vol.~84, No.~9, 909-955 (2004)}
%
%\pacs{} ]
\narrowtext
\twocolumn
%\vfill
%\pagebreak
%\widetext

% 0.
{\footnotesize
\contentsline {section}
 {\numberline {1.} Introduction}
{\pageref{s1}}
\contentsline {section}
 {\numberline {2.} Preliminaries}
{\pageref{s2}}
\contentsline {section}
 {\numberline {3.} Generalities }
{\pageref{s3}}
\contentsline {subsection}
 {\numberline {3.1.} Some remarks on the single-particle
excitation energies}
{\pageref{s3a}}
\contentsline {subsection}
 {\numberline {3.2.} Regular and singular contributions to the 
single-particle spectral function $A_{\sigma}({\Bf k};\varepsilon)$}
{\pageref{s3b}}
\contentsline {subsection}
 {\numberline {3.3.} On ratios of discontinuous functions}
{\pageref{s3c}}
\contentsline {subsection}
 {\numberline {3.4.} Possible discontinuities in 
$\varepsilon_{{\bF k};\sigma}^{<}$ and 
$\varepsilon_{{\bF k};\sigma}^{>}$ }
{\pageref{s3d}}
\contentsline {subsection}
 {\numberline {3.5.} On the points where 
$\varepsilon_{{\bF k};\sigma}^- = 
\varepsilon_{{\bF k};\sigma}^{<}$ 
(and $\varepsilon_{{\bF k};\sigma}^+ = 
\varepsilon_{{\bF k};\sigma}^{>}$)}
{\pageref{s3e}}
\contentsline {subsection}
{\numberline {3.6.} On the number of solutions of the quasiparticle 
equation at ${\Bf k} = {\Bf k}_{{\sc f};\sigma}^{\mp}$ corresponding 
to well-defined quasiparticles }
{\pageref{s3f}}
\contentsline {subsection}
 {\numberline {3.7.} On the positivity and the possible 
discontinuities of $\Delta\varepsilon_{{\bF k};\sigma}^{<}$ }
{\pageref{s3g}}
\contentsline {section}
 {\numberline {4.} Detailed considerations}
{\pageref{s4}}
\contentsline {subsection}
 {\numberline {4.1.} The interacting Hamiltonian and some 
basic details }
{\pageref{s4a}}
\contentsline {subsection}
 {\numberline {4.2} On the relationship between
discontinuities in ${\sf n}_{\sigma}({\Bf k})$ and 
${\Bf\nabla}_{\bF k} \varepsilon_{{\bF k};\sigma}^{<}$ }
{\pageref{s4b}}
\contentsline {subsection}
 {\numberline {4.3} Explicit calculation of the amount of 
discontinuity in ${\Bf\nabla}_{\bF k} 
\varepsilon_{{\bF k};\sigma}^{<}$ at ${\Bf k}={\Bf k}_{\star}$, 
with ${\Bf k}_{\star} \in 
{\rm FS}_{\sigma}\backslash {\cal S}_{{\sc f};\sigma}$ }
{\pageref{s4c}}
\contentsline {section}
 {\numberline {5.} Summary and discussion}
{\pageref{s5}}
\contentsline {section}
 {\numberline {{}} Acknowledgements}
{\pageref{ack}}
\contentsline {section}
 {\numberline {{}} {Appendix~A:} An approximate 
$\varepsilon_{\lowercase{\bF k};\sigma}^{<}$ }
{\pageref{app}}
\contentsline {section}
 {\numberline {{}} References}
{\pageref{ref}}
}
%.

\subsection*{\bf\S~1. \sc Introduction}
\label{s1}

Angle-resolved photoemission spectroscopy (ARPES) \cite{DHS03,SD95} 
provides information with regard to the single-particle spectral 
function $A_{\sigma}({\Bf k};\varepsilon)$ of systems for $\varepsilon 
< \mu$, where $\varepsilon$ is the external energy parameter, 
$\mu$ the chemical potential, ${\Bf k}$ the wave vector and $\sigma$ 
the spin index. The quantity measured experimentally is the 
photoelectron intensity $I_{\sigma}({\Bf k};\varepsilon)$ which at 
finite temperature $T$ and under some specified approximations is 
proportional to $f(\varepsilon-\mu) A_{\sigma}({\Bf k};\varepsilon)$, 
where $f(\varepsilon) \equiv 
1/\big(1+\exp(\varepsilon/[k_{\sc b} T])\big)$ is the Fermi 
distribution function \cite{DHS03,SD95}. The experimental results 
obtained by Zhou {\sl et al.} \cite{XJZ03} and other workers
\cite{TV99,PVB00,PDJ01,AK01,AL01}, to be discussed in the present 
paper, are deduced from the so-called `momentum distribution curve' 
(MDC); this curve is obtained by setting out $I_{\sigma}({\Bf k};
\varepsilon)$ along the desired directions in the ${\Bf k}$ space
for a fixed value of $\varepsilon$. A related curve, referred to as
the `energy distribution curve' (EDC), is obtained by plotting
$I_{\sigma}({\Bf k};\varepsilon)$ for a fixed ${\Bf k}$ along the
$\varepsilon$ axis. Amongst others, the determination of the peak 
positions in $A_{\sigma}({\Bf k};\varepsilon)$ from the corresponding 
MDC does not suffer from the ambiguities associated with 
$f(\varepsilon-\mu)$. 

In this paper we first consider the relationship between the 
energy dispersions as measured in angle-resolved photoemission 
experiments and the single-particle excitation energy dispersions 
as determined by solving the quasi-particle equation in terms 
of the self-energy operator; the latter accounts for the correlation
in the ground states (GSs) of the underlying systems. The formalism 
concerning the {\sl measured} energy dispersions that follows from our 
purely heuristic arguments turns out to coincide, in all details, with 
that developed by the present author in his recent investigations 
regarding the nature of the uniform metallic GSs of the conventional 
single-band Hubbard Hamiltonian \cite{BF02a} and more general 
single-band Hamiltonians \cite{BF03} in which the particle-particle 
interaction can be of arbitrary range. The specific aspects of the 
formalism developed in \cite{BF02a,BF03} that greatly suited the 
aims in the aforementioned investigations are that for the indicated 
GSs it formally yields 
\vspace{0.8pt}

\indent\hspace{2pt}
(i) the exact GS total energy, \\
\indent\hspace{0.0pt}
(ii) the exact Fermi energy, \\
\indent
(iii) the exact Fermi surface and \\
\indent
(iv) the exact momentum distribution function. 

\vspace{0.8pt}
\noindent
The formalism has made possible \cite{BF02a,BF03} to make exact 
predictions concerning the values that the GS momentum distribution 
function ${\sf n}_{\sigma}({\Bf k})$ can take for ${\Bf k}$ in the 
close neighbourhoods of the underlying Fermi surface 
${\cal S}_{{\sc f};\sigma}$. On the basis of this, it has been 
possible to expose \cite{BF03} the significant influence that 
the range of the two-body interaction potential can have on 
${\sf n}_{\sigma}({\Bf k})$ and consequently on the partitioning 
of the GS total energy into `kinetic' and `interaction' parts. 
This observation led us to the conclusion \cite{BF03} that the 
experimentally-inferred excess `kinetic' energy in the normal 
states of the copper-oxide based high-temperature superconductors
\cite{SS01,MPMKL02} (for other pertinent references see \cite{BF03}) 
may be signifying the importance of the long range of the two-body 
interaction potential in determining other distinctive aspects of 
the normal metallic states of these compounds that mark them as 
unconventional metals \cite{PWA97}. 

For the normal uniform metallic GSs of the single-band Hubbard 
Hamiltonian, we have established that, so long as these are Fermi 
liquids, for ${\Bf k}$ inside the Fermi sea and close to the Fermi 
surface, a single-particle energy dispersion $\varepsilon_{{\bF k};
\sigma}^{<}$, which in \cite{BF02a,BF03} was attributed to some
`fictitious' particles and in this paper we reason to be the one 
measured through the ARPES, is up to a multiplicative 
${\Bf k}$-independent constant equal to the energy dispersion 
$\varepsilon_{{\bF k};\sigma}$ of the Landau quasi-particles 
inside the Fermi sea (more precisely, the energy dispersion 
$\varepsilon_{{\bF k};\sigma}^-$ as obtained from the quasi-particle 
equation in Eq.~(\ref{e3}) below; see \S~3.1); similarly for 
$\varepsilon_{{\bF k};\sigma}^{>}$ (Eq.~(\ref{e23})) and
$\varepsilon_{{\bF k};\sigma}^+$ concerning ${\Bf k}$ outside the 
Fermi sea. For cases where the two-particle interaction potential 
is the long-range Coulomb potential, in \cite{BF03} it was however 
shown that no such strict relationship exists between 
$\varepsilon_{{\bF k};\sigma}^{<}$ (and similarly 
$\varepsilon_{{\bF k};\sigma}^{>}$) and $\varepsilon_{{\bF k};
\sigma}^-$ ($\varepsilon_{{\bF k};\sigma}^+$), not even for Fermi 
liquid metallic states. In this paper we rigorously demonstrate 
that in spite of these aspects, in cases where for a given ${\Bf k}=
{\Bf k}_{\star}$ the quasi-particle equation (Eq.~(\ref{e3}) below) 
has solutions $\varepsilon_{{\bF k}_{\star};\sigma}^-$ and 
$\varepsilon_{{\bF k}_{\star};\sigma}^+$ corresponding to 
well-defined Landau quasi-particles, irrespective of the nature of 
the two-body interaction potential $\varepsilon_{{\bF k}_{\star};
\sigma}^{<}$ identically coincides with $\varepsilon_{{\bF k}_{\star};
\sigma}^{-}$ (if $\varepsilon_{{\bF k};\sigma}^{<}$ is continuous 
at ${\Bf k}={\Bf k}_{\star}$; Eq.~(\ref{e27})) and 
$\varepsilon_{{\bF k}_{\star};\sigma}^{>}$ with 
$\varepsilon_{{\bF k}_{\star};\sigma}^{+}$ (if 
$\varepsilon_{{\bF k};\sigma}^{>}$ is continuous at ${\Bf k}
={\Bf k}_{\star}$; Eq.~(\ref{e30})); we present the appropriate 
expressions for $\varepsilon_{{\bF k}_{\star};\sigma}^-$ and 
$\varepsilon_{{\bF k}_{\star};\sigma}^+$ in terms of 
$\varepsilon_{{\bF k}_{\star}^{\mp};\sigma}^{<}$ and 
$\varepsilon_{{\bF k}_{\star}^{\mp};\sigma}^{>}$ respectively for 
cases in which either or both of $\varepsilon_{{\bF k};\sigma}^{<}$
and $\varepsilon_{{\bF k};\sigma}^{>}$ are discontinuous at ${\Bf k}
={\Bf k}_{\star}$; according to our analysis (\S~3.4), these 
functions {\sl cannot} both be continuous at ${\Bf k}={\Bf k}_{\star}$.

In the light of the above-mentioned similarities between 
$\varepsilon_{{\bF k};\sigma}^{<}$ and the possible solutions 
$\varepsilon_{{\bF k};\sigma}^-$ of the quasi-particle equation
associated with particle-like excitations (as opposed to collective 
low-lying excitations in non-Fermi-liquid metallic states) as well 
as the heuristic arguments underlying the definition of 
$\varepsilon_{{\bF k};\sigma}^{<}$ (see \S~2), in this paper we 
assert that the ARPES data concerning the dispersion of the 
single-particle excitation energies are to be compared with 
$\varepsilon_{{\bF k};\sigma}^{<}$. The fact that the behaviour 
of $\varepsilon_{{\bF k};\sigma}^{<}$ corresponding to Fermi-liquid 
metallic states of systems of fermions interacting through the 
long-range Coulomb potential in the neighbourhood of 
${\cal S}_{{\sc f};\sigma}$ demonstrably differs from that of the 
asymptotic solution (see \S~3.1) $\varepsilon_{{\bF k};\sigma}^-$ of 
the quasi-particle equation in the neighbourhood of ${\cal S}_{{\sc f};
\sigma}$ (see above) does not diminish the strength of our arguments. 
In this connection we should emphasize that our assertion has bearing 
on the energy dispersions {\sl as measured by means of the ARPES} and 
thus does not negate the fundamental significance of the possible 
asymptotic solutions of the quasi-particle equation, nor the essential 
role that these (explicitly, the asymptotic solutions corresponding to 
the exact quasi-particle solutions for ${\Bf k} \in {\cal S}_{{\sc f};
\sigma}$; \S~3.1) play in determining the low-temperature thermodynamic 
and transport properties of the Fermi-liquid metallic systems \cite{PN66}.

Our point of departure in this paper consists of the consideration
that the {\sl measured} (by means of the ARPES) single-particle energy 
dispersions coincide with the expectation value of $\varepsilon$
with respect to the normalized distribution of the single-particle 
excitations whose energies lie above $-E$ and below $\mu$, that is
$A_{\sigma}({\Bf k};\varepsilon)/\int_{-E}^{\mu} {\rm d}\varepsilon\;
A_{\sigma}({\Bf k};\varepsilon)$ (note the dependence on ${\Bf k}$ of 
the normalization amplitude); for systems with bounded single-particle 
energy spectrum (such as those described by the single-band Hubbard 
and the extended Hubbard Hamiltonians), $E$ can be identified with 
infinity (\S~2). This consideration regarding the measured energy 
dispersions provides the opportunity to introduce a quantitative size 
for the width of the main peak in the single-particle spectral 
function $A_{\sigma} ({\Bf k};\varepsilon)$ centred at $\varepsilon 
=\varepsilon_{{\bF k};\sigma}^{<}$, which is expressible in terms of 
$\int_{-E}^{\mu} {\rm d}\varepsilon\; A_{\sigma}({\Bf k};\varepsilon)$, 
$\int_{-E}^{\mu} {\rm d}\varepsilon\; \varepsilon^2\, A_{\sigma}
({\Bf k};\varepsilon)$ and $\varepsilon_{{\bF k};\sigma}^{<}$ (see 
Eq.~(\ref{e7})). According to this measure and under the conditions 
that we shall explicitly specify in the due place, the behaviour of 
the width of the single-particle spectral function centred at 
$\varepsilon_{{\bF k};\sigma}^{<}$ is directly correlated with the 
behaviour of ${\sf n}_{\sigma}({\Bf k})$; on transposing ${\Bf k}$ 
through a region where ${\sf n}_{\sigma}({\Bf k})$ undergoes a rapid 
decrease (increase), this behaviour is directly reflected in a 
concomitant decrease (increase) in the latter width. This, which for 
metallic states can be shown to be the characteristic aspect of the 
corresponding $A_{\sigma}({\Bf k};\varepsilon)$ for ${\Bf k}$ 
approaching the underlying ${\cal S}_{{\sc f};\sigma}$, 
\footnote{\label{f1}
On the basis of experimental observations, it has even been 
advocated \protect\cite{REA95,CDNR96} (see however the cautionary 
remarks in reference 8 of \protect\cite{CDNR96}) that Fermi 
surface be defined as the locus of points in the ${\Bf k}$ space 
at which $\|{\Bf\nabla}_{\bF k} {\sf n}_{\sigma}({\Bf k})\|$ is 
maximal. This definition has been utilized in  
\protect\cite{PLS98} where the ${\sf n}_{\sigma}({\Bf k})$ pertaining 
to the GS of the \tJ Hamiltonian has been calculated from a 
high-temperature series expansion. See also \protect\cite{BF02a}. }
thus turns out to be a generic property of $A_{\sigma}({\Bf k};
\varepsilon)$ pertaining to systems of interacting fermions for 
${\Bf k}$ in the neighbourhoods of all points (and not 
solely the points of ${\cal S}_{{\sc f};\sigma}$) where 
${\sf n}_{\sigma}({\Bf k})$ exhibits large local variation. 
According to this insight, the experimental observations in 
\cite{XJZ03,PVB00,AK01,AL01} with regard to a sharp decrease in 
the width of the peak in $A_{\sigma}({\Bf k};\varepsilon)$ centred 
at $\varepsilon=\varepsilon_{{\bF k};\sigma}^{<}$ for ${\Bf k}=
{\Bf k}_{\star}$ is indicative that ${\Bf k}_{\star}$ is a point 
at which the underlying GS momentum distribution functions undergo 
a sharp decrease, if not discontinuity. Our analysis of other 
aspects of the experimental observations reported in
\cite{XJZ03,TV99,PVB00,PDJ01,AK01,AL01} supports a discontinuity in 
the underlying ${\sf n}_{\sigma}({\Bf k})$, with ${\sf n}_{\sigma}
({\Bf k}_{\star}^-) > {\sf n}_{\sigma}({\Bf k}_{\star}^+)$, where 
$\| {\Bf k}_{\star}^- -{\Bf k}_{\star}^+\|$ is infinitesimally small
and ${\Bf k}_{\star}^+$ denotes the vector nearest to the nodal 
Fermi wave vector ${\Bf k}_{{\sc f};\sigma}$.

In the sections that follow, we combine our theoretical 
considerations with brief discussions of the pertinent 
experimental observations in 
\cite{XJZ03,TV99,PVB00,PDJ01,AK01,AL01}. Our theoretical 
treatment in this paper is restricted to $T=0$.

\subsection*{\bf\S~2. \sc Preliminaries}
\label{s2}

Assuming that, for a given ${\Bf k}$, the experimentally determined 
$A_{\sigma}({\Bf k};\varepsilon)$ is dominantly sharp at and 
symmetrical around $\varepsilon=\varepsilon_{{\bF k};\sigma}$ (later 
we shall identify this energy with $\varepsilon_{{\bF k};\sigma}^-$ 
for reasons that will be clarified), $\varepsilon_{{\bF k};\sigma}$ 
would coincide with the expectation value of $\varepsilon$ 
with respect to the energy distribution function $\hbar^{-1}
A_{\sigma}({\Bf k};\varepsilon)$. In this connection it is important 
to realize that $A_{\sigma}({\Bf k};\varepsilon)$ is positive 
semidefinite, as befits a proper distribution function; however, 
although $\hbar^{-1} A_{\sigma}({\Bf k};\varepsilon)$ is normalized 
to unity for $\varepsilon$ over $(-\infty,\infty)$, this is not the 
case for $\varepsilon$ restricted to the semi-infinite interval
$(-\infty,\mu]$. Thus, under the above-mentioned conditions, for 
$\varepsilon_{{\bF k};\sigma} < \mu$, corresponding to `occupied' 
single-particle states, one should have (see Eq.~(\ref{e39}) below)
\begin{equation}
\label{e1}
\varepsilon_{{\bF k};\sigma} \approx
\frac{\frac{1}{\hbar}\int_{-\infty}^{\mu} {\rm d}\varepsilon\;
\varepsilon A_{\sigma}({\Bf k};\varepsilon)}
{ {\sf n}_{\sigma}({\Bf k})},
\end{equation}
where 
\begin{equation}
\label{e2}
{\sf n}_{\sigma}({\Bf k}) \equiv \frac{1}{\hbar}
\int_{-\infty}^{\mu} {\rm d}\varepsilon\;
A_{\sigma}({\Bf k};\varepsilon)
\end{equation}
stands for the GS momentum 
distribution function pertaining to particles with spin index 
$\sigma$. The right-hand side (RHS) of Eq.~(\ref{e1}) identically 
coincides with the expression for the energy dispersion
$\varepsilon_{{\bF k};\sigma}^{<}$ as defined in
\footnote{\label{f2}
See, for instance, Eqs.~(6) and (39) in \protect\cite{BF03}. }
\cite{BF02a,BF03}. We note in passing that $\varepsilon_{{\bF k};
\sigma}^{<} < \mu$ for {\sl all} ${\Bf k}$ \cite{BF02a,BF03}. 
Evidently, the experimentally measured $A_{\sigma}({\Bf k};
\varepsilon)$ exhibits more structure than a single dominant 
peak, however so long as these additional structure correspond 
to high-energy processes (e.g. core-level peaks, which is not 
described by single-band models), for a relatively small range 
of ${\Bf k}$, such structure can only account for a rigid (i.e. 
${\Bf k}$-independent) shift of the RHS of Eq.~(\ref{e1}) with 
respect to the low-energy peak position of the experimentally 
determined $A_{\sigma}({\Bf k};\varepsilon)$.

Although $\varepsilon_{{\bF k};\sigma}^{<}$ (i.e., the RHS of 
Eq.~(\ref{e1})) is not formally a `quasiparticle' energy, in the 
sense of unconditionally being the solution of the quasiparticle 
equation (see later)
\begin{equation}
\label{e3}
\varepsilon_{\bF k} + 
\hbar\Sigma_{\sigma}({\Bf k};\varepsilon)
= \varepsilon,
\end{equation}
in which $\varepsilon_{\bF k}$ is the single-particle energy 
dispersion describing the non-interacting system (see Eq.~(\ref{e36}) 
below) and $\Sigma_{\sigma}({\Bf k};\varepsilon)$ the self-energy, 
in \cite{BF02a,BF03} it has been rigorously shown
\footnote{\label{f3}
See, e.g., Eq.~(40) in \protect\cite{BF03}. }
that, for ${\Bf k}$ approaching the interacting Fermi surface
${\cal S}_{{\sc f};\sigma}$, $\varepsilon_{{\bF k};\sigma}^{<}$
approaches the exact Fermi energy $\varepsilon_{\sc f}$; with
${\Bf k}_{{\sc f};\sigma} \in {\cal S}_{{\sc f};\sigma}$ and
\footnote{\label{f4}
For reasons that we have presented earlier \cite{BF99a,BF02a}, one can 
identify ${\Bf k}_{{\sc f};\sigma}^-$ with ${\Bf k}_{{\sc f};\sigma}$
(i.e., ${\cal S}_{{\sc f};\sigma}$ can be considered as
a proper subset of FS$_{\sigma}$); on the other hand, strict 
distinction has to be made between ${\Bf k}_{{\sc f};\sigma}^+$ 
and ${\Bf k}_{{\sc f};\sigma}$. The same applies to ${\Bf k}_{\star}$ 
that we frequently encounter in the text. See \S\S~3.1 and 3.2. }
${\Bf k}_{{\sc f};\sigma}^-$ (${\Bf k}_{{\sc f};\sigma}^+$) 
{\sl inside} ({\sl outside}) the Fermi sea FS$_{\sigma}$ 
(in this paper we denote the set of ${\Bf k}$ points complementary 
to FS$_{\sigma}$ with respect to the available wave-vector space
by $\ol{\rm FS}_{\sigma}$ so that ${\Bf k}_{{\sc f};\sigma}^+
\in \ol{\rm FS}_{\sigma}$) and infinitesimally close to 
${\Bf k}_{{\sc f};\sigma}$, we have \cite{BF02a,BF03}
\begin{equation}
\label{e4}
\varepsilon_{\sc f} =
\frac{ \int_{-\infty}^{\mu} {\rm d}\varepsilon\;
\varepsilon A_{\sigma}({\Bf k}_{{\sc f};\sigma}^{\mp};\varepsilon) }
{\int_{-\infty}^{\mu} {\rm d}\varepsilon\;
A_{\sigma}({\Bf k}_{{\sc f};\sigma}^{\mp};\varepsilon) }.
\end{equation}
The freedom in the choice of either of the two vectors 
${\Bf k}_{{\sc f};\sigma}^-$ and ${\Bf k}_{{\sc f};\sigma}^+$ 
in the above expression reflects the continuity of 
$\varepsilon_{{\bF k};\sigma}^{<}$ in a neighbourhood of 
${\cal S}_{{\sc f};\sigma}$ (for details see \S\S~3.4 and 4.2 below); 
our use of ${\Bf k}_{{\sc f};\sigma}^{\mp}$, rather than simply 
${\Bf k}_{{\sc f};\sigma}$ (the continuity of $\varepsilon_{{\bF k};
\sigma}^{<}$ at ${\Bf k}={\Bf k}_{{\sc f};\sigma}$ notwithstanding), 
is prompted by the fact that, in cases where ${\sf n}_{\sigma}
({\Bf k})$ is discontinuous at ${\Bf k}={\Bf k}_{{\sc f};\sigma}$, 
the expression on the RHS of Eq.~(\ref{e4}) is in its present form 
formally ill defined.

We point out that for metallic GSs, $\varepsilon=\varepsilon_{\sc f}$ 
satisfies Eq.~(\ref{e3}); Fermi surface ${\cal S}_{{\sc f};\sigma}$
is the locus of all ${\Bf k}$ points for which this equation is 
fulfilled for $\varepsilon=\varepsilon_{\sc f}$ and moreover 
$\varepsilon_{\bF k} +\hbar\Sigma_{\sigma}({\Bf k};\varepsilon_{\sc f})
-\varepsilon_{\sc f}$ has different signs for ${\Bf k}$ infinitesimally 
transposed in opposite directions along the normal to the 
constant-energy surface, i.e. ${\cal S}_{{\sc f};\sigma}$, 
described by $\varepsilon_{\bF k} + \hbar\Sigma_{\sigma}({\Bf k};
\varepsilon_{\sc f})=\varepsilon_{\sc f}$. 
\footnote{\label{f5}
For completeness, for a given ${\cal S}_{{\sc f};\sigma}$ there also
exists a set ${\cal S}_{{\sc f};\sigma}^+$, a generic point of which
we denote by ${\Bf k}_{{\sc f};\sigma}^+$ (see footnote
\protect\ref{f4}), for which Eq.~(\protect\ref{e3}) is satisfied at
$\varepsilon=\varepsilon_{\sc f}^+$, with $\varepsilon_{\sc f}^+$
infinitesimally greater than $\varepsilon_{\sc f}$. See \S\S~3.1 
and 3.2. }
Equation (\ref{e4}) presages the possibility that the RHS of 
Eq.~(\ref{e1}) may at least be a useful mathematical device for 
the purpose of estimating the energies of the low-lying 
single-particle excitations in interacting metallic systems.

As a measure both of the accuracy with which $\varepsilon_{{\bF k};
\sigma}^{<}$ describes the location of the main peak in $A_{\sigma}
({\Bf k};\varepsilon)$, $\varepsilon < \mu$, and of the width of 
the latter peak, one can consider the variance 
$\Delta\varepsilon_{{\bF k};\sigma}^{<}$ of $\varepsilon$. Since 
for the systems in which $\varepsilon_{\bF k}$ is unbounded from 
above, which for the uniform GSs considered in this paper is feasible 
only for those defined on the continuum (to be contrasted with those 
defined on lattice), one has $A_{\sigma}({\Bf k};\varepsilon) \sim 
\hbar^2 \Xi_{\sigma}({\Bf k})/(\pi^2 \vert\varepsilon\vert^3)$ for 
$\vert\varepsilon\vert \to \infty$ (see Eqs.~(239a) and (239b) in 
\cite{BF02b}), where $\Xi_{\sigma}({\Bf k})$ stands for a 
well-defined function, 
\footnote{\label{f6}
See Eq.~(227) in \protect\cite{BF02b} and Eq.~(A5) in 
\protect\cite{BF03}. By the assumption of the stability of the 
underlying GSs, $\Xi_{\sigma}({\Bf k})$ is non-negative. } 
it follows that for these systems $\int_{-E}^{\mu} {\rm d}\varepsilon\; 
\varepsilon^2\, A_{\sigma}({\Bf k};\varepsilon)$ is logarithmically 
divergent for $E\to\infty$. Consequently, for systems in which 
$\varepsilon_{\bF k}$ is unbounded from above it is necessary to 
employ a finite energy cutoff, $E$, in order for 
$\Delta\varepsilon_{{\bF k};\sigma}^{<}$ to be a meaningful quantity. 
\footnote{\label{f7}
It is evident that, by defining $\Delta\varepsilon_{{\bF k};\sigma}^{<}$
as the expectation value of, for instance, $\vert\varepsilon
-\varepsilon_{{\bF k};\sigma}^{<}\vert$, no need for the introduction
of an energy cut-off would arise, this is, however, at the expense of 
forsaking the simplicity of the expression for 
$\Delta\varepsilon_{{\bF k};\sigma}^{<}$ corresponding to the standard 
variance of $\varepsilon$.}
Thus, although for the latter systems the average value of 
$\varepsilon$, that is $\varepsilon_{{\bF k};\sigma}^{<}$, is bounded 
for $E\to\infty$, a consistent formalism for the determination of the 
position of the fundamental peak and the associated width in the 
single-particle spectral function requires introduction of a single
cut-off energy throughout. We thus define (see Eqs.~(\ref{e1}) 
and (\ref{e39}))
\begin{equation}
\label{e5}
\varepsilon_{{\bF k};\sigma}^{<}(E)
{:=} \frac{ \int_{-E}^{\mu} {\rm d}\varepsilon\;
\varepsilon\, A_{\sigma}({\Bf k};\varepsilon)}
{ \hbar\, {\sf n}_{\sigma}({\Bf k};E) },
\end{equation}
where (see Eq.~(\ref{e2}))
\begin{equation}
\label{e6}
{\sf n}_{\sigma}({\Bf k};E) {:=}
\frac{1}{\hbar} \int_{-E}^{\mu} {\rm d}\varepsilon\;
A_{\sigma}({\Bf k};\varepsilon).
\end{equation}
For the variance we consequently have
\begin{eqnarray}
\label{e7}
\Delta\varepsilon_{{\bF k};\sigma}^{<}(E)
&{:=}& 
\left(\frac{\int_{-E}^{\mu} {\rm d}\varepsilon\;
\big(\varepsilon-\varepsilon_{{\bF k};\sigma}^{<}(E)\big)^2\,
A_{\sigma}({\Bf k};\varepsilon)}
{\int_{-E}^{\mu} {\rm d}\varepsilon\;
A_{\sigma}({\Bf k};\varepsilon) } \right)^{1/2}\nonumber\\
&\equiv& 
\left(\frac{\int_{-E}^{\mu} {\rm d}\varepsilon\;
\varepsilon^2\, A_{\sigma}({\Bf k};\varepsilon)}
{\hbar\, {\sf n}_{\sigma}({\Bf k};E) }
- \big(\varepsilon_{{\bF k};\sigma}^{<}(E)\big)^2\right)^{1/2}\!\!\!\!.
\end{eqnarray}
For systems with bounded $\varepsilon_{\bF k}$, $E$ in the above 
expressions can be equated with $+\infty$. For systems with
unbounded $\varepsilon_{\bF k}$, it is necessary to set $E$ equal to 
some finite value; however, none of the main results in this paper 
crucially depends on the precise value chosen for $E$ so long as this 
value is greater than the absolute values of the energies of interest, 
specifically $\varepsilon_{{\bF k};\sigma}^{<}$. Consequently, in 
the remaining parts of this paper we leave out $E$ as the argument 
of the pertinent functions; for consistency with the results in 
\cite{BF02a,BF03} (which do not consider $\Delta\varepsilon_{{\bF k};
\sigma}^{<}$), at places we still write $-\infty$, rather than $-E$, 
in the lower boundaries of the integrals over $\varepsilon$.

It is interesting to note that according to the RHS of Eq.~(\ref{e7}) 
a smooth (or continuous) behaviour in $\int_{-E}^{\mu} 
{\rm d}\varepsilon\;\varepsilon^2\, A_{\sigma}({\Bf k};\varepsilon)$ 
and $\varepsilon_{{\bF k};\sigma}^{<}$ in the neighbourhood of a point 
${\Bf k}_{\star}$, where ${\sf n}_{\sigma}({\Bf k})$ undergoes a steep 
(or discontinuous) change, must be accompanied by a decrease or increase 
in $\Delta\varepsilon_{{\bF k};\sigma}^{<}$ (corresponding to a 
narrowing or broadening of the peak centred around 
$\varepsilon_{{\bF k}_{\star};\sigma}^{<}$) upon transposing ${\Bf k}$ 
from ${\Bf k}_{\star}^{-}$ to ${\Bf k}_{\star}^{+}$, depending on whether 
${\sf n}_{\sigma}({\Bf k}_{\star}^{-}) > 
{\sf n}_{\sigma}({\Bf k}_{\star}^{+})$ or 
${\sf n}_{\sigma}({\Bf k}_{\star}^{-}) < 
{\sf n}_{\sigma}({\Bf k}_{\star}^{+})$ 
respectively. This observation is based on the non-negativity of the 
argument of the square root function on the RHS of Eq.~(\ref{e7}) for 
{\sl all} ${\Bf k}$. In fact, in \S~3.7 we rigorously demonstrate 
that, at points where $\varepsilon_{{\bF k};\sigma}^{<}$ is continuous 
(such as the points of the underlying Fermi surface; see \S~3.4) and 
${\sf n}_{\sigma}({\Bf k})$ is discontinuous, 
$\Delta\varepsilon_{{\bF k};\sigma}^{<}$ must necessarily be 
discontinuous. In the light of this, the sudden decrease in the 
width of the MDC at ${\Bf k}={\Bf k}_{\star}$ (while transposing 
${\Bf k}$ towards the nodal Fermi point ${\Bf k}_{{\sc f};\sigma}$ 
along the nodal direction of the first Brillouin zone (1BZ)) in 
the experimental observations by Zhou {\sl et al.} \cite{XJZ03}
(corresponding to the hole concentration $x=0.063$) is indicative 
of a sharp decrease (if not discontinuity) in ${\sf n}_{\sigma}
({\Bf k})$ at ${\Bf k}={\Bf k}_{\star}$. The same applies for the 
similar observations in \cite{PVB00,AK01,AL01}.

It is important to realize that, for {\sl interacting} GSs,
$\Delta\varepsilon_{{\bF k};\sigma}^{<}$ does {\sl not} vanish 
for any ${\Bf k}$, not even for ${\Bf k}\to {\cal S}_{{\sc f};
\sigma}$. This follows from the fact that, with the exception 
of $\varepsilon =\varepsilon_{{\bF k};\sigma}^{<}$, 
$(\varepsilon-\varepsilon_{{\bF k};\sigma}^{<})^2$ is positive 
{\sl everywhere} along the $\varepsilon$ axis and that 
$A_{\sigma}({\Bf k};\varepsilon) \ge 0$, $\forall\varepsilon$. 
For non-interacting GSs, where $A_{\sigma}({\Bf k};\varepsilon)$
consists solely of a single $\delta$ function (see appendix A), 
the corresponding $\Delta\varepsilon_{{\bF k};\sigma}^{<}$ is 
naturally identically vanishing. 

In order to appreciate the significance of $\Delta\varepsilon_{{\bF k};
\sigma}^{<}$, it is appropriate to consider the following. Fermi-liquid 
metallic states are characterized by smooth quasiparticle energy 
dispersions $\varepsilon_{{\bF k};\sigma}$ in some neighbourhood 
of the underlying ${\cal S}_{{\sc f};\sigma}$ (see \S~3.1), a 
property that on general grounds one expects to be shared by 
$\varepsilon_{{\bF k};\sigma}^{<}$ (see \S~3.4). For conventional 
Fermi-liquids, ${\cal S}_{{\sc f};\sigma}$ constitutes the locus of 
the ${\Bf k}$ points in the vicinity of which the underlying 
${\sf n}_{\sigma}({\Bf k})$ undergoes its strongest variation (see 
footnote \ref{f1}). With these in mind, our above-mentioned 
statement with regard to the behaviour of 
$\Delta\varepsilon_{{\bF k};\sigma}^{<}$ is seen to be in full 
conformity with the fact that for the Fermi-liquid metallic states 
the width of the quasi-particle peak, centred at $\varepsilon=
\varepsilon_{{\bF k};\sigma}$, in the corresponding 
$A_{\sigma}({\Bf k};\varepsilon)$ also diminishes. It is important 
at this point to realize that the descend towards zero of 
$\vert{\rm Im}[\Sigma_{\sigma}({\Bf k};\varepsilon)]\vert$ (which 
in the case of the {\sl conventional} Fermi liquids is like 
$(\varepsilon_{{\bF k};\sigma}-\varepsilon_{\sc f})^2$) for 
${\Bf k}$ approaching ${\cal S}_{{\sc f};\sigma}$ does {\sl not} 
imply the same for $\Delta\varepsilon_{{\bF k};\sigma}^{<}$ (which 
otherwise would contradict our above statement with regard to 
the strict positivity of the latter quantity in the case of 
interacting GSs), for $\Delta\varepsilon_{{\bF k};\sigma}^{<}$ 
takes account of $A_{\sigma}({\Bf k};\varepsilon)$ in its entirety 
and not solely of the coherent contribution to 
$A_{\sigma}({\Bf k};\varepsilon)$ (i.e. ${\sf S}_{\sigma}^{-}
(\varepsilon)$ in Eq.~(\ref{e8}) below) which indeed has a 
vanishing width. 

In the light of the above observations, we conclude that the 
standard variance $\Delta\varepsilon_{{\bF k};\sigma}^{<}$ defined 
in Eq.~(\ref{e7}) provides a reliable representation of the width 
of the peak in the single-particle spectral function $A_{\sigma}
({\Bf k};\varepsilon)$ centred at $\varepsilon =\varepsilon_{{\bF k};
\sigma}^{<}$ for ${\Bf k}$ close to any point (which may or may not 
be a point of ${\cal S}_{{\sc f};\sigma}$) at which 
$\varepsilon_{{\bF k};\sigma}^{<}$ is continuous and the underlying 
${\sf n}_{\sigma}({\Bf k})$ undergoes a sharp or discontinuous change.

\subsection*{\bf\S~3. \sc Generalities}
\label{s3}

%\vspace{0.4cm}
\noindent{\bf 3.1. \small Some remarks on the single-particle
excitation energies}
\label{s3a}
\vspace{0.3cm}

For an arbitrary wave vector ${\Bf k}$ the quasi-particle equation, 
Eq.~(\ref{e3}), has in general {\sl no} solution.
\footnote{\label{f8}
We have in various papers elaborated on this subject matter, for which 
we refer the reader to \protect\cite{BF02b} (\S~3.4) and the references 
therein. Here and in the following when stating that, for an arbitrary 
${\Bf k}$, in general, Eq.~(\protect\ref{e3}) does not possess a 
solution, we do {\sl not} thereby consider the possible solutions 
of this equation on the non-physical Riemann sheets whose determination 
requires the knowledge of the analytically continued $\Sigma_{\sigma}
({\Bf k};\varepsilon)$ into these Riemann sheets (see, e.g.,
\cite{BF99b}). }
In spite of this, in cases where ${\Bf k}$ is located in a small
neighbourhood of a point, say ${\Bf k}_0$, where Eq.~(\ref{e3}) is 
satisfied {\sl and} the self-energy is sufficiently smooth with respect 
to ${\Bf k}$ (for $\varepsilon=\varepsilon_{{\bF k}_0;\sigma}$) and 
$\varepsilon$ (for ${\Bf k}={\Bf k}_0$), an energy dispersion 
$\varepsilon_{{\bF k};\sigma}$ is capable of being defined through 
an asymptotic series in terms of the asymptotic sequence 
$\{ ({\Bf k}-{\Bf k}_0)^{\ell}\, \| \, \ell = 0, 1,\dots\}$, with 
the zeroth-order term in this series coinciding with the exact 
solution $\varepsilon_{{\bF k}_0;\sigma}$ of Eq.~(\ref{e3}); we thus 
refer to $\varepsilon_{{\bF k};\sigma}$ as the {\sl asymptotic} 
solution of Eq.~(\ref{e3}), reflecting the fact that, for the ${\Bf k}$ 
at issue, Eq.~(\ref{e3}) has {\sl no} true solution. By the non-analytic 
nature of $\Sigma_{\sigma}({\Bf k};\varepsilon)$ at $({\Bf k},
\varepsilon) = ({\Bf k}_0,\varepsilon_{{\bF k}_0;\sigma})$ (see 
later), the aforementioned series can only be of finite order, 
that is there exists some finite integer $\ell_0$ such that the 
coefficient of the $\ell$th-order term in this series will be 
unbounded for $\ell \ge \ell_0$. Fermi-liquid metallic states are 
those for which the $\ell_0$ corresponding to all ${\Bf k}$ in the 
neighbourhood of the underlying Fermi surface ${\cal S}_{{\sc f};
\sigma}$ is strictly greater than unity 
\protect\cite{BF99a,BF02b,BF02a}; the effective mass, or the Fermi 
velocity, of the Landau `quasi-particle' at ${\Bf k}$ is related 
to the coefficient of the asymptotic term linear in $({\Bf k}-
{\Bf k}_0)$, with ${\Bf k}_0 \in {\cal S}_{{\sc f};\sigma}$; for 
anisotropic metallic GSs the latter coefficient is in general a 
non-diagonal Cartesian tensor. 

Concerning the above-mentioned non-analytic nature of 
$\Sigma_{\sigma}({\Bf k};\varepsilon)$ at $({\Bf k},\varepsilon)
=({\Bf k}_0,\varepsilon_{{\bF k}_0;\sigma})$, this follows from 
the fact that for interacting systems (specifically those in 
the thermodynamic limit) the solutions of the quasi-particle 
equation are not isolated points but accumulation points 
\protect\cite{BF99a,BF02b}; this aspect is directly related to 
the overcompleteness of the set of amplitudes of the 
single-particle excitations (the so-called Lehmann amplitudes) 
in the single-particle Hilbert space of interacting Hamiltonians 
(see \cite{BF02b}). 

\begin{figure}[t!]
\protect
\centerline{
\psfig{figure=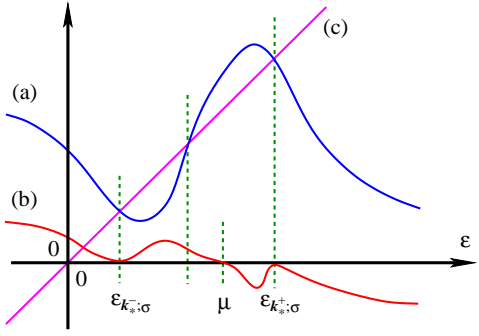,width=2.5in} }
\vskip 5pt
\caption{\label{fi1} \sf
Schematic representation of the conditions under which the
quasi-particle equation $\varepsilon_{\bF k} +\hbar \Sigma_{\sigma}
({\Bf k};\varepsilon) = \varepsilon$ has apparently two solutions 
at ${\Bf k}={\Bf k}_{\star}$, where ${\Bf k}_{\star}$ is located
strictly inside the Fermi sea FS$_{\sigma}$; the two solutions are 
denoted by $\varepsilon_{{\bF k}_{\star}^-;\sigma}$ and 
$\varepsilon_{{\bF k}_{\star}^+;\sigma}$, where 
$\varepsilon_{{\bF k}_{\star}^-;\sigma} < \mu$ and
$\varepsilon_{{\bF k}_{\star}^+;\sigma} > \mu$. The singularity of
${\sf n}_{\sigma}({\Bf k})$ at ${\Bf k}={\Bf k}_{\star}$ together 
with the normalization condition $\hbar^{-1} \int_{-\infty}^{\infty} 
{\rm d}\varepsilon\; A_{\sigma}({\Bf k};\varepsilon) =1$ imply 
that $\varepsilon_{{\bF k}_{\star}^{-};\sigma}$ and
$\varepsilon_{{\bF k}_{\star}^{+};\sigma}$ must be solutions of 
the quasi-particle equation at ${\Bf k}={\Bf k}_{\star}^{-}$ and
${\Bf k}={\Bf k}_{\star}^{+}$ respectively (see \S~3.2). Curve
(a) depicts $\varepsilon_{{\bF k}_{\star}^{\mp}} + \hbar 
{\rm Re}[\Sigma_{\sigma}({\Bf k}_{\star}^{\mp};\varepsilon)]$, curve 
(b) gives $\hbar {\rm Im}[\Sigma_{\sigma}({\Bf k}_{\star}^{\mp};
\varepsilon)]$, and curve 
(c) shows the function $f(\varepsilon) {:=} \varepsilon$. Although the 
functions are depicted schematically, care has been taken to ensure that
$\hbar {\rm Im}[\Sigma_{\sigma}({\Bf k}_{\star}^{\mp};\varepsilon)]$ 
satisfy the necessary requirements
${\rm Im}[\Sigma_{\sigma}({\Bf k}_{\star}^{\mp};\varepsilon)] \ge 0$
for $\varepsilon < \mu$,
${\rm Im}[\Sigma_{\sigma}({\Bf k}_{\star}^{\mp};\varepsilon)] \le 0$
for $\varepsilon > \mu$, and 
${\rm Im}[\Sigma_{\sigma}({\Bf k}_{\star}^{\mp};\mu)] = 0$ (the 
stability of the GS requires the more stringent condition
${\rm Im}[\Sigma_{\sigma}({\Bf k};\mu)] = 0$, $\forall {\Bf k}$).
Since ${\sf n}_{\sigma}({\Bf k})$ is assumed to be discontinuous 
at ${\Bf k}={\Bf k}_{\star}$, $\Sigma_{\sigma}({\Bf k}_{\star}^{\mp};
\varepsilon)$ is continuously differentiable with respect to
$\varepsilon$ in the neighbourhood of $\varepsilon
=\varepsilon_{{\bF k}_{\star}^{\mp};\sigma}$. Note that, since 
following the considerations in \S~3.2 for the case dealt with 
here, the spectral weights $Z_{{\bF k}_{\star}^{-};\sigma}$ and
$Z_{{\bF k}_{\star}^{+};\sigma}$ corresponding to the singular 
contributions ${\sf S}_{\sigma}^{-}(\varepsilon)$ and
${\sf S}_{\sigma}^{+}(\varepsilon)$ respectively are equal (we 
denote the common value by $Z_{{\bF k}_{\star};\sigma}$), it follows 
that ${\rm d}\Sigma_{\sigma}({\Bf k}_{\star}^-;
\varepsilon)/{\rm d}\varepsilon$ at $\varepsilon=
\varepsilon_{{\bF k}_{\star}^-;\sigma}$ is equal to
${\rm d}\Sigma_{\sigma}({\Bf k}_{\star}^+;
\varepsilon)/{\rm d}\varepsilon$ at $\varepsilon=
\varepsilon_{{\bF k}_{\star}^+;\sigma}$. Further, since in general 
$0 \le Z_{{\bF k}_{\star};\sigma} \le 1$, it follows that the latter 
derivatives are negative or at most zero (the latter corresponding
to $Z_{{\bF k}_{\star};\sigma}=1$). }
\end{figure}

In view of the specific applications in this paper, we mention that 
even for single-band models, based on a non-interacting energy 
dispersion (in this paper denoted by $\varepsilon_{\bF k}$), the 
possible solutions of Eq.~(\ref{e3}) for a given ${\Bf k}$ occur 
in pairs (for a more precise specification see footnote \ref{f9} 
below), of which one is associated with point ${\Bf k}^-$ and one
with point ${\Bf k}^+$ where ${\Bf k}^-$ and ${\Bf k}^+$ are 
infinitesimally close to ${\Bf k}$ (see Fig.~\ref{fi1}). This 
aspect, which is universally neglected in the literature for 
${\Bf k} \in {\cal S}_{{\sc f};\sigma}$ (see however \cite{BF99a} 
and specifically \cite{BF02a}; see also footnote \ref{f5} above), 
attains an extraordinary significance in cases where the 
above-mentioned ${\Bf k}$ is located at some finite distance from 
${\cal S}_{{\sc f};\sigma}$ (considering for the moment metallic 
GSs), and hence our explicit reference to $\varepsilon_{{\bF k};
\sigma}^{\mp} \equiv \varepsilon_{{\bF k}^{\mp};\sigma}$ in \S~1
(see Eq.~(\ref{e13}) below). This aspect will be fully clarified 
in \S~3.2 below.

\vspace{0.4cm}
\noindent{\bf 3.2. \small Regular and singular contributions
to the single-particle spectral function 
$A_{\sigma}({\Bf k};\varepsilon)$ }
\label{s3b}
\vspace{0.3cm}

In anticipation of our following discussions and in view of
Eq.~(\ref{e4}) it is instructive to introduce a decomposition
of $A_{\sigma}({\Bf k};\varepsilon)$ in terms of its regular
(`incoherent') and singular (`coherent') parts. To this end, let 
${\Bf k}_{\star}$ denote a wave vector at which ${\sf n}_{\sigma}
({\Bf k})$ is singular (not necessarily discontinuous). For 
${\Bf k}={\Bf k}_{\star}^{\mp}$, where ${\Bf k}_{\star}^{\mp}$ 
are infinitesimally close to ${\Bf k}_{\star}$ and whose 
distinction is implicitly determined through Eq.~(\ref{e9}) below 
and the accompanying conditions $\varepsilon_{{\bF k}_{\star}^{\mp};
\sigma}\, {\Ieq><}\, \mu$, one can write
\begin{equation}
\label{e8}
A_{\sigma}({\Bf k}_{\star}^{\mp};\varepsilon) =
{\sf R}_{\sigma}(\varepsilon) +
{\sf S}_{\sigma}^{\mp}(\varepsilon), 
\end{equation}
where ${\sf R}_{\sigma}(\varepsilon)$ (${\sf S}_{\sigma}^{\mp}
(\varepsilon)$) stands for a regular (singular) `function' of 
$\varepsilon$; for ${\Bf k}_{\star}$ coinciding with a regular 
point of ${\sf n}_{\sigma}({\Bf k})$, ${\sf S}_{\sigma}^{\mp}
(\varepsilon)$ would be identically vanishing. We point out that
the equality in Eq.~(\ref{e8}) is between two distributions
(as opposed to functions) so that it applies in some space 
of test functions. In the event that ${\sf n}_{\sigma}({\Bf k})$ 
is discontinuous at ${\Bf k}_{\star}$, in the simplest case 
\footnote{\label{f9}
In general, one may have ${\sf S}_{\sigma}^{\mp}(\varepsilon) = \hbar 
\sum_{j=1}^{M_{\sigma}^{\mp}} Z_{{\bF k}_{\star}^{\mp};\sigma}^{(j)}
\delta(\varepsilon-\varepsilon_{{\bF k}_{\star}^{\mp};\sigma}^{(j)})$ 
where, for the same reasons that $Z_{{\bF k}_{\star}^{-};\sigma} = 
Z_{{\bF k}_{\star}^{+};\sigma}$ (Eq.~(\protect\ref{e10})), 
$\{ Z_{{\bF k}_{\star}^{-};\sigma}^{(j)}\}$ and 
$\{ Z_{{\bF k}_{\star}^{+};\sigma}^{(j)}\}$ are interdependent. 
As considerations involving $M_{\sigma}^{\mp} > 1$ only unnecessarily 
complicate the analysis, in this paper we explicitly deal with
cases where $M_{\sigma}^{\mp} \le 1$. } 
one has
\begin{equation}
\label{e9}
{\sf S}_{\sigma}^{\mp}(\varepsilon) =
\hbar Z_{{\bF k}_{\star}^{\mp};\sigma}\,
\delta(\varepsilon - 
\varepsilon_{{\bF k}_{\star}^{\mp};\sigma}), 
\end{equation}
where $Z_{{\bF k}_{\star}^{\mp};\sigma} > 0$, 
$\varepsilon_{{\bF k}_{\star}^{\mp};\sigma}\, {\Ieq><}\, \mu$ 
(for the latter see the text following Eq.~(\ref{e11}) below), and
\begin{equation}
\label{e10}
Z_{{\bF k}_{\star}^-;\sigma} =
Z_{{\bF k}_{\star}^+;\sigma} \le 1.
\end{equation}
The strict equality on the left-hand side (LHS) of Eq.~(\ref{e10}) 
(whereby in the subsequent parts of this paper we denote
$Z_{{\bF k}_{\star}^{\mp};\sigma}$ by $Z_{{\bF k}_{\star};\sigma}$)
follows from the normalization condition 
\begin{equation}
\label{e11}
\frac{1}{\hbar} \int_{-\infty}^{\infty} 
{\rm d}\varepsilon\, A_{\sigma}({\Bf k};\varepsilon) = 1,
\end{equation}
which applies for {\sl all} ${\Bf k}$, including 
${\Bf k}_{\star}^-$ and ${\Bf k}_{\star}^+$. We point out that 
$\varepsilon_{{\bF k}_{\star}^{-};\sigma}$ and 
$\varepsilon_{{\bF k}_{\star}^{+};\sigma}$ satisfy Eq.~(\ref{e3}) 
for ${\Bf k}={\Bf k}_{\star}^{-}$ and ${\Bf k}={\Bf k}_{\star}^{+}$ 
respectively (see Fig.~\ref{fi1}) and, moreover, the expression 
for ${\sf S}_{\sigma}^{\mp}(\varepsilon)$ in Eq.~(\ref{e9}) is 
specific to cases where $\Sigma_{\sigma}({\Bf k}_{\star}^{\mp};
\varepsilon)$ is continuously differentiable with respect to 
$\varepsilon$ in the neighbourhood of $\varepsilon
=\varepsilon_{{\bF k}_{\star}^{\mp};\sigma}$ ({\it cf}. 
Eq.~(\ref{e80}) below); see \cite{BF99a}.

The normalization condition in Eq.~(\ref{e11}) in combination 
with the defining equation for ${\sf n}_{\sigma}({\Bf k})$ in 
Eq.~(\ref{e2}) imply that
\begin{equation}
\label{e12}
\frac{1}{\hbar} \int_{\mu}^{\infty}
{\rm d}\varepsilon\, A_{\sigma}({\Bf k};\varepsilon) = 
1 - {\sf n}_{\sigma}({\Bf k}).
\end{equation}
As is evident, the discontinuity of ${\sf n}_{\sigma}({\Bf k})$
at ${\Bf k}={\Bf k}_{\star}$ requires that 
$\varepsilon_{{\bF k}_{\star}^-;\sigma} < \mu$, for otherwise
${\sf S}_{\sigma}^{-}(\varepsilon)$ would not have any impact on 
the behaviour of ${\sf n}_{\sigma}({\Bf k})$, defined according 
to Eq.~(\ref{e2}). On the other hand, the discontinuity of 
${\sf n}_{\sigma}({\Bf k})$ at ${\Bf k}={\Bf k}_{\star}$ implies 
discontinuity of $1-{\sf n}_{\sigma}({\Bf k})$ at ${\Bf k}
={\Bf k}_{\star}$, which, following Eq.~(\ref{e12}), can only
materialize if the condition $\varepsilon_{{\bF k}_{\star}^+;\sigma} 
> \mu$ is satisfied \cite{BF02a}. 

Since for ${\Bf k}_{\star}$ located at some finite (i.e. 
non-vanishing) distance from ${\cal S}_{{\sc f};\sigma}$ the 
solutions $\varepsilon_{{\bF k}_{\star}^{-};\sigma}$ and 
$\varepsilon_{{\bF k}_{\star}^{+};\sigma}$ differ by some 
finite amount 
\footnote{\label{f10}
Since by the arguments presented in the text following 
Eq.~(\protect\ref{e12}) we have 
$\varepsilon_{{\bF k}_{\star}^-;\sigma} < \mu <
\varepsilon_{{\bF k}_{\star}^+;\sigma}$, it follows that
in the event that $\varepsilon_{{\bF k}_{\star}^-;\sigma}$
and $\varepsilon_{{\bF k}_{\star}^-;\sigma}$ would
differ only by an infinitesimal amount, ${\Bf k}_{\star}$
would either be a point of ${\cal S}_{{\sc f};\sigma}$ or 
infinitesimally close to it. }
(see \S\S~4.2 and 4.3 as well as Fig.~\ref{fi1}), it is of 
crucial importance that particular attention is given to the 
superscripts $-$ and $+$ in $\varepsilon_{{\bF k}_{\star}^{\mp};
\sigma}$ which as a matter of habit one is apt to ignore. We 
shall elaborate on the single-particle energies 
$\varepsilon_{{\bF k}^{\mp};\sigma}$ and the relationship 
between these and $\varepsilon_{{\bF k};\sigma}^{\Ieq><}$ in \S~3.5 
where we are equipped with the details necessary for this task. 
Suffice it for the moment to mention that the single-particle 
excitation energies of single-band models can be viewed as 
consisting of at least (see footnote \ref{f9}) two branches, 
conveniently denoted by $\varepsilon_{{\bF k}^{\mp};\sigma}$ with 
$\varepsilon_{{\bF k}^{-};\sigma} < \mu$ and 
$\varepsilon_{{\bF k}^{+};\sigma} > \mu$. The energy dispersions 
$\varepsilon_{{\bF k};\sigma}^{\Ieq><}$ dealt with in this paper 
are `caricatures' of $\varepsilon_{{\bF k}^{\mp};\sigma}$ which, 
however, when continuous at the points of discontinuity of 
${\sf n}_{\sigma}({\Bf k})$ (see \S~3.4), coincide with 
$\varepsilon_{{\bF k}^{\mp};\sigma}$ ({\it cf}. Eqs.~(\ref{e27}) 
and (\ref{e30}) below). From this perspective, it is more 
transparent to denote $\varepsilon_{{\bF k}^{\mp};\sigma}$ by 
$\varepsilon_{{\bF k};\sigma}^{\mp}$. In this paper we shall 
therefore alternately use the following convention:
\begin{equation}
\label{e13}
\varepsilon_{{\bF k};\sigma}^{\mp} \equiv
\varepsilon_{{\bF k}^{\mp};\sigma}, \;\;\;
\forall {\Bf k}.
\end{equation}
We emphasize that, for a general ${\Bf k}$, Eq.~(\ref{e3}) has 
no solution so that, in general $\varepsilon_{{\bF k}^{\mp};\sigma}$, 
or $\varepsilon_{{\bF k};\sigma}^{\mp}$, should be considered in 
the manner specified in \S~3.1.

From Eqs.~(\ref{e8}) and (\ref{e9}) and the condition
$\varepsilon_{{\bF k}_{\star};\sigma}^{\mp} \, {\Ieq><}\, \mu$ 
which implies ${\sf S}_{\sigma}^{\pm}(\varepsilon) \equiv 0$ for 
$\varepsilon\, {\Ieq><}\, \mu$, for cases where ${\sf n}_{\sigma}
({\Bf k})$ is discontinuous at ${\Bf k}={\Bf k}_{\star}$ one obtains
\begin{equation}
\label{e14}
{\sf R}_{\sigma}(\varepsilon) \equiv 
A_{\sigma}({\Bf k}_{\star}^{\pm};\varepsilon)\;\;\;
\mbox{\rm for}\;\;\; \varepsilon \, {\Ieq><}\, \mu. 
\end{equation}
From this result as specialized to ${\Bf k}_{\star} = 
{\Bf k}_{{\sc f};\sigma} \in {\cal S}_{{\sc f}; \sigma}$ 
and Eq.~(\ref{e4}), for interacting systems we immediately obtain
\begin{equation}
\label{e15}
\varepsilon_{\sc f} = 
\frac{ \int_{-\infty}^{\mu} {\rm d}\varepsilon\;
\varepsilon {\sf R}_{\sigma}(\varepsilon) }
{\int_{-\infty}^{\mu} {\rm d}\varepsilon\;
{\sf R}_{\sigma}(\varepsilon) } \;\;\;
\mbox{\rm for}\;\;\; {\Bf k}_{\star} \in {\cal S}_{{\sc f};\sigma}.
\end{equation}
This result is of experimental significance, as the possible
singular contribution to $A_{\sigma}({\Bf k}_{{\sc f};\sigma}^{\mp};
\varepsilon)$ is not capable of being resolved in real experiments. 
It is also of significance to our subsequent considerations as it 
shows that, for ${\Bf k}$ sufficiently close to ${\cal S}_{{\sc f};
\sigma}$, there is no restriction on the actual shape of the 
corresponding ${\sf R}_{\sigma}(\varepsilon)$ in order for the 
expression on the RHS of Eq.~(\ref{e1}) to amount to a reliable 
description of the dispersion of the low-lying single-particle 
excitations in interacting systems, this in surprising contrast 
with the heuristic arguments in \S~2 that initially motivated our 
introduction of the expression in Eq.~(\ref{e1}). For illustration, 
assuming that ${\sf n}_{\sigma}({\Bf k})$ is discontinuous at 
${\Bf k}_{\star}={\Bf k}_{{\sc f};\sigma} \in {\cal S}_{{\sc f};
\sigma}$ and denoting $Z_{{\bF k}_{\star};\sigma}$ by 
$Z_{{\bF k}_{{\sc f};\sigma}}$, from Eqs.~(\ref{e2}), (\ref{e8}) 
and (\ref{e9}) we obtain the following relationship:
\footnote{\label{f11}
With reference to our remark in footnote \protect\ref{f9}, we note 
that for cases where ${\sf S}_{\sigma}^{-}(\varepsilon)$ has the more 
general form as presented in that footnote, the counterpart of the 
result in Eq.~(\protect\ref{e16}) would in principle differ from 
that which stands in Eq.~(\protect\ref{e16}). However, the 
considerations in \S~3.6 demonstrate that the $M_{\sigma}^{\mp}$ 
corresponding to ${\Bf k}_{\star} \in {\cal S}_{{\sc f};\sigma}$ can 
at most be equal to unity. In other words, Eq.~(\protect\ref{e16}) is 
more general than the simple form for ${\sf S}_{\sigma}^{-}(\varepsilon)$ 
in Eq.~(\protect\ref{e9}) would suggest. For completeness, in 
Eq.~(\protect\ref{e16}), ${\Bf k}_{{\sc f};\sigma}^{\mp} {:=} 
{\Bf k}_{{\sc f};\sigma} \mp \kappa\, 
\hat{\Bf n}({\Bf k}_{{\sc f};\sigma})$, $\kappa\downarrow 0$, where 
$\hat{\Bf n}({\Bf k}_{{\sc f};\sigma})$ is the unit vector normal to 
${\cal S}_{{\sc f};\sigma}$ at ${\Bf k}={\Bf k}_{{\sc f};\sigma}$, 
pointing from the inside to outside the Fermi sea. }
\begin{equation}
\label{e16}
Z_{{\bF k}_{{\sc f};\sigma}} = 
{\sf n}_{\sigma}({\Bf k}_{{\sc f};\sigma}^-) -
{\sf n}_{\sigma}({\Bf k}_{{\sc f};\sigma}^+),
\end{equation}
which is a well-known Migdal \cite{ABM57,JML60} theorem. 

\vspace{0.4cm}
\noindent{\bf 3.3. \small On ratios of discontinuous functions}
\label{s3c}
\vspace{0.3cm}

A result that proves considerably useful in our considerations 
in this paper is based on some simple observations. Let 
\begin{equation}
\label{e17}
h(x) {:=} \frac{f(x)}{g(x)},
\end{equation}
and assume that $g(x)$ is discontinuous at $x=x_0$ so that 
$\Delta_g \not=0$, where 
\begin{equation}
\label{e18}
\Delta_g {:=} g(x_0^-)-g(x_0^+),
\end{equation}
in which $x_0^-$ and $x_0^+$ are infinitesimally close to $x_0 
\in (x_0^-,x_0^+)$. Assuming $h(x)$ to be continuous at $x=x_0$, 
i.e. $h(x_0^{\mp}) = h(x_0)$, one readily deduces that 
\begin{equation}
\label{e19}
\frac{\Delta_f}{\Delta_g} = h(x_0), 
\end{equation}
where 
\begin{equation}
\label{e20}
\Delta_f {:=} f(x_0^-) - f(x_0^+). 
\end{equation}
Note that, according to Eq.~(\ref{e19}), unless $h(x_0)=0$, 
discontinuity of $g(x)$ at $x=x_0$ implies that of $f(x)$ 
at $x=x_0$, that is $\Delta_f \not= 0$. A number of applications 
of the result in Eq.~(\ref{e19}), with remarkable consequences, 
will be encountered in the following sections.

Another result that will prove useful to our later analysis
concerns a situation in which $h(x)$ as defined in Eq.~(\ref{e17})
is discontinuous at $x=x_0$, similar to $f(x)$ and $g(x)$. Let
\begin{equation}
\label{e21}
\Delta_h {:=} h(x_0^-) - h(x_0^+).
\end{equation}
Through some straightforward algebra one obtains
\begin{equation}
\label{e22}
\Delta_h = \frac{ \Delta_f - h(x_0^-) \Delta_g}{g(x_0^+)}.
\end{equation}
One observes that the requirement $\Delta_h=0$ immediately
leads to the result in Eq.~(\ref{e19}).

\vspace{0.4cm}
\noindent{\bf 3.4. \small Possible discontinuities in 
$\varepsilon_{{\bF k};\sigma}^{<}$ and 
$\varepsilon_{{\bF k};\sigma}^{>}$ }
\label{s3d}
\vspace{0.3cm}

Some of the crucial results in this paper rely on the condition
of continuity of $\varepsilon_{{\bF k};\sigma}^{<}$ at such points 
as ${\Bf k}={\Bf k}_{\star}$ where ${\sf n}_{\sigma}({\Bf k})$ is 
discontinuous. Here we investigate whether the two conditions can 
be compatible. To this end we introduce $\varepsilon_{{\bF k};
\sigma}^{>}$, defined as (see Eq.~(\ref{e48}) below)
\begin{equation}
\label{e23}
\varepsilon_{{\bF k};\sigma}^{>} {:=}
\frac{\int_{\mu}^{\infty} {\rm d}\varepsilon\, 
\varepsilon A_{\sigma}({\Bf k};\varepsilon)}
{\int_{\mu}^{\infty} {\rm d}\varepsilon\,
A_{\sigma}({\Bf k};\varepsilon)}.
\end{equation}
It can be easily shown that 
$\varepsilon_{{\bF k};\sigma}^{<}$ and $\varepsilon_{{\bF k};
\sigma}^{>}$ satisfy the following exact relationship 
\cite{BF02a,BF03}
\begin{equation}
\label{e24}
{\sf n}_{\sigma}({\Bf k}) \varepsilon_{{\bF k};\sigma}^{<}
+ \big(1-{\sf n}_{\sigma}({\Bf k})\big)
\varepsilon_{{\bF k};\sigma}^{>} = 
\varepsilon_{\bF k} + \hbar\Sigma_{\sigma}^{\sc hf}({\Bf k}),
\;\;\; \forall {\Bf k},
\end{equation}
where $\Sigma_{\sigma}^{\sc hf}({\Bf k})$ stands for the
Hartree-Fock self-energy.
Let now ${\sf n}_{\sigma}({\Bf k})$ be discontinuous at
${\Bf k}={\Bf k}_{\star}$. With 
\begin{equation}
\label{e25}
Z_{{\bF k}_{\star};\sigma} {:=} 
{\sf n}_{\sigma}({\Bf k}_{\star}^-) -
{\sf n}_{\sigma}({\Bf k}_{\star}^+),
\end{equation}
where ${\Bf k}_{\star}^+$ is the nearest of the two points 
${\Bf k}_{\star}^-$ and ${\Bf k}_{\star}^+$ to ${\cal S}_{{\sc f};
\sigma}$ and (by assumption) $Z_{{\bF k}_{\star};\sigma} > 0$, 
from Eq.~(\ref{e24}) we readily obtain
\begin{eqnarray}
\label{e26}
&&Z_{{\bF k}_{\star};\sigma} \big(
\varepsilon_{{\bF k}_{\star}^-;\sigma}^{>} -
\varepsilon_{{\bF k}_{\star}^-;\sigma}^{<} \big)
= {\sf n}_{\sigma}({\Bf k}_{\star}^+)
\big( \varepsilon_{{\bF k}_{\star}^-;\sigma}^{<} -
\varepsilon_{{\bF k}_{\star}^+;\sigma}^{<} \big) \nonumber\\
&&\;\;\;\;\;\;\;\;\;\;\;\;\;\;\;\;\;\;\;\;\;\;\;\;\;\;
+\big(1 - {\sf n}_{\sigma}({\Bf k}_{\star}^+) \big)
\big( \varepsilon_{{\bF k}_{\star}^-;\sigma}^{>} -
\varepsilon_{{\bF k}_{\star}^+;\sigma}^{>} \big).
\end{eqnarray}
For metallic states, where $\varepsilon_{{\bF k}_{\star}^-;
\sigma}^{>}$ and $\varepsilon_{{\bF k}_{\star}^-;\sigma}^{<}$ 
are up to infinitesimal corrections equal to $\varepsilon_{\sc f}$
for ${\Bf k}_{\star} = {\Bf k}_{{\sc f};\sigma} \in
{\cal S}_{{\sc f};\sigma}$ \cite{BF02a,BF03}, the LHS of 
Eq.~(\ref{e26}) is vanishing for ${\Bf k}_{\star} \in 
{\cal S}_{{\sc f};\sigma}$ so that, by the arguments presented in 
\cite{BF02a}, Eq.~(\ref{e26}) establishes the continuity of 
$\varepsilon_{{\bF k};\sigma}^{<}$ and 
$\varepsilon_{{\bF k};\sigma}^{>}$ at ${\Bf k} = {\Bf k}_{\star} 
\in {\cal S}_{{\sc f};\sigma}$ (see also \S~4.1). 

For ${\Bf k}_{\star}$ at a finite distance from ${\cal S}_{{\sc f};
\sigma}$, the LHS of Eq.~(\ref{e26}) is strictly positive 
(see footnote \ref{f10}), from which and from Eq.~(\ref{e26}) it 
follows that {\sl at least} one of the two functions 
$\varepsilon_{{\bF k};\sigma}^{<}$ and $\varepsilon_{{\bF k};
\sigma}^{>}$ must be discontinuous at ${\Bf k}={\Bf k}_{\star}$. 
For ${\Bf k}_{\star}$ sufficiently close to ${\cal S}_{{\sc f};\sigma}$, 
on account of the aforementioned continuity of $\varepsilon_{{\bF k};
\sigma}^{<}$ and $\varepsilon_{{\bF k};\sigma}^{>}$ at all points of 
${\cal S}_{{\sc f};\sigma}$ and on account of the fact that these 
functions are required to achieve the value $\varepsilon_{\sc f}$ at 
the latter points, it is most likely that, for a possible discontinuity 
of $\varepsilon_{{\bF k};\sigma}^{<}$ at ${\Bf k}={\Bf k}_{\star}$,
one has $\varepsilon_{{\bF k}_{\star}^-;\sigma}^{<} <
\varepsilon_{{\bF k}_{\star}^+;\sigma}^{<}$ and that, for a possible 
discontinuity of $\varepsilon_{{\bF k};\sigma}^{>}$ at ${\Bf k}
={\Bf k}_{\star}$, $\varepsilon_{{\bF k}_{\star}^-;\sigma}^{>} >
\varepsilon_{{\bF k}_{\star}^+;\sigma}^{>}$. Raising the status of 
these observations to facts, an inspection of the signs of the terms 
on both sides of Eq.~(\ref{e26}) reveals that the possibility of 
a continuous $\varepsilon_{{\bF k};\sigma}^{>}$ and discontinuous 
$\varepsilon_{{\bF k};\sigma}^{<}$ at ${\Bf k}={\Bf k}_{\star}$ 
is ruled out; in contrast, a similar inspection reveals that a 
continuous $\varepsilon_{{\bF k};\sigma}^{<}$ and discontinuous 
$\varepsilon_{{\bF k};\sigma}^{>}$ at ${\Bf k}={\Bf k}_{\star}$ is 
potentially feasible. On the same grounds, it is in principle 
possible that both $\varepsilon_{{\bF k};\sigma}^{<}$ and 
$\varepsilon_{{\bF k};\sigma}^{>}$ are discontinuous at ${\Bf k}
={\Bf k}_{\star}$ although, in such an event, $\varepsilon_{{\bF k};
\sigma}^{>}$ is required to undergo a larger amount of discontinuity 
(as implied by Eq.~(\ref{e26})) than is necessary for the case 
in which $\varepsilon_{{\bF k};\sigma}^{<}$ is continuous at 
${\Bf k}={\Bf k}_{\star}$.

It is interesting to point out that the energy dispersion as 
measured by Zhou {\sl et al.} \cite{XJZ03} corresponding to 
(La$_{2-x}$Sr$_x$)CuO$_4$ in the extreme underdoped limit, where 
$x=0.03$, appears to exhibit a finite amount of discontinuity at 
${\Bf k}={\Bf k}_{\star}$. On identifying the measured energy 
dispersion with $\varepsilon_{{\bF k};\sigma}^{<}$, it is observed
that it indeed satisfies the relationship 
$\varepsilon_{{\bF k}_{\star}^-;\sigma}^{<} < 
\varepsilon_{{\bF k}_{\star}^+;\sigma}^{<}$ suggested above on
general grounds.

\vspace{0.4cm}
\noindent{\bf 3.5. \small On the points where 
$\varepsilon_{{\bF k};\sigma}^- = 
\varepsilon_{{\bF k};\sigma}^{<}$ 
(and $\varepsilon_{{\bF k};\sigma}^+ = 
\varepsilon_{{\bF k};\sigma}^{>}$)}
\label{s3e}
\vspace{0.3cm}

In our above considerations we have been able to deduce a number 
of properties associated with $\varepsilon=\varepsilon_{\sc f}$ and 
${\Bf k}={\Bf k}_{{\sc f};\sigma} \in {\cal S}_{{\sc f};\sigma}$
by specifically relying on the property of continuity of 
$\varepsilon_{{\bF k};\sigma}^{<}$ at ${\Bf k}= {\Bf k}_{{\sc f};
\sigma} \in {\cal S}_{{\sc f};\sigma}$ (see \S~3.4). Here we 
explicitly assume $\varepsilon_{{\bF k};\sigma}^{<}$ to be 
continuous also at ${\Bf k} \not\in {\cal S}_{{\sc f};\sigma}$, 
specifically at ${\Bf k} = {\Bf k}_{\star}$ where ${\sf n}_{\sigma}
({\Bf k})$ is discontinuous; following the analysis in \S~3.4,
$\varepsilon_{{\bF k};\sigma}^{>}$ must therefore by necessity be 
discontinuous at ${\Bf k}={\Bf k}_{\star}$. Let the latter 
${\Bf k}_{\star}$ be further located inside the Fermi sea of the GS 
of the system under consideration; for the ensuing discussions it is 
{\sl not} necessary that this GS be metallic. The general expression 
for $\varepsilon_{{\bF k};\sigma}^{<}$ can be cast into the form 
$\Delta_f/\Delta_g = h(x_0)$ (see Eq.~(\ref{e19}) above), with $x_0$ 
here identified with ${\Bf k}_{\star}$, $f$ with the numerator of 
the expression on the RHS of Eq.~(\ref{e1}) and $g$ with the 
denominator (and thus $h$ with $\varepsilon_{{\bF k};\sigma}^{<}$). 
From this and Eqs.~(\ref{e8}) and (\ref{e9}) it follows that 
$\Delta_f$ is to be identified with $Z_{{\bF k}_{\star};\sigma} 
\varepsilon_{{\bF k}_{\star};\sigma}^-$ and $\Delta_g$ with
$Z_{{\bF k}_{\star};\sigma}$, so that by Eq.~(\ref{e19}) we have 
the following exact result:
\footnote{\label{f12}
With reference to our remark in footnote \protect\ref{f9}, we 
point out that Eq.~(\protect\ref{e27}) is specific to cases in 
which $M_{\sigma}^{\mp}=1$. }
\begin{equation}
\label{e27}
\varepsilon_{{\bF k}_{\star};\sigma}^{<} = 
\varepsilon_{{\bF k}_{\star};\sigma}^-,
\end{equation}
where $\varepsilon_{{\bF k}_{\star};\sigma}^{-}$ is the exact 
solution of Eq.~(\ref{e3}) and, as mentioned above, ${\Bf k}_{\star}$ 
a point of discontinuity of ${\sf n}_{\sigma}({\Bf k})$ at which,
moreover, $\varepsilon_{{\bF k};\sigma}^{<}$ is explicitly assumed to 
be continuous. Equation (\ref{e27}) is to be contrasted with the more 
specific exact result in Eq.~(\ref{e4}) above. In this connection we 
should point out that the validity of Eq.~(\ref{e4}) does {\sl not} 
depend on whether ${\sf n}_{\sigma}({\Bf k})$ is discontinuous at 
${\Bf k}={\Bf k}_{{\sc f};\sigma} \in {\cal S}_{{\sc f};\sigma}$. 
This suggests that the validity of Eq.~(\ref{e27}) may similarly 
not depend on the assumed discontinuity (as opposed to the more 
general condition of singularity) of ${\sf n}_{\sigma}({\Bf k})$ 
at the ${\Bf k}={\Bf k}_{\star}$ under consideration; at the time 
of writing this paper, we are not in a position to make a definite 
statement on this aspect. In this connection note that in general 
${\sf n}_{\sigma}({\Bf k})$ is singular (not necessarily 
discontinuous) for {\sl all} ${\Bf k} \in 
{\cal S}_{{\sc f};\sigma}$ \cite{BF02a,BF03}.

It is interesting to note that in cases where $\varepsilon_{{\bF k};
\sigma}^{<}$ is, similar to ${\sf n}_{\sigma}({\Bf k})$, discontinuous 
at ${\Bf k}={\Bf k}_{\star}$, application of Eq.~(\ref{e22}) followed 
by some trivial algebra yields the following exact expression
\begin{equation}
\label{e28}
\varepsilon_{{\bF k}_{\star};\sigma}^{-} 
= \frac{1}{Z_{{\bF k}_{\star};\sigma}} 
\Big( {\sf n}_{\sigma}({\Bf k}_{\star}^-)\,
\varepsilon_{{\bF k}_{\star}^-;\sigma}^{<} -
{\sf n}_{\sigma}({\Bf k}_{\star}^+)\,
\varepsilon_{{\bF k}_{\star}^+;\sigma}^{<} \Big),
\end{equation}
from which one recovers the result in Eq.~(\ref{e27}) for
$\varepsilon_{{\bF k}_{\star}^-;\sigma}^{<}$ approaching
$\varepsilon_{{\bF k}_{\star}^+;\sigma}^{<}$. In cases where 
${\sf n}_{\sigma}({\Bf k}_{\star}^-)/Z_{{\bF k}_{\star};\sigma}
\approx 1$, implying thus ${\sf n}_{\sigma}({\Bf k}_{\star}^+)
\approx 0$, from Eq.~(\ref{e28}) it follows that 
$\varepsilon_{{\bF k}_{\star};\sigma}^{-} \approx 
\varepsilon_{{\bF k}_{\star}^-;\sigma}^{<}$.

In the light of the results in Eqs.~(\ref{e4}) and (\ref{e27}), it 
is important to point out that, for Fermi-liquid metallic states
of the single-band Hubbard Hamiltonian (if such states at all
exist), we have shown \cite{BF02a} that the gradient of the
quasi-particle energy dispersion $\varepsilon_{{\bF k};\sigma}^{-}$
(or the Fermi velocity times $\hbar$) for ${\Bf k}$ inside the 
Fermi sea and in the close neighbourhood of ${\cal S}_{{\sc f};
\sigma}$ differs from ${\Bf\nabla}_{\bF k} \varepsilon_{{\bF k};
\sigma}^{<}$ by a multiplicative constant, $\lambda$, which we 
have shown \cite{BF02a} to approach unity
\footnote{\label{f13}
In \protect\cite{BF02a} for isotropic Fermi-liquid metallic states of
fermions of bare mass $m_{\rm e}$ interacting through a short-range
potential we have obtained $\lambda = (m_{\sigma}^*/m_{\rm e})
{\sf n}_{\sigma}(k_{{\sc f};\sigma}^-)/[2 {\sf n}_{\sigma}(k_{{\sc
f};\sigma}^-) - 1]$, where $m_{\sigma}^*$ stands for the
renormalized (or effective) mass. For the sole purpose of illustration,
making the assumption that ${\sf n}_{\sigma}(k_{{\sc f};\sigma}^-)
\approx (1+Z_{k_{{\sc f};\sigma}})/2$ and employing the data concerning
$Z_{k_{{\sc f};\sigma}}$ and $m_{\rm e}/m_{\sigma}^*$, corresponding 
to the {\sl Coulomb-interacting} homogeneous electron-gas system, as
presented in respectively Tables 5.6 and 5.7 of the book by Mahan
\protect\cite{GDM90}, from $\lambda \approx m_{\sigma}^*
(1 + Z_{k_{{\sc f};\sigma}})/(2 m_{\rm e} Z_{k_{{\sc f};\sigma}})$
for $(r_{\rm s};\lambda)$ we obtain: $(0;1)$ [exact], $(1;1.12)$,
$(2;1.16)$, $(3;1.19)$ and $(4;1.22)$. Here $r_{\rm s}$ stands for
the dimensionless Wigner-Seitz density parameter, with $r_{\rm s}
\to 0$ corresponding to the uncorrelated limit. }
on reducing the on-site interaction energy $U$. It follows that in 
general the validity of (\ref{e27}) does not extend beyond the 
points ${\Bf k}$ where ${\sf n}_{\sigma}({\Bf k})$ is discontinuous; 
in other words, in general the asymptotic series expansions of 
$\varepsilon_{{\bF k};\sigma}^-$ and $\varepsilon_{{\bF k};
\sigma}^{<}$ centred around a ${\Bf k}$, say ${\Bf k}_{\star}$, 
at which ${\sf n}_{\sigma}({\Bf k})$ is discontinuous (so that
$\varepsilon=\varepsilon_{{\bF k}_{\star};\sigma}^{-}$ satisfies 
Eq.~(\ref{e3}) at ${\Bf k}={\Bf k}_{\star}^-$; see \S~3.1), only 
agree to the leading order and deviate at higher orders in 
$({\Bf k}-{\Bf k}_{\star})$; analogously for $\varepsilon_{{\bF k};
\sigma}^+$ and $\varepsilon_{{\bF k};\sigma}^{>}$ in the event that 
$\varepsilon_{{\bF k};\sigma}^{>}$ is continuous at ${\Bf k}=
{\Bf k}_{\star}$ (see however the last but one paragraph in
\S~3.4). This aspect becomes clearly evident by considering 
the two-body interaction potential to be the long-range Coulomb 
potential. In this case, ${\Bf\nabla}_{\bF k} \varepsilon_{{\bF k};
\sigma}^{<}$ is logarithmically divergent for ${\Bf k}$ approaching 
the points of discontinuity of ${\sf n}_{\sigma}({\Bf k})$ (see \S~4; 
see also \cite{BF03}), a property that may or may not be shared by 
${\Bf\nabla}_{\bF k} \varepsilon_{{\bF k};\sigma}^{\mp}$; for 
instance, for Fermi-liquid metallic states, ${\Bf\nabla}_{\bF k} 
\varepsilon_{{\bF k};\sigma}^{\mp}$ is by definition \cite{BF99a} 
bounded, satisfying 
\begin{equation}
\label{e29}
\left. {\Bf\nabla}_{\bF k} \varepsilon_{{\bF k};
\sigma}^{-}\right|_{{\bF k}={\bF k}_{{\sc f};\sigma}^-} =
\left. {\Bf\nabla}_{\bF k} \varepsilon_{{\bF k};
\sigma}^{+}\right|_{{\bF k}={\bF k}_{{\sc f};\sigma}^+}
\;\;\;\mbox{\rm for}\;\;\;
{\Bf k}_{{\sc f};\sigma}\in {\cal S}_{{\sc f};\sigma}. 
\end{equation}

The fact that the two important ${\Bf k}$ points that feature in 
the experimental observations as reported in
\cite{XJZ03,TV99,PVB00,PDJ01,AK01,AL01}, namely the nodal Fermi 
point ${\Bf k}_{{\sc f};\sigma}$ and ${\Bf k}_{\star}$, are 
relatively very close to one another (typically, 
$\|{\Bf k}_{{\sc f};\sigma}-{\Bf k}_{\star}\|$ amounts to approximately 
$5$\% of $\|{\Bf k}_{{\sc f};\sigma}\|$), implies, following 
Eqs.~(\ref{e4}) and (\ref{e27}), that the fundamental distinction between 
$\varepsilon_{{\bF k};\sigma}^-$ and $\varepsilon_{{\bF k};\sigma}^{<}$
must be of minor observational consequence for ${\Bf k}$ inside the 
interval between ${\Bf k}_{\star}$ and ${\Bf k}_{{\sc f};\sigma}$ 
along the diagonal direction of the 1BZ compared with outside. In other 
words, in the light of Eqs.~(\ref{e4}) and (\ref{e27}), it is expected 
that, for ${\Bf k}$ inside the latter interval (between 
${\Bf k}_{\star}$ and ${\Bf k}_{{\sc f};\sigma}$), 
$\varepsilon_{{\bF k};\sigma}^-$ and $\varepsilon_{{\bF k};\sigma}^{<}$ 
should to a good approximation coincide. In the specific case at hand, 
where one of our main conclusions drawn from the experimental 
observations in \cite{XJZ03,TV99,PVB00,PDJ01,AK01,AL01}, specifically 
those by Zhou {\sl et al.} \cite{XJZ03}, is that the underlying 
${\sf n}_{\sigma}({\Bf k})$ is continuous at the nodal points of 
the Fermi surfaces of the compounds studied (see \S~4.1), the 
above-mentioned differences between $\varepsilon_{{\bF k};\sigma}^-$ 
and $\varepsilon_{{\bF k};\sigma}^{<}$ should be minimal for 
${\Bf k}$ in the close neighbourhoods of the nodal Fermi points.

Concerning the `universality' in the nodal Fermi velocities as observed 
by Zhou {\sl et al.} \cite{XJZ03}, since in all cases (whether 
different cuprate compounds are concerned or a specific compound at 
different levels of hole doping) the corresponding Fermi energies are
chosen as the origin of energy, it follows that the above-mentioned 
universality of the nodal Fermi velocities is indicative of the 
universality of the energy difference $\varepsilon_{\sc f}-
\varepsilon_{\star}$, where $\varepsilon_{\star} {:=} 
\varepsilon_{{\bF k}_{\star};\sigma}^- \equiv 
\varepsilon_{{\bF k}_{\star};\sigma}^{<}$. This in turn is 
suggestive of the possibility that the root cause of the discontinuity 
in ${\sf n}_{\sigma}({\Bf k})$ at ${\Bf k}={\Bf k}_{\star}$ should
be external to the electronic degrees of freedom, such as longitudinal 
optical phonons which have been suggested \cite{PVB00,AL01} as 
inducing the `kink', or `break', in the observed single-particle 
energy dispersions in the investigated cuprates. For an extensive 
discussion of this aspect see \S~5 where we argue that, whereas phonons 
may be vital in bringing about the observed `kinks' in the energy 
dispersions, they {\sl cannot} be the immediate cause but possibly a 
secondary determinant through their modification of the electronic 
states in such a way that ${\sf n}_{\sigma}({\Bf k})$ is rendered
discontinuous at ${\Bf k}={\Bf k}_{\star}$ inside the underlying Fermi 
seas and that $\varepsilon_{\sc f}-\varepsilon_{\star}$ is close to 
$50$ to $70$ meV. It is important to point out that, with the exception 
of {\sl rigid} parabolic bands, all {\sl rigid} tight-binding bands 
give rise to relatively strong variation of the bare mass of fermions 
as a function of doping; such variation has undoubtedly consequences 
for the corresponding dressed mass and thus the interacting Fermi 
velocity which, in such cases as considered here, can be artificial 
and undesirable. The experimental observations by Zhou {\sl et al.} 
\cite{XJZ03} are thus suggestive of the possibility that rigid 
tight-binding bands may not be appropriate (specifically) while 
dealing with the cuprate compounds. The unsuitability of rigid 
tight-binding bands (as well as rigid on-site interaction energies) 
in general is the essential argument underlying the dynamic Hubbard 
model proposed by Hirsch \cite{JEH01} (see also \cite{JEH03} and the 
references therein) and their shortcomings specifically in 
applications concerning the cuprate compounds has been explicitly 
emphasized in \cite{BF02a}.

For completeness, in cases in which $\varepsilon_{{\bF k};\sigma}^{>}$ 
is continuous at ${\Bf k}={\Bf k}_{\star}$ where ${\sf n}_{\sigma}
({\Bf k})$ is discontinuous (so that, by the observations in \S~3.4, 
$\varepsilon_{{\bF k};\sigma}^{<}$ must necessarily be discontinuous 
at ${\Bf k}={\Bf k}_{\star}$; see however the last but one paragraph
in \S~3.4), from Eq.~(\ref{e23}) and the application of the result 
in Eq.~(\ref{e19}) we have (see Eqs.~(\ref{e13}) and (\ref{e27}))
\begin{equation}
\label{e30}
\varepsilon_{{\bF k}_{\star};\sigma}^{>} = 
\varepsilon_{{\bF k}_{\star};\sigma}^+.
\end{equation}
By relaxing the condition of the continuity of 
$\varepsilon_{{\bF k};\sigma}^{>}$ at ${\Bf k}={\Bf k}_{\star}$,
along the same lines as those leading to Eq.~(\ref{e28}) we obtain
\begin{equation}
\label{e31}
\varepsilon_{{\bF k}_{\star};\sigma}^{+} 
= \frac{1}{Z_{{\bF k}_{\star};\sigma}} 
\Big( {\sf n}_{\sigma}({\Bf k}_{\star}^-)\,
\varepsilon_{{\bF k}_{\star}^-;\sigma}^{>} -
{\sf n}_{\sigma}({\Bf k}_{\star}^+)\,
\varepsilon_{{\bF k}_{\star}^+;\sigma}^{>} \Big),
\end{equation}
from which Eq.~(\ref{e30}) is recovered for
$\varepsilon_{{\bF k}_{\star}^-;\sigma}^{>}$ approaching
$\varepsilon_{{\bF k}_{\star}^+;\sigma}^{>}$. In cases where 
${\sf n}_{\sigma}({\Bf k}_{\star}^-)/Z_{{\bF k}_{\star};\sigma}
\approx 1$, implying thus ${\sf n}_{\sigma}({\Bf k}_{\star}^+)
\approx 0$, from Eq.~(\ref{e31}) it follows that 
$\varepsilon_{{\bF k}_{\star};\sigma}^{+} \approx 
\varepsilon_{{\bF k}_{\star}^-;\sigma}^{>}$.

The results in Eqs.~(\ref{e27}) and (\ref{e30}), which cannot 
simultaneously hold (see \S~3.4), gain special significance by 
realizing the fact that $\varepsilon_{{\bF k};\sigma}^{<}$ and 
$\varepsilon_{{\bF k};\sigma}^{>}$ correspond to a single-particle 
spectral function ${\cal A}_{\sigma}({\Bf k};\varepsilon)$ 
introduced in \cite{BF02a,BF03}, defined for all ${\Bf k}$ as 
follows
\begin{equation}
\label{e32}
{\cal A}_{\sigma}({\Bf k};\varepsilon) 
{:=} \hbar \big\{ {\sf n}_{\sigma}({\Bf k})\,
\delta(\varepsilon-\varepsilon_{{\bF k};\sigma}^{<})
+ [ 1 - {\sf n}_{\sigma}({\Bf k})]\,
\delta(\varepsilon-\varepsilon_{{\bF k};\sigma}^{>}) \big\}.
\end{equation}
In \cite{BF02a,BF03} a number of salient properties of 
${\cal A}_{\sigma}({\Bf k};\varepsilon)$ have been explicitly 
demonstrated. For instance, on replacing $A_{\sigma}({\Bf k};
\varepsilon)$ by ${\cal A}_{\sigma}({\Bf k};\varepsilon)$ in 
the expression for the total energy of the GS of the interacting 
Hamiltonian $\wh{H}$, one obtains exactly the same value, that is 
the exact GS total energy. Although in \cite{BF02a,BF03} 
${\cal A}_{\sigma}({\Bf k};\varepsilon)$ was associated with 
some fictitious particles, having some direct significance to the
properties of the single-particle excitations in the neighbourhood 
of ${\cal S}_{{\sc f};\sigma}$, Eqs.~(\ref{e27}) and (\ref{e30}) 
unequivocally show that ${\cal A}_{\sigma}({\Bf k};\varepsilon)$ 
has special significance also for all ${\Bf k}$ outside 
${\cal S}_{{\sc f};\sigma}$ where ${\sf n}_{\sigma}({\Bf k})$
is discontinuous (as we have indicated earlier, this significance
may also extend to all points outside ${\cal S}_{{\sc f};\sigma}$
where ${\sf n}_{\sigma}({\Bf k})$ is merely singular and not 
specifically discontinuous). With this in mind, Eq.~(\ref{e32}) 
shows that, at such points as ${\Bf k}={\Bf k}_{\star}$ where
${\sf n}_{\sigma}({\Bf k})$ undergoes a discontinuous change, this 
is accompanied by an equally discontinuous redistribution of the 
spectral weights of the peaks in the single-particle spectral
function centred at $\varepsilon = \varepsilon_{{\bF k};\sigma}^{-}$ 
and $\varepsilon=\varepsilon_{{\bF k};\sigma}^{+}$ (see 
Eqs.~(\ref{e27}), (\ref{e28}), (\ref{e30}) and (\ref{e31})). 
Concerning the general subject of the spectral weight redistribution 
in interacting systems as described by the Hubbard Hamiltonian, we 
refer the reader to \cite{HL67}, and for related considerations 
concerning specific strongly correlated systems to 
\cite{EMS91,EO94,EOMS94}. We emphasize once more that Eqs.~(\ref{e27}) 
and (\ref{e30}) are local (i.e. they apply to isolated points, such 
as ${\Bf k}_{\star}$ is, and not to open domains of the ${\Bf k}$ 
space) so that they do {\sl not} imply equality of 
${\Bf\nabla}_{\bF k} \varepsilon_{{\bF k};\sigma}^{\mp}$ with 
${\Bf\nabla}_{\bF k} \varepsilon_{{\bF k};\sigma}^{\Ieq><}$ at 
${\Bf k}={\Bf k}_{\star}$.

\vspace{0.4cm}
\noindent{\bf 3.6. \small On the number of solutions of the 
quasiparticle equation at ${\Bf k} = {\Bf k}_{{\sc f};\sigma}^{\mp}$ 
corresponding to well-defined quasiparticles }
\label{s3f}
\vspace{0.3cm}

In \S~3.4 we have presented details which rigorously demonstrate 
the continuity of $\varepsilon_{{\bF k};\sigma}^{<}$ for ${\Bf k}$ 
in the neighbourhood of ${\cal S}_{{\sc f};\sigma}$ (see also \S~4). 
A corollary to this result in conjunction with Eq.~(\ref{e19}) is 
that, in cases where the ${\sf S}_{\sigma}^{-}(\varepsilon)$ 
corresponding to ${\Bf k}_{\star}={\Bf k}_{{\sc f};\sigma}^{-}$ is 
not identically vanishing, one must have 
\begin{equation}
\label{e33}
\varepsilon_{\sc f} = 
\frac{ \int_{-\infty}^{\mu} {\rm d}\varepsilon\;
\varepsilon {\sf S}_{\sigma}^{-}(\varepsilon) }
{ \int_{-\infty}^{\mu} {\rm d}\varepsilon\;
{\sf S}_{\sigma}^{-}(\varepsilon) }.
\end{equation}
This result establishes that, for ${\Bf k}_{\star}={\Bf k}_{{\sc f};
\sigma}^{-}$, Eq.~(\ref{e3}) cannot possess more than one solution 
corresponding to a well-defined quasi-particle; that is to say, in 
cases where ${\sf n}_{\sigma}({\Bf k})$ is discontinuous at ${\Bf k}
={\Bf k}_{\star}$ and ${\Bf k}_{\star} \in {\cal S}_{{\sc f};\sigma}$, 
the expression for ${\sf S}_{\sigma}^{\mp}(\varepsilon)$ in 
Eq.~(\ref{e9}) is the most general expression.
\footnote{\label{f14} 
With reference to our remark in footnote \protect\ref{f9}, this
implies that, for ${\Bf k}_{\star} \in {\cal S}_{{\sc f};\sigma}$, 
$M_{\sigma}^{\mp} \le 1$. }
The proof of this statement is as follows. Let ${\sf S}_{\sigma}^{-}
(\varepsilon) = \hbar Z_1 \delta(\varepsilon-\varepsilon_1) +
\hbar Z_2 \delta(\varepsilon-\varepsilon_2)$, where $\varepsilon_1, 
\varepsilon_2 < \mu$ and $Z_1 + Z_2 \le 1$ (the following arguments 
can be trivially generalized for any number of similar terms 
contributing to ${\sf S}_{\sigma}^{-}(\varepsilon)$). From 
Eq.~(\ref{e33}) it follows that 
\begin{eqnarray}
\varepsilon_{\sc f} = 
\frac{ \varepsilon_1 Z_1 + \varepsilon_2 Z_2 }{Z_1 + Z_2}, 
\nonumber
\end{eqnarray}
which, through the fact that in the case under consideration 
$\varepsilon_1 = \varepsilon_{\sc f}$ (or $\varepsilon_2 = 
\varepsilon_{\sc f}$), implies that $Z_2 = 0$ ($Z_1=0$). This 
proves our above assertion.  In other words, under the 
above-mentioned conditions, Eq.~(\ref{e3}) has at most one solution 
at ${\Bf k}_{\star}^{-}$ (and similarly for ${\Bf k}_{\star}^{+}$) 
corresponding to a well-defined quasiparticle.

\vspace{0.4cm}
\noindent{\bf 3.7. \small On the positivity and the possible
discontinuities of $\Delta\varepsilon_{{\bF k};\sigma}^{<}$ }
\label{s3g}
\vspace{0.3cm}

The continuity of $\varepsilon_{{\bF k};\sigma}^{<}$ at a point, 
say ${\Bf k}_{\star}$, implies that, for $\Delta\varepsilon_{{\bF k};
\sigma}^{<}$ to be continuous at this point, it is necessary and
sufficient that the first term in the second expression on the RHS 
of Eq.~(\ref{e7}) be continuous at ${\Bf k}={\Bf k}_{\star}$. The 
continuity of the latter term at ${\Bf k}={\Bf k}_{\star}$ in cases 
where ${\sf n}_{\sigma}({\Bf k})$ is discontinuous at ${\Bf k}
={\Bf k}_{\star}$, following the results in Eqs.~(\ref{e19}) and 
(\ref{e27}), implies that, for $\Delta\varepsilon_{{\bF k};
\sigma}^{<}$ to be continuous at ${\Bf k}={\Bf k}_{\star}$, one 
must have
\begin{equation}
\label{e34}
(\Delta\varepsilon_{{\bF k}_{\star};\sigma}^{<})^2
= \frac{ Z_{{\bF k}_{\star};\sigma} 
(\varepsilon_{{\bF k}_{\star};\sigma}^-)^2 }
{ Z_{{\bF k}_{\star};\sigma} }
- (\varepsilon_{{\bF k}_{\star};\sigma}^-)^2 \equiv 0.
\end{equation}
This result is in contradiction with the fact that, for interacting 
systems, $\Delta\varepsilon_{{\bF k};\sigma}^{<} > 0$ (see \S~2). 
It thus follows that our assumption with regard to the continuity 
of both $\varepsilon_{{\bF k};\sigma}^{<}$ and 
$\Delta\varepsilon_{{\bF k};\sigma}^{<}$ at ${\Bf k}={\Bf k}_{\star}$, 
where ${\sf n}_{\sigma}({\Bf k})$ is discontinuous, must be false. 
We thus arrive at the general conclusion that, at points (which may 
or may not be points of ${\cal S}_{{\sc f};\sigma}$) where 
${\sf n}_{\sigma}({\Bf k})$ is discontinuous, $\varepsilon_{{\bF k};
\sigma}^{<}$ and $\Delta\varepsilon_{{\bF k};\sigma}^{<}$ 
cannot both be continuous; at ${\Bf k}={\Bf k}_{{\sc f};\sigma}
\in {\cal S}_{{\sc f};\sigma}$ where $\varepsilon_{{\bF k};\sigma}^{<}$ 
is continuous (see \S~3.4), $\Delta\varepsilon_{{\bF k};\sigma}^{<}$ 
must be discontinuous in cases where ${\sf n}_{\sigma}({\Bf k})$ 
is discontinuous.

\subsection*{\bf\S~4. \sc Detailed considerations}
\label{s4}

%\vspace{0.4cm}
\noindent{\bf 4.1. \small The interacting Hamiltonian and some 
basic details }
\label{s4a}
\vspace{0.3cm}

We consider the following interacting Hamiltonian: 
\begin{equation}
\label{e35}
\wh{H} = \wh{H}_0 + {\sf g}\,\wh{\sf H}_1,
\end{equation}
where
\begin{equation}
\label{e36}
\wh{H}_0 =
\sum_{{\bF k},\sigma}
\varepsilon_{{\bF k}}\,
{\hat c}_{{\bF k};\sigma}^{\dag}
{\hat c}_{{\bF k};\sigma},
\end{equation}
\begin{equation}
\label{e37}
\wh{\sf H}_1 =
\frac{1}{2 \Omega} \sum_{\sigma,\sigma'}\,
\sum_{{\bF k},{\bF p},{\bF q}}
{\bar w}(\| {\Bf q}\|)\,
{\hat c}_{{\bF k}+{\bF q};\sigma}^{\dag}
{\hat c}_{{\bF p}-{\bF q};\sigma'}^{\dag}
{\hat c}_{{\bF p};\sigma'}
{\hat c}_{{\bF k};\sigma}.
\end{equation}
In Eq.~(\ref{e35}), ${\sf g}$ (which has the dimension of energy) 
is the coupling constant of interaction, $\varepsilon_{\bF k}$ 
is a spin-degenerate single-particle energy dispersion (which may 
or may not be isotropic), ${\hat c}_{{\bF k};\sigma}^{\dag}$ and
${\hat c}_{{\bF k};\sigma}$ are the canonical creation and
annihilation operators respectively for fermions with spin index
$\sigma$, ${\sf g}\, {\bar w}(\|{\Bf q}\|) \equiv
{\bar v}(\|{\Bf q}\|)$ is the Fourier transform of the two-body
interaction potential $v({\Bf r}-{\Bf r}')$ (assumed to be isotropic),
and $\Omega = L^d$ is the volume of the (macroscopic) $d$-dimensional
hypercubic box occupied by the system. The wave-vector sums in
Eqs.~(\ref{e36}) and (\ref{e37}) are over a regular lattice (the
underlying lattice constant being equal to $2\pi/L$) covering in
principle the entire reciprocal space. On effecting the
substitutions
\begin{equation}
\label{e38}
{\sf g} \rightharpoonup U,\;\;\;\;
{\bar w}(\| {\Bf q}\|) \rightharpoonup
\frac{\Omega}{N_{\sc l}},
\end{equation}
and restricting the wave-vector sums to $N_{\sc l}$ wave vectors
uniformly distributed over the 1BZ associated with the Bravais 
lattice spanned by $\{ {\Bf R}_{j} \| j=1, \dots, N_{\sc l} \}$, 
the Hamiltonian in Eq.~(\ref{e35}) transforms into the conventional 
single-band Hubbard Hamiltonian $\wh{\cal H}$ corresponding to 
the on-site interaction energy $U$; in cases where ${\Bf k}+{\Bf q}$ 
and ${\Bf p}-{\Bf q}$ on the RHS of Eq.~(\ref{e37}) lie outside the 
1BZ, these vectors are to be identified with the vectors inside the 
1BZ obtained from ${\Bf k}+{\Bf q}$ and ${\Bf p}-{\Bf q}$ through 
Umklapp processes. On relaxing the replacements in Eq.~(\ref{e38}), 
while maintaining the above-mentioned restrictions concerning the 
wave vectors, the Hamiltonian in Eq.~(\ref{e35}) coincides with 
an `extended' Hubbard Hamiltonian.

For the $N$-particle uniform GS $\vert\Psi_{N;0}\rangle$ of the 
Hamiltonian in Eq.~(\ref{e35}) we have \cite{BF02a,BF03}
\begin{equation}
\label{e39}
\frac{ \int_{-\infty}^{\mu} {\rm d}\varepsilon\;
\varepsilon A_{\sigma}({\Bf k};\varepsilon) }
{\int_{-\infty}^{\mu} {\rm d}\varepsilon\;
A_{\sigma}({\Bf k};\varepsilon) }
{=:} \varepsilon_{{\bF k};\sigma}^{<}
\equiv \varepsilon_{\bF k} + {\sf g}\, \xi_{{\bF k};\sigma},
\;\;\; \forall {\Bf k},
\end{equation}
where
\begin{equation}
\label{e40}
\xi_{{\bF k};\sigma} \equiv
\frac{\beta_{{\bF k};\sigma}^{<}}{{\sf n}_{\sigma}({\Bf k})},
\end{equation}
in which
\begin{eqnarray}
\label{e41}
&&\beta_{{\bF k};\sigma}^{<}
\equiv \frac{-1}{\Omega} \sum_{\sigma'}
\sum_{{\bF p}',{\bF q}'} {\bar w}(\|{\Bf q}'\|)
\nonumber\\
&&\;\;\;\;\;\;
\times\langle\Psi_{N;0}\vert
{\hat c}_{{\bF k};\sigma}^{\dag}
{\hat c}_{{\bF p}'+{\bF q}';\sigma'}^{\dag}
{\hat c}_{{\bF k}+{\bF q}';\sigma}
{\hat c}_{{\bF p}';\sigma'}
\vert\Psi_{N;0}\rangle.
\end{eqnarray}
We should emphasize that, although $\varepsilon_{{\bF k};\sigma}^{<}
< \mu$, nonetheless $\varepsilon_{{\bF k};\sigma}^{<}$ is defined
over the entire available ${\Bf k}$ space; this we have made explicit
in Eq.~(\ref{e39}) through $\forall {\Bf k}$. We note in passing that, 
similar to $\varepsilon_{{\bF k};\sigma}^{<}$ and following the same 
techniques as employed in \cite{BF02a,BF03,BF02b}, for systems
corresponding to bounded $\varepsilon_{\bF k}$, one can readily 
express $\Delta\varepsilon_{{\bF k};\sigma}^{<}(E\to\infty)$, as 
defined in Eq.~(\ref{e7}) above, in terms of GS correlation functions. 

Along the same lines as used in \cite{BF02a,BF03} it can be shown
\footnote{\label{f15}
To this end, one has to use Eq.~(18) in \protect\cite{BF03}, which
holds for {\sl all} ${\Bf k}$, and the fact that $\varepsilon_{\bF k}$
and $\Sigma_{\sigma}^{\sc hf}({\Bf k})$, the Hartree-Fock self-energy,
are continuous for all ${\Bf k}$. }
that, for the set ${\cal S}$ of $\tilde{\Bf k}$ points for which
$\varepsilon_{\tilde{\bF k};\sigma}^{<}$ coincides with
$\varepsilon_{\sc f}$, the function
\begin{equation}
\label{e42}
\zeta_{{\bF k};\sigma} {:=} \xi_{{\bF k};\sigma}
- \frac{1}{\sf g} (\varepsilon_{\sc f} -
\varepsilon_{\tilde{\bF k}}),\;\;\;\
\tilde{\Bf k} \in {\cal S},
\end{equation}
approaches zero for $\|{\Bf k}-\tilde{\Bf k}\| \to 0$. This 
observation is significant in that it shows that, for ${\Bf k} \in
{\cal S}_{{\sc f};\sigma}$ where $\varepsilon_{{\bF k};\sigma}^{<}
= \varepsilon_{\sc f}$ and ${\sf n}_{\sigma}({\Bf k})$ is generically
(but not necessarily) discontinuous, the energy dispersion 
$\varepsilon_{{\bF k};\sigma}^{<}$ is continuous in the 
neighbourhood of ${\cal S}_{{\sc f};\sigma}$ (see also \S~3.4). 
With reference to Eq.~(\ref{e19}), it follows that, in cases where
${\sf n}_{\sigma}({\Bf k})$ is discontinuous at ${\cal S}_{{\sc f};
\sigma}$, $\beta_{{\bF k};\sigma}^{<}$ must undergo a discontinuity 
commensurate with that in ${\sf n}_{\sigma}({\Bf k})$ at 
${\cal S}_{{\sc f};\sigma}$ (and indeed at {\sl any} set ${\cal S}$ 
introduced above) for which $\varepsilon_{{\bF k};\sigma}^{<}$ 
attains the value $\varepsilon_{\sc f}$ (see \S~3.4); in appendix 
A this aspect is made explicit within an approximate framework. In 
what follows, unless we explicitly indicate otherwise, we assume 
$\varepsilon_{{\bF k};\sigma}^{<}$ to be continuous for all 
${\Bf k}$.

The experimental ARPES data in \cite{XJZ03,TV99,PVB00,PDJ01,AK01,AL01}
indicate that the measured $\varepsilon_{{\bF k};\sigma}^-$ (i.e. 
$\varepsilon_{{\bF k};\sigma}^{<}$ from our present standpoint) 
behave linearly as function of ${\Bf k}-{\Bf k}_{{\sc f};\sigma}$ 
for a finite range of ${\Bf k}$ away from ${\Bf k}_{{\sc f};\sigma}$; 
focusing on the data in \cite{XJZ03}, this is the case for
$\|{\Bf k}-{\Bf k}_{{\sc f};\sigma}\|$ varying between zero and 
approximately $5$\% of $\|{\Bf k}_{{\sc f};\sigma}\|$ for 
${\Bf k}_{{\sc f};\sigma}$ a nodal Fermi wave vector along the 
$(0,0) - (\pi,\pi)$ direction of the underlying square-shaped 1BZ. 
With $\varepsilon_{\bF k}$ being linear in ${\Bf k}-{\Bf k}_{{\sc f};
\sigma}$ for such a relatively small interval of ${\Bf k}$ vectors 
(excluding the neighbourhoods of the saddle points of 
$\varepsilon_{\bF k}$), we interpret the experimental observations 
as implying that
\begin{equation}
\label{e43}
\zeta_{{\bF k};\sigma} \sim
{\Bf b}({\Bf k}_{{\sc f};\sigma}^-) \cdot
({\Bf k}-{\Bf k}_{{\sc f};\sigma})\;\;\;
\mbox{\rm for}\;\;\; {\Bf k} \to {\Bf k}_{{\sc f};\sigma},
\end{equation}
where ${\Bf k}_{{\sc f};\sigma}$ is a nodal Fermi wave vector and where 
we assume ${\Bf k}$ to lie inside the Fermi sea. From Eqs.~(\ref{e39}), 
(\ref{e42}) and (\ref{e43}) it follows that (see Eq.~(\ref{e64}) below) 
\begin{eqnarray}
\label{e44}
{\Bf v}_{{\bF k}_{{\sc f};\sigma}^-;\sigma}^{<}
&{:=}& \frac{1}{\hbar}\left. {\Bf\nabla}_{\bF k}
\varepsilon_{{\bF k};\sigma}^{<}\right|_{{\bF k}
={\bF k}_{{\sc f};\sigma}^-} \nonumber\\
&=& \frac{1}{\hbar} \Big(\left. {\Bf\nabla}_{\bF k}
\varepsilon_{\bF k}\right|_{{\bF k}
={\bF k}_{{\sc f};\sigma}} + {\sf g}\,
{\Bf b}({\Bf k}_{{\sc f};\sigma}^-) \Big).
\end{eqnarray}
With reference to our earlier discussions in \S~3.5, we may express 
the Fermi velocity ${\Bf v}_{{\bF k}_{{\sc f};\sigma}^-;\sigma}$ 
as encountered within the framework of the Landau Fermi-liquid 
theory as
\begin{equation}
\label{e45}
{\Bf v}_{{\bF k}_{{\sc f};\sigma}^-;\sigma} = \frac{1}{\lambda}\,
{\Bf v}_{{\bF k}_{{\sc f};\sigma}^-;\sigma}^{<},
\end{equation}
where $\lambda$ is a constant of the order of unity (see footnote 
\ref{f13}). Experiments by Zhou {\sl et al.} \cite{XJZ03} 
indicate that ${\Bf v}_{{\bF k}_{{\sc f};\sigma}^-;\sigma}^{<}$, 
or ${\Bf v}_{{\bF k}_{{\sc f};\sigma}^-;\sigma}$, very weakly 
depends on the doping level. 

The observation in Eq.~(\ref{e43}) is of far-reaching consequence.
This follows from a combination of two facts. Firstly, on ${\Bf k}$
approaching ${\cal S}_{{\sc f};\sigma}$ from inside or outside the
Fermi sea, the following equation has to be satisfied 
\cite{BF02a,BF03}:
\begin{equation}
\label{e46}
\Gamma_{\sigma}({\Bf k}) = \Lambda_{\sigma}({\Bf k}),
\end{equation}
where
\begin{equation}
\label{e47}
\Lambda_{\sigma}({\Bf k}) {:=}
\frac{ {\sf n}_{\sigma}({\Bf k})}
{1 - {\sf n}_{\sigma}({\Bf k}) },\;\;\;
\Gamma_{\sigma}({\Bf k}) {:=}
\frac{\mu-\varepsilon_{{\bF k};\sigma}^{<}}
{\varepsilon_{{\bF k};\sigma}^{>}-\mu}.
\end{equation}
For $\varepsilon_{{\bF k};\sigma}^{>}$, defined in Eq.~(\ref{e23})
above, we have the following equivalent expression \cite{BF03}
\begin{equation}
\label{e48}
\varepsilon_{{\bF k};\sigma}^{>} \equiv 
\varepsilon_{\bF k} +
\frac{\hbar\Sigma_{\sigma}^{\sc hf}({\Bf k}) - {\sf g}\,
\beta_{{\bF k};\sigma}^{<}}{1 - {\sf n}_{\sigma}({\Bf k})},
\end{equation}
where, as mentioned earlier, $\Sigma_{\sigma}^{\sc hf}({\Bf k})$
stands for the Hartree-Fock self-energy. It can be shown \cite{BF03} 
that, for ${\Bf k}\to {\Bf k}_{{\sc f};\sigma} \in 
{\cal S}_{{\sc f};\sigma}$, one has
\begin{equation}
\label{e49}
\varepsilon_{{\bF k};\sigma}^{<} \sim \mu
+ {\Bf a}({\Bf k}_{{\sc f};\sigma})
\cdot ({\Bf k}-{\Bf k}_{{\sc f};\sigma})
+ {\sf g}\, \zeta_{{\bF k};\sigma},
\end{equation}
\begin{equation}
\label{e50}
\varepsilon_{{\bF k};\sigma}^{>} \sim \mu
+ {\Bf a}({\Bf k}_{{\sc f};\sigma})
\cdot ({\Bf k}-{\Bf k}_{{\sc f};\sigma})
- {\sf g}\, \Lambda_{\sigma}({\Bf k})\,
(\zeta_{{\bF k};\sigma} -\eta_{{\bF k};\sigma}),
\end{equation}
where
\begin{equation}
\label{e51}
{\Bf a}({\Bf k}_{{\sc f};\sigma}) {:=}
\left. {\Bf\nabla}_{\bF k}
\varepsilon_{\bF k}\right|_{{\bF k}={\bF k}_{{\sc f};\sigma}},
\end{equation}
\begin{equation}
\label{e52}
\eta_{{\bF k};\sigma} {:=}
\frac{ \hbar\Sigma_{\sigma}^{\sc hf}({\Bf k}) -
(\varepsilon_{\sc f} -
\varepsilon_{{\bF k}_{{\sc f};\sigma}})}
{ {\sf g}\, {\sf n}_{\sigma}({\Bf k})}.
\end{equation}
It can further be shown that \cite{BF03}
\begin{equation}
\label{e53}
\eta_{{\bF k};\sigma} \sim 0\;\;\;
\mbox{\rm for}\;\; {\Bf k} \to {\Bf k}_{{\sc f};\sigma}
\in {\cal S}_{{\sc f};\sigma}.
\end{equation}
Following this and Eq.~(\ref{e50}), one arrives at our earlier
statement concerning $\varepsilon_{{\bF k};\sigma}^{>} \sim \mu$
for ${\Bf k}\to {\Bf k}_{{\sc f};\sigma} \in {\cal S}_{{\sc f};
\sigma}$, and similarly for $\varepsilon_{{\bF k};\sigma}^{<}$, as is 
apparent from Eqs.~(\ref{e49}) and (\ref{e42}) (see \S~3.4).

Let ${\cal D}$ denote the open domain of wave vectors over which 
${\sf n}_{\sigma}({\Bf k})$ is continuous. For ${\Bf k}\in
{\cal D}$, applying ${\Bf\nabla}_{\bF k}$ on both sides of 
Eq.~(\ref{e24}), we have 
\begin{eqnarray}
\label{e54}
&&(\varepsilon_{{\bF k};\sigma}^{<}-
\varepsilon_{{\bF k};\sigma}^{>})\, 
{\Bf\nabla}_{\bF k} {\sf n}_{\sigma}({\Bf k}) \nonumber\\
&&\;\;\;
+{\sf n}_{\sigma}({\Bf k}) 
{\Bf\nabla}_{\bF k} \varepsilon_{{\bF k};\sigma}^{<}
+\big(1-{\sf n}_{\sigma}({\Bf k}) \big)
{\Bf\nabla}_{\bF k} \varepsilon_{{\bF k};\sigma}^{>}
\nonumber\\
&&\;\;\;\;\;\;\;\;\;\;\;\;\;\;\;\;\;\;\;\;\;\;\;\;\;\;\;
= {\Bf\nabla}_{\bF k} \varepsilon_{\bF k}
+\hbar {\Bf\nabla}_{\bF k}\Sigma_{\sigma}^{\sc hf}({\Bf k}),
\;\;\; {\Bf k} \in {\cal D}.
\end{eqnarray}
For ${\Bf k}$ in an infinitesimal neighbourhood of 
${\cal S}_{{\sc f};\sigma}$ where $\varepsilon_{{\bF k};\sigma}^{<}$ 
and $\varepsilon_{{\bF k};\sigma}^{>}$ take the value $\mu$ (up 
to infinitesimal corrections), from Eq.~(\ref{e54}) we observe that, 
for a sufficiently smooth $\varepsilon_{\bF k}$, a possible singular 
behaviour in ${\Bf\nabla}_{\bF k} \varepsilon_{{\bF k};\sigma}^{<}$ 
and ${\Bf\nabla}_{\bF k} \varepsilon_{{\bF k};\sigma}^{>}$ is 
determined by that in ${\Bf\nabla}_{\bF k}\Sigma_{\sigma}^{\sc hf}
({\Bf k})$ on the RHS of Eq.~(\ref{e54}). 
\footnote{\label{f16}
Let $\hat{\Bf n}({\Bf k}_{{\sc f};\sigma})$ denote the unit 
vector normal to ${\cal S}_{{\sc f};\sigma}$ at ${\Bf k}_{{\sc f};
\sigma} \in {\cal S}_{{\sc f};\sigma}$ pointing to the exterior 
of the Fermi sea FS$_{\sigma}$. Since, for ${\Bf k}$ in a close 
neighbourhood of ${\Bf k}_{{\sc f};\sigma}$, one has
$\hat{\Bf n}({\Bf k}_{{\sc f};\sigma}) \cdot {\Bf\nabla}_{\bF k} 
\varepsilon_{{\bF k};\sigma}^{<} \, {\Ieq<>}\, 0$ for ${\Bf k} \in 
{\rm FS}_{\sigma}, \overline{\rm FS}_{\sigma}$ and
$\hat{\Bf n}({\Bf k}_{{\sc f};\sigma}) \cdot {\Bf\nabla}_{\bF k} 
\varepsilon_{{\bF k};\sigma}^{>} \, {\Ieq><}\, 0$ for ${\Bf k} \in 
{\rm FS}_{\sigma}, \overline{\rm FS}_{\sigma}$, one observes that, 
in cases where ${\Bf\nabla}_{\bF k} \Sigma_{\sigma}^{\sc hf}({\Bf k})$ 
is divergent for ${\Bf k}\to {\Bf k}_{{\sc f};\sigma} \in 
{\cal S}_{{\sc f};\sigma}$, some useful information can be 
immediately deduced from Eq.~(\protect\ref{e54}). For instance, for 
cases in which the interaction potential is as long-ranged as the 
Coulomb potential and ${\sf n}_{\sigma}({\Bf k})$ is discontinuous 
at ${\Bf k}={\Bf k}_{{\sc f};\sigma}$, one can explicitly show that 
$\hat{\Bf n}({\Bf k}_{{\sc f};\sigma}) \cdot {\Bf\nabla}_{\bF k} 
\Sigma_{\sigma}^{\sc hf}({\Bf k}) \to +\infty$ as ${\Bf k}\to 
{\Bf k}_{{\sc f};\sigma}$ so that, from Eq.~(\protect\ref{e54}), one 
directly deduces that, firstly, $\hat{\Bf n}({\Bf k}_{{\sc f};\sigma}) 
\cdot {\Bf\nabla}_{\bF k} \varepsilon_{{\bF k};\sigma}^{<} \to 
\pm\infty$, $\hat{\Bf n}({\Bf k}_{{\sc f};\sigma}) \cdot       
{\Bf\nabla}_{\bF k} \varepsilon_{{\bF k};\sigma}^{>} \to \mp\infty$, 
for ${\Bf k} \in {\rm FS}_{\sigma}, \overline{\rm FS}_{\sigma}$ 
and, secondly, the balance between these diverging contributions 
must be such that the resultant function on the LHS of
Eq.~(\protect\ref{e54}) approaches $+\infty$ as ${\Bf k} \to 
{\Bf k}_{{\sc f};\sigma}$, irrespective of whether ${\Bf k} \in 
{\rm FS}_{\sigma}$ or ${\Bf k} \in \overline{\rm FS}_{\sigma}$. }
This aspect is explicitly reflected in the defining expression
for $\eta_{{\bF k};\sigma}$ in Eq.~(\ref{e52}) which through 
Eq.~(\ref{e50}) determines the behaviour of $\varepsilon_{{\bF k};
\sigma}^{>}$ for ${\Bf k}\to {\Bf k}_{{\sc f};\sigma} \in 
{\cal S}_{{\sc f};\sigma}$. In other regions of the ${\Bf k}$ space 
where $\varepsilon_{{\bF k};\sigma}^{>} > \varepsilon_{{\bF k};
\sigma}^{<}$, a possible singular behaviour in ${\Bf\nabla}_{\bF k} 
\Sigma_{\sigma}^{\sc hf}({\Bf k})$ in some neighbourhood can
in principle entirely or partly be accounted for by a similar 
behaviour in ${\Bf\nabla}_{\bF k} {\sf n}_{\sigma}({\Bf k})$. 
\footnote{\label{f17}
In this connection, we point out that according to Belyakov
\protect\cite{VAB61} (see also \cite{SM80}) 
$\hat{\Bf n}({\Bf k}_{{\sc f};\sigma}) \cdot 
{\Bf\nabla}_{\bF k} {\sf n}_{\sigma}({\Bf k})$ logarithmically 
diverges as ${\Bf k} \to {\Bf k}_{{\sc f};\sigma}$ for 
${\sf n}_{\sigma}({\Bf k})$ pertaining to the isotropic GS of 
fermions interacting through a short-range potential.
For some relevant details see \S~6 in \protect\cite{BF99a}. }
The above observations will prove useful in our subsequent 
considerations below.

For the interaction potential $v({\Bf r}-{\Bf r}')$ identified
with the long-range Coulomb potential (or a potential as 
long ranged as this), it can be shown that, so long as 
$Z_{{\bF k}_{{\sc f};\sigma}} > 0$, in consequence of 
$\Sigma_{\sigma}^{\sc hf}({\Bf k})$ on the RHS of Eq.~(\ref{e52}),
for ${\Bf k} \to {\Bf k}_{{\sc f};\sigma}^{-}$ (i.e. for ${\Bf k}$
approaching ${\Bf k}_{{\sc f};\sigma}$ from {\sl inside} the Fermi
sea), one has
\begin{equation}
\label{e55}
\vert \eta_{{\bF k};\sigma} \vert \sim
\| {\Bf c}({\Bf k}_{{\sc f};\sigma}^-)\|
\vert \ln\|{\Bf k}-{\Bf k}_{{\sc f};\sigma}\| \vert\,
\| {\Bf k}-{\Bf k}_{{\sc f};\sigma}\|,
\end{equation}
where ${\Bf c}({\Bf k}_{{\sc f};\sigma}^-)$ stands for some constant 
vector. The combination of Eqs.~(\ref{e43}) and (\ref{e55}) implies 
that, to leading order in $({\Bf k}-{\Bf k}_{{\sc f};\sigma})$ for 
${\Bf k}\to {\Bf k}_{{\sc f};\sigma}^{-}$, one has according to 
Eq.~(\ref{e50})
\begin{equation}
\label{e56}
\varepsilon_{{\bF k};\sigma}^{>} \sim
\mu + {\sf g}\, \Lambda_{\sigma}({\Bf k}) \,
\eta_{{\bF k};\sigma},
\end{equation}
and consequently
\begin{equation}
\label{e57}
\Gamma_{\sigma}({\Bf k}) \sim 0\;\;\;
\mbox{\rm for}\;\; {\Bf k}\to {\Bf k}_{{\sc f};\sigma}^{-} \;
\Longrightarrow\;
{\sf n}_{\sigma}({\Bf k}_{{\sc f};\sigma}^{-}) =0.
\end{equation}
Since ${\sf n}_{\sigma}({\Bf k}_{{\sc f};\sigma}^{-}) = 0$ corresponds
to the peculiar condition $A_{\sigma}({\Bf k}_{{\sc f};\sigma}^{-};
\varepsilon) \equiv 0$, for {\sl all} $\varepsilon\in (-\infty,\mu]$
(see Eq.~(\ref{e2}) above), we discard the condition in Eq.~(\ref{e55}) 
for being unphysical in the present context where Eq.~(\ref{e43}) is 
held as valid on account of experimental observations. It thus follows
that, for the case of Coulomb-interacting particles, $Z_{{\bF k}_{{\sc f};
\sigma}}$ cannot be positive, whereby Eq.~(\ref{e55}) is rendered 
obsolete. Assuming $Z_{{\bF k}_{{\sc f};\sigma}}$ to be positive but 
relatively small, it is not inconceivable that, for ${\Bf k}$ sufficiently 
close to ${\Bf k}_{{\sc f};\sigma}$, $\vert\zeta_{{\bF k};\sigma}\vert$ 
behaves similarly to $\vert\eta_{{\bF k};\sigma}\vert$ as presented 
in Eq.~(\ref{e55}) which experiments by Zhou {\sl et al.} \cite{XJZ03} 
may not have resolved. Consequently, at this stage the possibility of 
Eq.~(\ref{e55}), and thus of $Z_{{\bF k}_{{\sc f};\sigma}} > 0$, 
cannot be unequivocally rejected. We point out that here we are 
relying on our standpoint that experimentally one measures 
$\varepsilon_{{\bF k};\sigma}^{<}$ rather than the `solution' to 
Eq.~(\ref{e3}). For completeness, in \cite{BF03} (\S~6.1.3) the 
consequences of $\vert\zeta_{{\bF k};\sigma}\vert$ satisfying an 
asymptotic relation similar to that in Eq.~(\ref{e55}) have been 
investigated in some detail. In \cite{BF03} (\S~6.1.3) it has been 
emphasized that, for fermions interacting through the long-range 
Coulomb potential, Fermi-liquid metallic states belong to the 
category of states for which, as ${\Bf k} \to {\Bf k}_{{\sc f};
\sigma} \in {\cal S}_{{\sc f};\sigma}$, $\eta_{{\bF k};\sigma}$ 
satisfies Eq.~(\ref{e55}) and $\zeta_{{\bF k};\sigma}$ fulfils a 
functionally similar expression.

We conclude that the assumption of the validity of Eq.~(\ref{e43}) 
can be compatible with a picture of fermions interacting through the 
long-range Coulomb potential (or a potential as long ranged as this)
only if $Z_{{\bF k}_{{\sc f};\sigma}} = 0$. This condition is satisfied 
(not exclusively) within the framework where the normal metallic states 
of the copper-oxide based superconductors along the nodal directions 
of the 1BZ are of a marginal Fermi-liquid type \cite{VLSRAR89}. This 
viewpoint enjoys direct experimental support through the angle-resolved 
photoemission experiments by Valla {\sl et al.} \cite{TV99} along 
the nodal directions of the 1BZ concerning the optimally-doped compound 
Bi$_2$Sr$_2$CaCu$_2$O$_{8+\delta}$.

The experimental results by Zhou {\sl et al.} \cite{XJZ03} further
show that, at ${\Bf k}={\Bf k}_{\star}$, with $\| {\Bf k}_{\star} -
{\Bf k}_{{\sc f};\sigma} \|$ approximately $5$\% of
$\|{\Bf k}_{{\sc f};\sigma}\|$, $\varepsilon_{{\bF k};\sigma}^{<}$ 
is singular, the singularity appearing to be a discontinuity in the
{\sl slope} of $\varepsilon_{{\bF k};\sigma}^{<}$ in the overdoped 
regime and some stronger singularity in the underdoped regime. The 
strongest singularity that can be expected from $\varepsilon_{{\bF k};
\sigma}^{<}$ is a discontinuity (see \S~3.4), followed by (in the 
order of significance), for two-body interaction potentials as 
long ranged as the Coulomb potential, a logarithmic divergence in 
${\Bf\nabla}_{\bF k}\varepsilon_{{\bF k};\sigma}^{<}$ for ${\Bf k}$ 
approaching the points of discontinuity of ${\sf n}_{\sigma}({\Bf k})$. 
We believe that the observations by Zhou {\sl et al.} \cite{XJZ03} 
are highly supportive of the viewpoint that, at least in the underdoped 
regime, the long-range of the Coulomb potential can be of considerable 
significance in determining the (unconventional) physical properties 
of these compounds. For clarity, for the hole doping fraction $x=0.03$
concerning (La$_{2-x}$Sr$_x$)CuO$_4$ (LSCO), one observes a behaviour 
in the measured energy dispersion \cite{XJZ03} which is reminiscent 
of a discontinuity superimposed by a contribution whose gradient 
is logarithmically divergent at ${\Bf k}={\Bf k}_{\star}$.

\vspace{0.4cm}
\noindent{\bf 4.2. \small On the relationship between 
discontinuities in ${\sf n}_{\sigma}({\Bf k})$ and 
${\Bf\nabla}_{\bF k} \varepsilon_{{\bF k};\sigma}^{<}$ }
\label{s4b}
\vspace{0.3cm}

Here we demonstrate that singularity (not necessarily discontinuity) 
of ${\sf n}_{\sigma}({\Bf k})$ at a point is directly reflected in 
the behaviour of ${\Bf\nabla}_{\bF k} \varepsilon_{{\bF k};\sigma}^{<}$ 
at the same point. We shall explicitly consider the case where 
${\sf n}_{\sigma}({\Bf k})$ is discontinuous in the interior of 
FS$_{\sigma}\backslash {\cal S}_{{\sc f};\sigma}$; concerning the
behaviour of ${\Bf\nabla}_{\bF k}\varepsilon_{{\bF k};\sigma}^{<}$ 
for ${\Bf k} \to {\Bf k}_{{\sc f};\sigma} \in {\cal S}_{{\sc f};
\sigma}$, this is implicit in our earlier work \cite{BF02a,BF03} 
which for completeness we recapitulate in the following. In this 
connection, note that ${\sf n}_{\sigma}({\Bf k})$ is always singular 
(not necessarily discontinuous) on ${\cal S}_{{\sc f};\sigma}$ 
\cite{BF02a,BF03}.

For ${\Bf k} \in {\cal S}_{{\sc f};\sigma}$, $\varepsilon_{{\bF k};
\sigma}^{<}$ and $\varepsilon_{{\bF k};\sigma}^{>}$ attain the value
$\varepsilon_{\sc f}$ (up to infinitesimal corrections) so that the 
exact inequalities $\varepsilon_{{\bF k};\sigma}^{<} < \mu$ and 
$\varepsilon_{{\bF k};\sigma}^{>} > \mu$ \cite{BF02a,BF03} imply that, 
for $\varepsilon_{\bF k}$ a smooth function of ${\Bf k}$ in the 
neighbourhood of ${\cal S}_{{\sc f};\sigma}$, it is required that 
${\Bf\nabla}_{\bF k}\xi_{{\bF k};\sigma}$ undergo at least 
\footnote{\label{f18}
This aspect is related to the restriction $0 < \gamma \le 1$ for the 
parameter $\gamma$ introduced in \protect\cite{BF02a}; $\gamma > 1$ 
is excluded for cases where $a_{\sigma} > 0$. }
a {\sl discontinuous} change for ${\Bf k}$ transposed through 
${\cal S}_{{\sc f};\sigma}$ from infinitesimally inside to 
infinitesimally outside the Fermi sea. For instance, for a specific 
class of uniform metallic GSs of the single-band Hubbard Hamiltonian 
(to which class the Fermi-liquid GSs belong) we have earlier shown 
\cite{BF02a} (see Eqs.~(105) and (106) herein) that, with 
$\hat{\Bf n}({\Bf k}_{{\sc f};\sigma})$ denoting the outward unit 
vector (pointing from inside to outside the Fermi sea) normal to 
${\cal S}_{{\sc f};\sigma}$ at ${\Bf k}={\Bf k}_{{\sc f};\sigma}$ 
one has 
\begin{equation}
\label{e58}
\left.
\hat{\Bf n}({\Bf k}_{{\sc f};\sigma}) \cdot
{\Bf\nabla}_{\bF k} \xi_{{\bF k};\sigma}
\right|_{{\bF k}={\bF k}_{{\sc f};\sigma}^+}
= -\frac{1}{U} a_{\sigma} - b_{\sigma}^{-},
\end{equation}
in which
\begin{equation}
\label{e59}
b_{\sigma}^{-} \equiv
\left.
\hat{\Bf n}({\Bf k}_{{\sc f};\sigma}) \cdot
{\Bf\nabla}_{\bF k} \xi_{{\bF k};\sigma}
\right|_{{\bF k}={\bF k}_{{\sc f};\sigma}^-},
\end{equation}
and (see Eq.~(\ref{e51}) above)
\begin{equation}
\label{e60}
a_{\sigma} {:=}
\hat{\Bf n}({\Bf k}_{{\sc f};\sigma}) \cdot
{\Bf a}({\Bf k}_{{\sc f};\sigma}) \equiv
\hat{\Bf n}({\Bf k}_{{\sc f};\sigma}) \cdot
(\hbar {\Bf v}_{{\sc f};\sigma}^{(0)}).
\end{equation}
In \cite{BF02a} it has been shown that the stability of the
GS under consideration requires that 
\begin{equation}
\label{e61}
b_{\sigma}^- > \frac{a_{\sigma}}{U\Lambda_{\sigma}^{-}} \ge 0,
\end{equation}
where 
\begin{equation}
\label{e62}
\Lambda_{\sigma}^{-} \equiv
\Lambda_{\sigma}({\Bf k}_{{\sc f};\sigma}^-)
\equiv \frac{ {\sf n}_{\sigma}({\Bf k}_{{\sc f};\sigma}^-)}
{1 - {\sf n}_{\sigma}({\Bf k}_{{\sc f};\sigma}^-)} \ge 1.
\end{equation}
From Eqs.~(\ref{e59}) and (\ref{e58}), one observes that on ${\Bf k}$ 
crossing ${\cal S}_{{\sc f};\sigma}$ (from inside to outside the Fermi 
sea), indeed the derivative of $\xi_{{\bF k};\sigma}$ with respect to
${\Bf k}$ in the direction normal to ${\cal S}_{{\sc f};\sigma}$ 
undergoes a discontinuous change (this amounts to $2 b_{\sigma}^- 
+ a_{\sigma}/{\sf g} > a_{\sigma} (2/\Lambda_{\sigma}^- + 1) \ge 0$), 
from the {\sl positive} value $b_{\sigma}^{-}$ at ${\Bf k}
={\Bf k}_{{\sc f};\sigma}^-$ to the {\sl negative} value $-(b_{\sigma}^- 
+a_{\sigma}/{\sf g})$ at ${\Bf k}={\Bf k}_{{\sc f};\sigma}^+$. Similar 
behaviour is shared by the $\xi_{{\bF k};\sigma}$ pertaining to systems 
of fermions interacting through two-body potentials of arbitrary range. 
An analogous discontinuity takes place in $\hat{\Bf n}({\Bf k}_{{\sc f};
\sigma}) \cdot {\Bf\nabla}_{\bF k}\varepsilon_{{\bF k};\sigma}^{>}$
on transposing ${\Bf k}$ through ${\cal S}_{{\sc f};\sigma}$. 
For Fermi-liquid metallic GSs of systems in which the two-body 
interaction potential is of shorter range than the Coulomb potential, 
the functions $\varepsilon_{{\bF k};\sigma}^{<}$ and 
$\varepsilon_{{\bF k};\sigma}^{>}$ are such that, on transposing 
${\Bf k}$ from inside to outside the Fermi sea, the continuous and 
differentiable extension of $\varepsilon_{{\bF k};\sigma}^{\Ieq><}$ 
coincides with $\varepsilon_{{\bF k};\sigma}^{\Ieq<>}$ in the close 
neighbourhood of ${\cal S}_{{\sc f};\sigma}$ (see Fig.~3 in 
\cite{BF02a}). On the basis of this fact and a number of available 
numerical results, in \cite{BF02a} we arrived at the conclusion that, 
in the case of the single-band Hubbard Hamiltonian described in terms 
of the nearest-neighbour hopping integral $t$ and the on-site interaction 
energy $U$, the corresponding $\varepsilon_{{\bF k};\sigma}^{<}$ and 
$\varepsilon_{{\bF k};\sigma}^{>}$ fail to posses the latter property 
for $U/t\, \Ieq{\sim}{>}\, 8$ (see footnote \ref{f27} below). 

Having described the singular behaviour of $\varepsilon_{{\bF k};
\sigma}^<$ in the neighbourhood of ${\cal S}_{{\sc f};\sigma}$,
we now set out to demonstrate that a discontinuity in 
${\sf n}_{\sigma}({\Bf k})$ at ${\Bf k} = {\Bf k}_{\star} \in
{\rm FS}_{\sigma}\backslash {\cal S}_{{\sc f};\sigma}$ (i.e.
${\Bf k}_{\star}$ is strictly located in the interior of the 
Fermi sea FS$_{\sigma}$) gives rise to the following result:
\begin{equation}
\label{e63}
{\Bf v}_{{\bF k}_{\star}^{+};\sigma}^{<} =
{\Bf v}_{{\bF k}_{\star}^{-};\sigma}^{<}
-\frac{Z_{{\bF k}_{\star};\sigma} }
{ {\sf n}_{\sigma}({\Bf k}_{\star}^-) }\,
{\Bf v}_{{\bF k}_{\star}^-;\sigma},
\end{equation}
where (see Eq.~(\ref{e13}) above)
\begin{eqnarray}
\label{e64}
{\Bf v}_{{\bF k}_{\star}^{\mp};\sigma}^{<} &{:=}& \left.
\frac{1}{\hbar} {\Bf\nabla}_{\bF k} 
\varepsilon_{{\bF k};\sigma}^{<} 
\right|_{{\bF k}={\bF k}_{\star}^{\mp}}, \\
\label{e65}
{\Bf v}_{{\bF k}_{\star}^-;\sigma} &{:=}& \left.
\frac{1}{\hbar} {\Bf\nabla}_{\bF k} 
\varepsilon_{{\bF k}^-;\sigma}
\right|_{{\bF k}={\bF k}_{\star}^-}.
\end{eqnarray}
We present the proof of the result in Eq.~(\ref{e63}) in \S~4.3 
below. Note that $0 \le Z_{{\bF k}_{\star};\sigma}/{\sf n}_{\sigma}
({\Bf k}_{\star}^-) \le 1$. 
\footnote{\label{f19}
It is important to realize that, in contrast with the case where 
a possible discontinuity in ${\sf n}_{\sigma}({\Bf k})$ at 
${\Bf k}={\Bf k}_{{\sc f};\sigma} \in {\cal S}_{{\sc f};\sigma}$ 
corresponds to $Z_{{\bF k}_{{\sc f};\sigma}} > 0$ (see 
Eq.~(\protect\ref{e16})), there is no {\sl a priori} restriction 
on the sign of $Z_{{\bF k}_{\star};\sigma}$ (viewed as the value of
${\sf n}_{\sigma}({\bF k}_{\star}^-) - {\sf n}_{\sigma}
({\bF k}_{\star}^+)$ and {\sl not} as spectral weight) for ${\Bf k}_{\star}$ 
outside ${\cal S}_{{\sc f};\sigma}$. This is appreciated by 
realizing the fact that all excitations corresponding to the 
wave vector ${\Bf k}_{\star}$ {\sl outside} ${\cal S}_{{\sc f};
\sigma}$ can, by the very definition of ${\cal S}_{{\sc f};\sigma}$, 
only correspond to many-body states whose energies are greater 
than the energy of the GS of the system under consideration, whereby a
possible negative $Z_{{\bF k}_{\star};\sigma}$ cannot signify
an instability of the latter GS. This aspect is already reflected 
in the fact that, for ${\Bf k}_{\star}$ located at a finite distance 
from ${\cal S}_{{\sc f};\sigma}$, the choice of what we denote by 
${\Bf k}_{\star}^-$ and ${\Bf k}_{\star}^+$ is in principle arbitrary. 
That in our present considerations $Z_{{\bF k}_{\star};\sigma} > 0$, 
is related to the fact that we have chosen ${\Bf k}_{\star}^+$ to 
be the closer of the two vectors ${\Bf k}_{\star}^-$ and 
${\Bf k}_{\star}^+$ to the nodal Fermi point ${\Bf k}_{{\sc f};
\sigma}$ and that, in order for $\varepsilon_{{\bF k};\sigma}^{-}$,
or $\varepsilon_{{\bF k};\sigma}^{<}$, to attain the required value 
$\varepsilon_{\sc f}$ at ${\Bf k}={\Bf k}_{{\sc f};\sigma}$, it 
must (monotonically) increase for ${\Bf k}$ transposed from 
${\Bf k}_{\star}$ to ${\Bf k}_{{\sc f};\sigma}$. }
For two-body interaction potentials of shorter range than the 
Coulomb potential, ${\Bf v}_{{\bF k}_{\star}^-;\sigma} \approx 
{\Bf v}_{{\bF k}_{\star}^+;\sigma}^{<}$; therefore, on account of 
Eq.~(\ref{e63}), one has
\begin{equation}
\label{e66}
{\Bf v}_{{\bF k}_{\star}^-;\sigma}^{<}
\approx \left(
1 +\frac{Z_{{\bF k}_{\star};\sigma}}{{\sf n}_{\sigma}
({\Bf k}_{\star}^-)} \right)\,
{\Bf v}_{{\bF k}_{\star}^+;\sigma}^{<}\;
{\Ieq\sim<}\; 2\, {\Bf v}_{{\bF k}_{\star}^+;\sigma}^{<}.
\end{equation}
Introducing, in analogy to the problem of electrons coupled to 
phonons \cite{GDM90} (see specifically \S~6.4 herein), the 
`mass-enhancement factor' $\lambda_{\star}$ as follows
\begin{equation}
\label{e67}
{\Bf v}_{{\bF k}_{\star}^-;\sigma}^{<} =
(1+\lambda_{\star})\,
{\Bf v}_{{\bF k}_{\star}^+;\sigma}^{<}, 
\end{equation} 
from Eq.~(\ref{e66}) we have
\begin{equation}
\label{e68}
\lambda_{\star} \approx 
\frac{Z_{{\bF k}_{\star};\sigma}}{{\sf n}_{\sigma}
({\Bf k}_{\star}^-)} \le 1.
\end{equation}
The result $1< 1+\lambda_{\star} \, {\Ieq\sim<}\, 2$ (specifically 
$1+\lambda_{\star} \approx 2$) conforms with the experimental 
observations concerning the cuprate compounds \cite{AL01} (see 
Fig.~4 herein where $\lambda'$ is what here we have 
denoted by $\lambda_{\star}$) (see also \cite{ZXS02}).

For ${\Bf v}_{{\bF k}_{\star}^-;\sigma}$ we have
\footnote{\label{f20}
We point out that the significance of ${\Bf k}_{\star}^-$ as the 
subscript of ${\Bf v}_{{\bF k}_{\star}^-;\sigma}$ lies in the 
finite difference between $\varepsilon_{{\bF k}_{\star}^-;\sigma}
\equiv \varepsilon_{{\bF k}_{\star};\sigma}^-$ (which lies strictly
below $\mu$) and $\varepsilon_{{\bF k}_{\star}^+;\sigma} \equiv
\varepsilon_{{\bF k}_{\star};\sigma}^+$ (which lies strictly above 
$\mu$) for ${\Bf k}_{\star}$ located at some finite distance from 
${\cal S}_{{\sc f};\sigma}$ (see \S~4.3) and {\sl not} in the 
possibility that ${\Bf\nabla}_{\bF k} \Sigma_{\sigma}({\Bf k};
\varepsilon_{{\bF k}_{\star}^-;\sigma})$ is a discontinuous 
function of ${\Bf k}$ at ${\Bf k}={\Bf k}_{\star}$. In view of the
latter, we should emphasize that far from dismissing a discontinuity
in ${\Bf\nabla}_{\bF k}\Sigma_{\sigma}({\Bf k};
\varepsilon_{{\bF k}_{\star}^-;\sigma})$, at ${\Bf k}={\Bf k}_{\star}$, 
as being {\sl a priori} infeasible, our statement here only reflects 
the confines of our considerations in this paper. We hope to return 
to this subject matter in a future publication. }
\begin{equation}
\label{e69}
{\Bf v}_{{\bF k}_{\star}^-;\sigma} \equiv 
Z_{{\bF k}_{\star};\sigma} \Big( \frac{1}{\hbar}
\left. {\Bf\nabla}_{\bF k}\varepsilon_{\bF k} 
\right|_{{\bF k}={\bF k}_{\star}^-}
+\left. {\Bf\nabla}_{\bF k} \Sigma_{\sigma}({\Bf k};
\varepsilon_{{\bF k}_{\star}^-;\sigma})
\right|_{{\bF k}={\bF k}_{\star}^-}\Big)
\end{equation}
whose existence depends on the existence of the second term 
in parentheses on the RHS of Eq.~(\ref{e69}). In this
connection it is important to realize that a finite positive
$Z_{{\bF k}_{\star};\sigma}$ is not sufficient for the last 
term in Eq.~(\ref{e69}) to be bounded. With reference to the 
convention adopted in Eq.~(\ref{e13}), we write
\begin{equation}
\label{e70}
{\Bf v}_{{\bF k}_{\star}^-;\sigma} \equiv
{\Bf v}_{{\bF k}_{\star};\sigma}^- 
{:=} \frac{1}{\hbar} {\Bf\nabla}_{\bF k} 
\varepsilon_{{\bF k};\sigma}^{-}\vert_{{\bF k}
={\bF k}_{\star}^-}. 
\end{equation}
In a similar fashion,
\begin{equation}
\label{e71}
{\Bf v}_{{\bF k}_{\star}^+;\sigma} \equiv
{\Bf v}_{{\bF k}_{\star};\sigma}^+ 
{:=} \frac{1}{\hbar} {\Bf\nabla}_{\bF k} 
\varepsilon_{{\bF k};\sigma}^{+}\vert_{{\bF k}
={\bF k}_{\star}^+}.
\end{equation}
For Fermi-liquid metallic states where ${\sf n}_{\sigma}({\Bf k})$ 
is discontinuous at ${\Bf k}\in {\cal S}_{{\sc f};\sigma}$, 
${\Bf v}_{{\bF k};\sigma}^-$ and ${\Bf v}_{{\bF k};\sigma}^+$ are 
collinear and point in the same direction for {\sl all} ${\Bf k}\in 
{\cal S}_{{\sc f};\sigma}$. For the uniform GSs of the single-band 
Hubbard Hamiltonian, in \cite{BF02a} strong evidence is presented 
indicating that for $U/t \,{\Ieq\sim>}\, 8$, ${\Bf v}_{{\bF k};
\sigma}^-$ and ${\Bf v}_{{\bF k};\sigma}^+$, ${\Bf k}\in 
{\cal S}_{{\sc f};\sigma}$, are not collinear so that these states 
cannot be Fermi liquids (see also footnote \ref{f27} below).

\begin{figure}[t!]
\protect
\centerline{
\psfig{figure=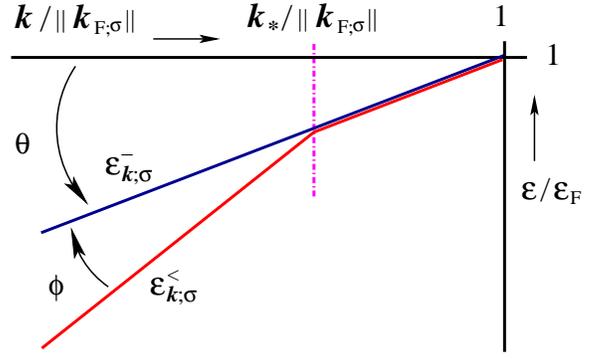,width=3.0in} }
\vskip 9pt
\caption{\label{fi2} \sf
Schematic representation of $\varepsilon_{{\bF k};\sigma}^-$, the 
asymptotic solution of the quasi-particle equation
(Eq.~(\protect\ref{e3})), and $\varepsilon_{{\bF k};\sigma}^{<}$
as defined in Eq.~(\protect\ref{e39}), for ${\Bf k}$ inside the 
Fermi sea and close to the Fermi wave vector ${\Bf k}_{{\sc f};
\sigma}$, with ${\Bf k}_{{\sc f};\sigma}-{\Bf k}$ pointing in the 
direction of the outward normal $\hat{\Bf n}({\Bf k}_{{\sc f};
\sigma})$ to the Fermi surface ${\cal S}_{{\sc f};\sigma}$ at 
${\Bf k}={\Bf k}_{{\sc f};\sigma}$. It is assumed that
${\sf n}_{\sigma}({\Bf k})$ is continuous at 
${\Bf k}={\Bf k}_{{\sc f};\sigma}$ (i.e.
$Z_{{\bF k}_{{\sc f};\sigma}}=0$) while it undergoes a finite 
discontinuity at ${\Bf k}= {\Bf k}_{\star}$ (i.e. 
$Z_{{\bF k}_{\star};\sigma} > 0$). Following Eqs.~(\protect\ref{e4})
and (\protect\ref{e27}), $\varepsilon_{{\bF k};\sigma}^-$ 
and $\varepsilon_{{\bF k};\sigma}^{<}$ coincide at 
${\Bf k}={\Bf k}_{{\sc f};\sigma}$ and 
${\Bf k}={\Bf k}_{\star}$ respectively and, owing to the
smallness of $\|{\Bf k}_{\star}-{\Bf k}_{{\sc f};\sigma}\|/
\|{\Bf k}_{{\sc f};\sigma}\|$ (in practice, of the order
of $5$\%), the possible difference between 
$\varepsilon_{{\bF k};\sigma}^-$ and $\varepsilon_{{\bF k};\sigma}^{<}$
cannot be experimentally discernible for ${\Bf k}$ in the interval
between ${\Bf k}_{\star}$ and ${\Bf k}_{{\sc f};\sigma}$. We have 
$\theta = \tan^{-1}\big(\gamma\,\hat{\Bf n}({\Bf k}_{{\sc f};\sigma})
\cdot {\Bf\nabla}_{\bF k} \varepsilon_{{\bF k};\sigma}^{-} 
\vert_{{\bF k}={\bF k}_{{\sc f};\sigma}^-} \big)$, in which
$\gamma\equiv \|{\Bf k}_{{\sc f};\sigma}\|/\varepsilon_{\sc f}$,
and $\phi \approx -Z_{{\bF k}_{\star};\sigma}
\theta/{\sf n}_{\sigma}({\Bf k}_{\star}^-)$; counterclockwise is 
the direction in which we count angles as positive. By viewing 
$\varepsilon_{{\bF k};\sigma}^{<}$ as the single-particle energy 
dispersion as measured through ARPES and excluding the results 
corresponding to the underdoped cuprates, our schematic representation
(which is based on Eq.~(\protect\ref{e63}); see also 
Eqs.~(\protect\ref{e74}) and (\protect\ref{e75})) is in excellent 
qualitative agreement with the experimental observations by Zhou 
{\sl et al.} \protect\cite{XJZ03}. For fermions interacting through 
the Coulomb interaction potential, $\hat{\Bf n}({\Bf k}_{{\sc f};
\sigma}) \cdot {\Bf\nabla}_{\bF k} \varepsilon_{{\bF k};\sigma}^{<}$ 
acquires a logarithmic divergence at ${\Bf k}={\Bf k}_{\star}$
(see Fig.~\protect\ref{fi3}). Taking this aspect into account, as 
well as a finite discontinuity in $\varepsilon_{{\bF k};\sigma}^{<}$ 
at ${\Bf k}={\Bf k}_{\star}$, which is feasible (\S~3.4), the 
corresponding schematic representation would similarly be in 
excellent (qualitative) agreement with the experimental observations 
by Zhou {\sl et al.} \protect\cite{XJZ03} concerning cuprates in 
the underdoped regime, as exemplified by (La$_{2-x}$Sr$_x$)CuO$_4$ 
for $x=0.03$. }
\end{figure}

Assuming $\|{\Bf v}_{{\bF k}_{\star};\sigma}^{-}\|$ to be non-zero 
and bounded, from Eq.~(\ref{e63}) it follows that ${\Bf \nu}\cdot 
{\Bf v}_{{\bF k}_{\star}^+;\sigma}^{<}$ differs from ${\Bf \nu}\cdot 
{\Bf v}_{{\bF k}_{\star}^-;\sigma}^{<}$ for any arbitrary vector 
${\Bf \nu}$ outside the plane normal to ${\Bf v}_{{\bF k}_{\star};
\sigma}^-$. Let $\hat{\Bf n}({\Bf k}_{\star})$ denote the outward 
unit vector normal to ${\cal S}_{{\sc f};\sigma}$ at 
${\Bf k}_{{\sc f};\sigma} \in {\cal S}_{{\sc f};\sigma}$, with 
${\Bf k}_{{\sc f};\sigma}$ the point at which the extension of the 
radius vector ${\Bf k}_{\star}$ meets ${\cal S}_{{\sc f};\sigma}$. 
For ${\Bf k}_{\star}$ inside the Fermi sea and sufficiently close 
to ${\Bf k}_{{\sc f};\sigma}$, the inner products of the three 
vectors in Eq.~(\ref{e63}) with $\hat{\Bf n}({\Bf k}_{\star})$ 
are all non-negative so that on account of Eq.~(\ref{e63}) in 
general we have
\begin{equation}
\label{e72}
0<\, \hat{\Bf n}({\Bf k}_{\star}) \cdot 
{\Bf v}_{{\bF k}_{\star}^+;\sigma}^{<}\,
<\, \hat{\Bf n}({\Bf k}_{\star}) \cdot 
{\Bf v}_{{\bF k}_{\star}^-;\sigma}^{<}. 
\end{equation}
The discontinuity in ${\Bf v}_{{\bF k};\sigma}^{<}$ at ${\Bf k}
={\Bf k}_{\star}$, where by assumption ${\sf n}_{\sigma}({\Bf k})$ 
is discontinuous, is seen to show the same trend as observed in the 
experimental results by Zhou {\sl et al.} \cite{XJZ03} (and other
workers \cite{TV99,PVB00,PDJ01,AK01,AL01}). To clarify this aspect, 
we point out that, by Eq.~(\ref{e4}), $\varepsilon_{{\bF k};\sigma}^-$ 
and $\varepsilon_{{\bF k};\sigma}^{<}$ coincide at ${\Bf k}=
{\Bf k}_{{\sc f};\sigma}$. Following Eq.~(\ref{e27}), whose validity 
depends on the discontinuity of ${\sf n}_{\sigma}({\Bf k})$ and 
continuity of $\varepsilon_{{\bF k};\sigma}^{<}$ at ${\Bf k}=
{\Bf k}_{\star}$, $\varepsilon_{{\bF k};\sigma}^-$ and 
$\varepsilon_{{\bF k};\sigma}^{<}$ also coincide at ${\Bf k}=
{\Bf k}_{\star}$. This situation is schematically illustrated in 
Fig.~\ref{fi2} where the distinctions between the two energy 
dispersions is depicted as being minute over the entire interval 
between ${\Bf k}_{\star}$ and ${\Bf k}_{{\sc f};\sigma}$ along a
nodal direction of the 1BZ. With this picture in mind, the conformity 
of the result in Eq.~(\ref{e72}) (and indeed Eq.~(\ref{e63})) with the 
experimental results by Zhou {\sl et al.} \cite{XJZ03} (excluding
those corresponding to underdoped samples) becomes evident.

A convenient (though not necessarily an exact) representation of 
Eq.~(\ref{e63}) in terms of two angles $\theta$ and $\phi$ is 
obtained as follows. We take the inner product of both sides of 
Eq.~(\ref{e63}) times $\hbar$ with $\hat{\Bf n}({\Bf k}_{\star}) 
\equiv \hat{\Bf n}({\Bf k}_{{\sc f};\sigma})$ and multiply both 
sides of the resulting expression by 
\begin{equation}
\label{e73}
\gamma {:=} 
\frac{ \|{\Bf k}_{{\sc f};\sigma} \|}{\varepsilon_{\sc f}}
\end{equation}
so as to obtain a relationship between dimensionless quantities. 
Assuming 
$\|{\Bf k}_{{\sc f};\sigma}-{\Bf k}_{\star}\|/\|{\Bf k}_{{\sc f};
\sigma} \|$ to be small, for similarly small 
$\|{\Bf k}-{\Bf k}_{\star}\|/\|{\Bf k}_{{\sc f};\sigma}\|$ one 
can consider $\varepsilon_{{\bF k};\sigma}^-$ as being a linear 
function of ${\Bf k}$, passing, along the nodal direction of the
1BZ, through the point $({\Bf k},\varepsilon) = ({\Bf k}_{{\sc f};
\sigma},\varepsilon_{\sc f})$ at angle $\theta$ with respect 
to the wave-vector axis (see Fig.~\ref{fi2}), where
\begin{equation}
\label{e74}
\theta = 
\tan^{-1}\big(\gamma\,\hat{\Bf n}({\Bf k}_{{\sc f};\sigma})\cdot 
{\Bf\nabla}_{\bF k}\varepsilon_{{\bF k};\sigma}^-\vert_{{\bF k}
={\bF k}_{{\sc f};\sigma}^-} \big).
\end{equation}
Here $0< \theta < \pi/2$, following the convention of counting 
angles as positive when they imply counterclockwise rotation 
with respect to point $(1,1)$ in Fig.~\ref{fi2}. Following 
Eqs.~(\ref{e4}) and (\ref{e27}), to a good approximation 
${\Bf v}_{{\bF k};\sigma}^-$ should be equal to ${\Bf v}_{{\bF k};
\sigma}^{<}$ in the interval between ${\Bf k}_{\star}$ and the nodal 
Fermi point ${\Bf k}_{{\sc f};\sigma}$; in the case of particles 
interacting through the long-range interaction Coulomb potential, 
the agreement between ${\Bf v}_{{\bF k};\sigma}^-$ and 
${\Bf v}_{{\bF k};\sigma}^{<}$ in the neighbourhood of ${\Bf k}
={\Bf k}_{{\sc f};\sigma}$ is the better the smaller the value of 
$Z_{{\bF k}_{{\sc f};\sigma}}$, ideally when $Z_{{\bF k}_{{\sc f};
\sigma}}=0$ (see \S~4.1). For ${\Bf k}$ satisfying $\|{\Bf k}\| < 
\|{\Bf k}_{\star}\|$, ${\Bf v}_{{\bF k};\sigma}^-$ and ${\Bf v}_{{\bF k};
\sigma}^{<}$ depart according to Eq.~(\ref{e63}). Since this aspect 
is a direct consequence of $Z_{{\bF k}_{\star};\sigma} > 0$ (see 
Eq.~(\ref{e63})), it follows that in the neighbourhood of ${\Bf k}
={\Bf k}_{\star}$ the behaviour of ${\Bf v}_{{\bF k};\sigma}^{<}$
in relation to ${\Bf v}_{{\bF k};\sigma}^-$ very crucially depends 
on the range of the two-body interaction potential. In view of the 
experimental results obtained by Zhou {\sl et al.} \cite{XJZ03}, we 
believe that the influence of the long range of interaction on 
${\Bf v}_{{\bF k};\sigma}^{<}$ is unequivocally present for samples 
in the underdoped regime; for the samples in the optimally-doped
and overdoped regimes, this aspect is not as unequivocal as in the 
case of the underdoped samples. It is reasonable to believe that this 
feature has its root in the magnitude of $Z_{{\bF k}_{\star};\sigma}$ 
which for an increasing level of (hole) doping should be decreasing, 
rendering thus the influence of the long range of the two-body Coulomb 
potential at ${\Bf k}={\Bf k}_{\star}$ less effective. In order to 
maintain a non-vanishing amount of discontinuity in ${\Bf v}_{{\bF k};
\sigma}^{<}$ at ${\Bf k}={\Bf k}_{\star}$, however, following 
Eq.~(\ref{e63}), the decrease in $Z_{{\bF k}_{\star};\sigma}$ should 
be accompanied by a concomitant decrease in ${\sf n}_{\sigma}
({\Bf k}_{\star}^-)$ in such a way that the rate of decrease
in $Z_{{\bF k}_{\star};\sigma}/{\sf n}_{\sigma}({\Bf k}_{\star}^-)$ 
(this decrease is evident from the the experimental results 
in \cite{XJZ03}) for the increase in the level of hole doping is 
not as strong as would be the case for a doping independent 
${\sf n}_{\sigma}({\Bf k}_{\star}^-)$. 

For an interaction potential of shorter range than the Coulomb 
potential, it is expected that the behaviour of $\varepsilon_{{\bF k};
\sigma}^{<}$ is semilinear
\footnote{\label{f21}
That is, it is linear, but its extension does not pass through the 
`origin' $(1,1)$ in Fig.~\protect\ref{fi2}. As we shall discuss in
some length in \S~5 below, interestingly the experimental energy 
dispersions in Fig.~1a of \protect\cite{XJZ03}, when linearly
extrapolated from the region $\|{\Bf k}\| < \|{\Bf k}_{\star}\|$ (i.e. 
at `high' binding energies) to ${\Bf k}={\Bf k}_{{\sc f};\sigma}$, 
all turn out to meet the energy axis at $75$ meV above the Fermi 
energy $\varepsilon_{\sc f}$.  }
for sufficiently small values of $\|{\Bf k}-{\Bf k}_{\star}\|$ in the 
region $\|{\Bf k}\| < \|{\Bf k}_{\star}\|$. This semi-linear line 
stands at angle $\phi$ with respect to the linear line depicting 
$\varepsilon_{{\bF k};\sigma}^-$ in Fig.~\ref{fi2}, for which we have
\begin{eqnarray}
\label{e75}
\phi &\equiv& \theta -
\tan^{-1}\big(\gamma\,
\hat{\Bf n}({\Bf k}_{{\sc f};\sigma})\cdot
{\Bf\nabla}_{\bF k} 
\varepsilon_{{\bF k};\sigma}^{<}\vert_{{\bF k}=
{\bF k}_{{\sc f};\sigma}^{-}} \big) 
\nonumber\\
&\approx& \frac{ -Z_{{\bF k}_{\star};\sigma} }
{ {\sf n}_{\sigma}({\Bf k}_{\star}^-) }\, \theta.
\end{eqnarray} 
Here $-\pi < \phi < 0$, following the convention stated above in 
connection with $\theta$ in Eq.~(\ref{e74}). 

\begin{figure}[t!]
\protect
\centerline{
\psfig{figure=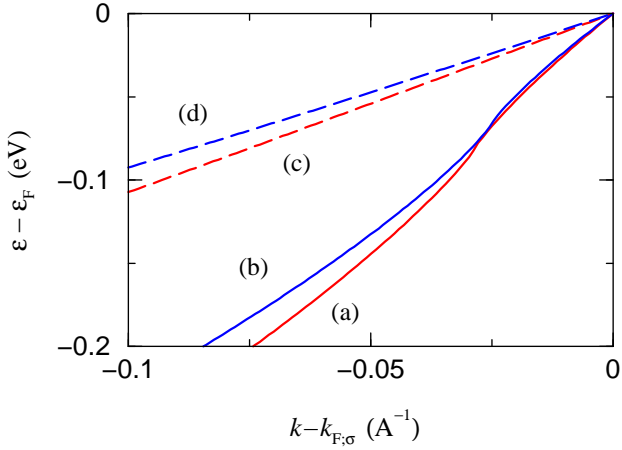,width=3.25in} }
\vskip 5pt
\caption{\label{fi3} \sf
The energy dispersion $\varepsilon_{{\bF k};\sigma}^{<} \equiv 
\varepsilon_{\bF k} + {\sf g}\, \xi_{{\bF k};\sigma}$
(Eq.~(\protect\ref{e39})) for $n {:=} N/N_{\sc l}$ equal to $1$ 
(curve (a), corresponding to the extreme limit of hole underdoping) 
and $0.7$ (curve (b), corresponding to an overdoped cuprate compound) 
along a diagonal direction of a square 1BZ as calculated within 
the framework in which ${\sf g}\, \xi_{{\bF k};\sigma}$ is replaced 
by $\hbar\Sigma_{\sigma}^{\sc hf}({\Bf k})$ (for the details
concerning this choice see the \S~4.1 and appendix A; see also 
Fig.~\protect\ref{fi2}) evaluated in terms of the long-range Coulomb 
potential and a GS momentum distribution function ${\sf n}_{\sigma}
({\Bf k})$ of the form ${\sf n}_{\sigma}(x {\Bf k}_{{\sc f};\sigma}) 
= 1-h_0\Theta(x-x_0)-Z_{{\bF k}_{{\sc f};\sigma}} \Theta(x-1)$ 
where $0\le x_0 \le 1$ and $h_0 = (1-Z_{{\bF k}_{{\sc f};\sigma}}) 
(x_{\sc z}^2-1)/(x_{\sc z}^2 -x_0^2)$; here ${\Bf k}_{\sc z}\equiv
x_{\sc z}\, {\Bf k}_{{\sc f};\sigma}$ stands for the vector from 
the centre of the 1BZ to the zone boundary in the direction of 
${\Bf k}_{{\sc f};\sigma} \in {\cal S}_{{\sc f};\sigma}$. Full account 
has been taken of the Umklapp of wave vectors that in the process of 
wave-vector integration move outside the 1BZ. The calculations are 
performed for $x_0=0.95$ and $Z_{{\bF k}_{{\sc f};\sigma}}=0$, 
$\forall {\Bf k}_{{\sc f};\sigma}$, so that by normalization
$h_0=0.968$ for $n=1$ and $h_0=0.976$ for $n=0.7$. For the energy 
dispersions $\varepsilon_{\bF k}$ (curve (c) for $n=1$ and curve 
(d) for $n=0.7$) we have chosen the tight-binding expression 
$\varepsilon_{\bF k} = -2 t [ \cos(a_0 k_x) + \cos(a_0 k_y)]$ where 
$k_x$ and $k_y$ stand for the Cartesian coordinates of ${\Bf k}$ and 
$a_0=3.895 \times 10^{-10}$~m for the lattice constant. For the 
Fourier transform of the Coulomb potential we have 
$\bar{v}(\|{\Bf q}\|) \equiv {\sf g}\, \bar{w}(\|{\Bf q}\|)$ where 
$\bar{w}(\|{\Bf q}\|) = 2\pi a_0/\|{\Bf q}\|$ and ${\sf g} = 
e^2/(4\pi\epsilon_0\epsilon_{\rm r} a_0)$ in which $e^2$ stands for 
the square of the electron charge, $\epsilon_0 = 8.854\dots\times 
10^{-12}$ F~m$^{-1}$ for the vacuum permittivity and $\epsilon_{\rm r}$ 
the relative dielectric constant of the background. With $e_0 {:=} 
\hbar^2/(m_{\star} a_0^2)$, the calculations presented here correspond 
to ${\sf g} = 3 e_0$. Here $m_{\star}$ denotes the effective mass 
associated with $\varepsilon_{\bF k}$ (to be distinguished from that 
associated with $\varepsilon_{{\bF k};\sigma}^{<}$) at ${\Bf k}
={\Bf k}_{{\sc f};\sigma}$ along a diagonal direction of the 1BZ, 
for which we have chosen $m_{\star} = 4 m_e$ (for both densities 
$n=1$ and $n=0.7$), where $m_e$ stands for the bare electron mass 
(for the above values of $a_0$ and $m_{\star}$, $e_0=0.1255$~eV; 
the choice ${\sf g} = 3 e_0$ thus amounts to $\epsilon_{\rm r} = 
9.814$). Consequently, $t= e_0 \kappa_{\sc f}/[2\sqrt{2} 
\sin(\kappa_{\sc f}/\sqrt{2})]$, where $\kappa_{\sc f} {:=} a_0 
\|{\Bf k}_{{\sc f};\sigma}\|$. For $n=1$ we have $\kappa_{\sc f} = 
2.221$ ($\kappa_{\sc f}/[\sqrt{2}\pi]=0.5$) and for $n=0.7$, 
$\kappa_{\sc f} = 1.965$ ($\kappa_{\sc f}/[\sqrt{2}\pi]=0.442$), 
corresponding to $t=0.785 e_0$ for $n=1$ and $t=0.706 e_0$ for 
$n=0.7$. A density-dependent $t$ is chosen to avoid a strongly
density-dependent $m_{\star}$. }
\end{figure}

In Fig.~\ref{fi3} we present results concerning 
$\varepsilon_{{\bF k};\sigma}^{<} \equiv \varepsilon_{\bF k} + 
{\sf g}\,\xi_{{\bF k};\sigma}$ within an approximate framework 
in which $\xi_{{\bF k};\sigma}$ has the functional form of the 
Hartree-Fock self-energy (see appendix A; Eq.~(\ref{ea21})), 
evaluated in terms of a momentum distribution function 
${\sf n}_{\sigma}({\Bf k})$ which is continuous at the nodal 
Fermi point ${\Bf k}_{{\sc f};\sigma}$ and discontinuous at 
${\Bf k}={\Bf k}_{\star}$. The results in Fig.~\ref{fi3} concern 
a planar square-lattice model (for the relevant parameters see 
the caption of the figure) and the two solid curves correspond to 
two densities of particles representing the underdoped and 
overdoped regions of the hole-doped cuprates. We obtain almost 
identical results as in Fig.~\ref{fi3} by employing an isotropic 
model defined on the continuum and for one, two and three spatial 
dimensions;
\footnote{\label{f22}
With $\bar{v}(\|{\Bf q}\|) = {\sf g}\, \bar{w}(\|{\Bf q}\|)$ and 
${\sf g} = e^2/(4\pi \epsilon_0 \epsilon_{\rm r} a_0)$ (for the 
specification of the quantities encountered here, see the caption 
of Fig.~\protect\ref{fi3}), we have employed the following 
$\bar{w}(\|{\Bf q}\|)$: 
For $d=3$,
$\bar{w}(\|{\Bf q}\|) = 4\pi a_0/\|{\Bf q}\|^2$;
for $d=2$, 
$\bar{w}(\|{\Bf q}\|) = 2\pi a_0/\|{\Bf q}\|$;
for $d=1$,
$\bar{w}(\|{\Bf q}\|) = a_0 \exp(\|{\Bf q}\|/q_0)\,
{\rm E}_1(\|{\Bf q}\|/q_0)$, where ${\rm E}_1(z)$ is the 
exponential-integral function and $q_0 > 0$ is a constant parameter 
of dimensions of reciprocal metre. }
our observations concerning $d=1$ are particularly relevant in light 
of the fact that in the superconducting state, or in the advanced 
stages of the pseudogap phase, the experimental observations along 
the nodal direction of the 1BZ should be influenced by the reduced 
wave-vector space available to gapless excitations. Not surprisingly, 
our numerical calculations for $d=2$ reveal that similarly all 
qualitative aspects of the results in Fig.~\ref{fi3} remain intact 
by modeling ${\sf n}_{\sigma}({\Bf k})$ in such a way that the 
discontinuity in ${\sf n}_{\sigma}({\Bf k})$ is anisotropic, taking 
its maximum value at ${\Bf k}={\Bf k}_{\star}$ along the diagonal 
directions of the 1BZ and diminishing for directions away from the 
diagonal directions.

Considering the fact that, in real materials, charged fermions interact 
through the long-range Coulomb potential, the experimental results by 
Zhou {\sl et al.} \cite{XJZ03}, as well as those by other workers 
\cite{TV99,PVB00,PDJ01,AK01,AL01}, should be viewed as consisting of 
a superposition of the results in Figs.~\ref{fi2} and \ref{fi3}; here 
we are leaving aside the possibility of a discontinuity in 
$\varepsilon_{{\bF k};\sigma}^{<}$ at ${\Bf k}={\Bf k}_{\star}$ which 
is theoretically feasible (see \S~3.4) and appears to be realized in 
the underdoped (La$_{2-x}$Sr$_x$)CuO$_4$ \cite{XJZ03}, specifically 
at $x=0.03$. 

\vspace{0.4cm}
\noindent{\bf 4.3. \small Explicit calculation of the amount of 
discontinuity in ${\Bf\nabla}_{\bF k} \varepsilon_{{\bF k};\sigma}^{<}$
at ${\Bf k}={\Bf k}_{\star}$, with ${\Bf k}_{\star} \in 
{\rm FS}_{\sigma}\backslash {\cal S}_{{\sc f};\sigma}$ }
\label{s4c}
\vspace{0.3cm}

Here we present the details underlying the derivation of the
result in Eq.~(\ref{e63}) above. For definiteness, unless we 
explicitly indicate otherwise, throughout this section we assume 
$\varepsilon_{{\bF k};\sigma}^{<}$ to be continuous at ${\Bf k}
={\Bf k}_{\star}$ where ${\sf n}_{\sigma}({\Bf k})$ is considered 
discontinuous (see \S~3.4).

\begin{figure}[t!]
\protect
\centerline{
\psfig{figure=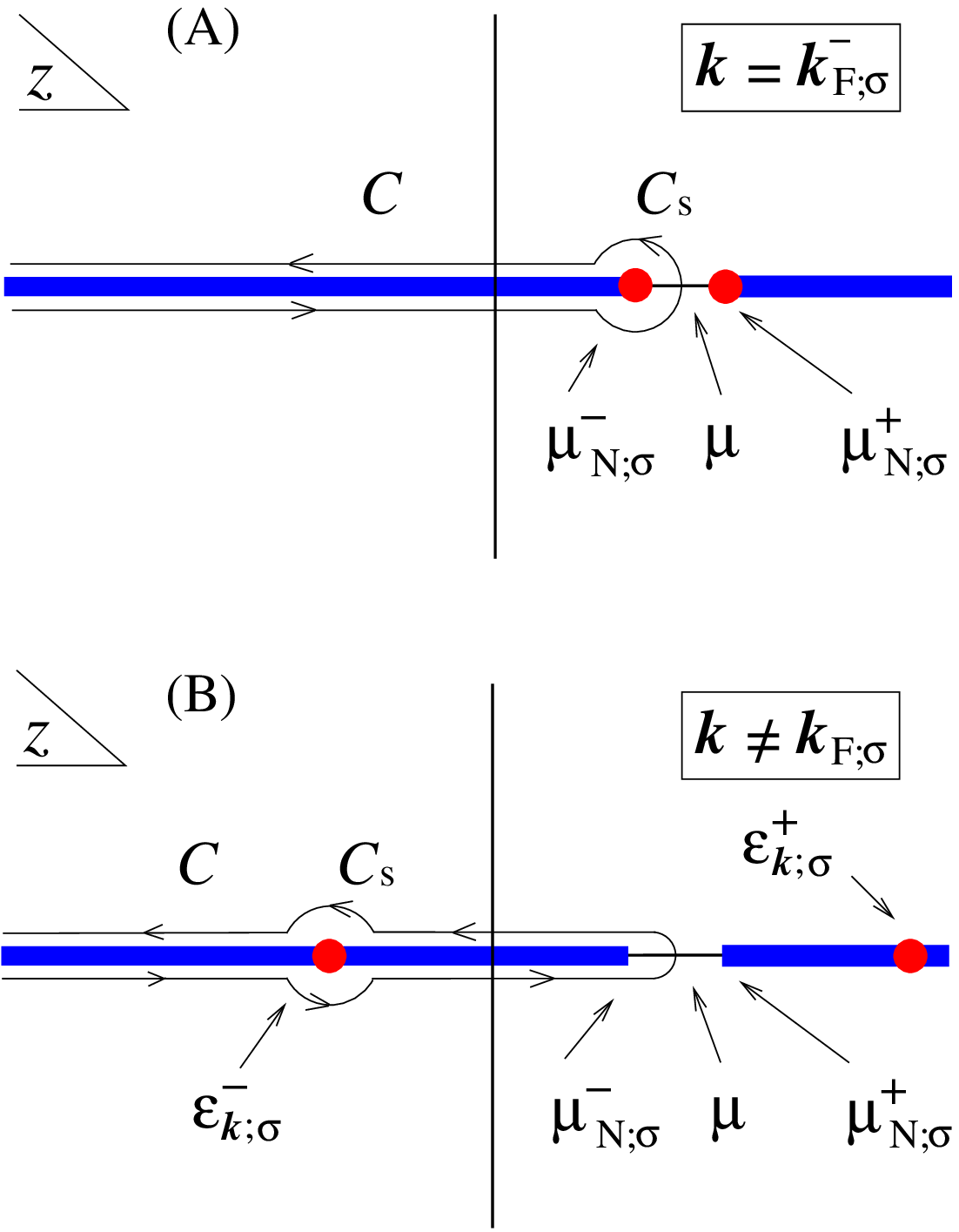,width=2.5in} }
\vskip 5pt
\caption{\label{fi4} \sf
The analytic structure of the single-particle Green function 
$\wt{G}_{\sigma}({\Bf k};z)$ on the physical Riemann sheet of 
the complex $z$ plane for cases where 
(A) ${\Bf k} = {\Bf k}_{{\sc f};\sigma}^-$, with 
${\Bf k}_{{\sc f};\sigma} \in {\cal S}_{{\sc f};\sigma}$ and 
(B) ${\Bf k} \in {\rm FS}_{\sigma}$ and at a finite non-vanishing
distance from ${\cal S}_{{\sc f};\sigma}$. It is assumed that 
in the latter case for the ${\Bf k}$ at issue the quasi-particle 
equation $\varepsilon_{\bF k} + \hbar\Sigma_{\sigma}({\Bf k};
\varepsilon) = \varepsilon$ is satisfied for $({\Bf k},\varepsilon)
=({\Bf k}^{\mp},\varepsilon_{{\bF k}^{\mp};\sigma})$ (see 
Fig.~\protect\ref{fi1}) in which $\varepsilon_{{\bF k}^{\mp};\sigma}
\equiv \varepsilon_{{\bF k};\sigma}^{\mp}$ (Eq.~(\protect\ref{e13})); 
for metallic states and ${\Bf k}_{{\sc f};\sigma} \in 
{\cal S}_{{\sc f};\sigma}$, this equation is satisfied for $({\Bf k},
\varepsilon)=({\Bf k}_{{\sc f};\sigma}^-,\varepsilon_{\sc f})$. 
The solid lines on the real axes whose end points coincide with 
$\mu_{N;\sigma}^- {:=} E_{N_{\sigma},N_{\bar\sigma};0} -
E_{N_{\sigma}-1,N_{\bar\sigma};0}$ and $\mu_{N;\sigma}^+ {:=} 
E_{N_{\sigma}+1,N_{\bar\sigma};0} - E_{N_{\sigma},N_{\bar\sigma};0}$, 
represent the branch cuts of $\wt{G}_{\sigma}({\Bf k};z)$; here 
$E_{M_{\sigma},N_{\bar\sigma};0}$, with $M_{\sigma} = N_{\sigma}-1, 
N_{\sigma}, N_{\sigma}+1$, stands for the total energy of the 
$(M_{\sigma}+N_{\bar\sigma})$-particle GS of the interacting 
Hamiltonian $\wh{H}$ (for $\sigma =\uparrow$, $\bar\sigma=\downarrow$, 
and vice versa). For metallic GSs, $\mu_{N;\sigma}^- = 
\varepsilon_{\sc f}$ and $\mu_{N;\sigma}^+$ differs only 
infinitesimally from $\mu_{N;\sigma}^-$; the chemical potential $\mu$ 
satisfies $\mu_{N;\sigma}^- <\mu <\mu_{N;\sigma}^+$. In evaluating 
the integrals of $z^m \wt{G}_{\sigma}({\Bf k};z)$, $m=0,1$, over 
the contour $C$ we employ the decomposition $\int_{C} {\rm d}z = 
\int_{C\backslash C_{\rm s}} {\rm d}z +\int_{C_{\rm s}} {\rm d}z$ 
which amounts to the decomposition, according to 
Eq.~(\protect\ref{e8}), of $A_{\sigma}({\Bf k};\varepsilon)$ into 
its regular and singular contributions respectively. }
\end{figure}

The expression for $\varepsilon_{{\bF k};\sigma}^{<}$ as presented 
in Eq.~(\ref{e39}) is not suited for the determination of 
${\Bf\nabla}_{\bF k}\varepsilon_{{\bF k};\sigma}^{<}$, on account 
of the fact that Eq.~(\ref{e39}) describes $\varepsilon_{{\bF k};
\sigma}^{<}$ in terms of a distribution (i.e. $A_{\sigma}({\Bf k};
\varepsilon)$; see text following Eq.~(\ref{e8}) above) while the 
available algebraic machinery required in our approach is effective 
for functions. To proceed we therefore define
\begin{equation}
\label{e76}
I_{m;\sigma}({\Bf k}) {:=}
\frac{1}{\hbar}\int_{\cal C} \frac{ {\rm d} z}{2\pi i}\;
z^m\, \wt{G}_{\sigma}({\Bf k};z),\;\;\; m = 0,1,
\end{equation}
in which the contour ${\cal C}$ has been depicted in Fig.~\ref{fi4}. 
For definiteness, we consider ${\cal C}$ to initiate and terminate 
at the point of infinity in the complex $z$ plane, corresponding 
to $E=\infty$ in Eqs.~(\ref{e5}) and (\ref{e6}). It can be readily 
verified that 
\begin{eqnarray}
\label{e77}
I_{0;\sigma}({\Bf k}) &\equiv& 
{\sf n}_{\sigma}({\Bf k}), \\
\label{e78}
I_{1;\sigma}({\Bf k}) &\equiv&
{\sf n}_{\sigma}({\Bf k})\,
\varepsilon_{{\bF k};\sigma}^{<}.
\end{eqnarray}
Making use of these results, from the defining expression in 
Eq.~(\ref{e39}) we have
\begin{equation}
\label{e79}
{\Bf\nabla}_{\bF k}\varepsilon_{{\bF k};\sigma}^{<}
= \frac{ {\Bf\nabla}_{\bF k} I_{1;\sigma}({\Bf k})
-\varepsilon_{{\bF k};\sigma}^{<}
{\Bf\nabla}_{\bF k} I_{0;\sigma}({\Bf k})}
{ {\sf n}_{\sigma}({\Bf k}) }. 
\end{equation}
This expression has the generic form of $h(x)$ defined in 
Eq.~(\ref{e17}) so that, in determining the amount of possible 
discontinuity in ${\Bf\nabla}_{\bF k}\varepsilon_{{\bF k};
\sigma}^{<}$, we shall rely on the result in Eq.~(\ref{e22}).

We assume ${\sf n}_{\sigma}({\Bf k})$ to be discontinuous at 
${\Bf k}={\Bf k}_{\star}$. In what follows we consider the case 
in which ${\Bf k}_{\star}$ lies strictly inside the Fermi sea, that 
is at a non-vanishing distance from ${\cal S}_{{\sc f};\sigma}$. On 
account of the assumed discontinuity of ${\sf n}_{\sigma}({\Bf k})$ 
at ${\Bf k}={\Bf k}_{\star}$, following Eq.~(\ref{e27}) we can 
replace $\varepsilon_{{\bF k};\sigma}^{<}$ on the RHS of 
Eq.~(\ref{e79}) by $\varepsilon_{{\bF k}^-;\sigma}\equiv
\varepsilon_{{\bF k};\sigma}^-$.
\footnote{\label{f23}
With reference to our remarks in footnote \protect\ref{f9},
here we are implicitly assuming that $M_{\sigma}^{\mp}=1$. }
The assumption with regard to the discontinuity of ${\sf n}_{\sigma}
({\Bf k})$ at ${\Bf k}={\Bf k}_{\star}$ implies that 
${\rm d}\Sigma_{\sigma}({\Bf k}_{\star}^-;
\varepsilon)/{\rm d}\varepsilon$ is bounded in the neighbourhood 
of $\varepsilon=\varepsilon_{{\bF k}_{\star};\sigma}^-$ (similarly 
for ${\rm d}\Sigma_{\sigma}({\Bf k}_{\star}^+;
\varepsilon)/{\rm d}\varepsilon$ in the neighbourhood of 
$\varepsilon=\varepsilon_{{\bF k}_{\star};\sigma}^+$ (see
Fig.~\ref{fi1})). Consequently, for ${\Bf k}$ in the neighbourhood 
of ${\Bf k}_{\star}$ we can write
\begin{eqnarray}
\label{e80}
\wt{\Sigma}_{\sigma}({\Bf k};z) &=&
\Sigma_{\sigma}({\Bf k};\varepsilon_{{\bF k}_{\star};\sigma}^{\mp})
+ \beta_{{\bF k};\sigma}\, 
(z - \varepsilon_{{\bF k}_{\star};\sigma}^{\mp}) \nonumber\\
&+& o(z-\varepsilon_{{\bF k}_{\star};\sigma}^{\mp})
\;\;\;\mbox{\rm for}\;\;\;
z \to \varepsilon_{{\bF k}_{\star};\sigma}^{\mp},
\end{eqnarray}
where $\beta_{{\bF k};\sigma}$ (not to be confused with 
$\beta_{{\bF k};\sigma}^{<}$ introduced in Eq.~(\ref{e40}) above) 
is a function which is bounded at ${\Bf k}={\Bf k}_{\star}$ and 
$o(z-\varepsilon_{{\bF k}_{\star};\sigma}^{\mp})$ denotes a function 
which asymptotically is subdominant with respect to 
$(z-\varepsilon_{{\bF k}_{\star};\sigma}^{\mp})$ for $z \to 
\varepsilon_{{\bF k}_{\star};\sigma}^{\mp}$. In Eq.~(\ref{e80}), 
$\wt{\Sigma}_{\sigma}({\Bf k};z)$ denotes the analytic continuation of 
$\Sigma_{\sigma}({\Bf k};\varepsilon)$ into the physical Riemann sheet 
of the complex $z$ plane, from which $\Sigma_{\sigma}({\Bf k};\varepsilon)$ 
is obtained according to $\Sigma_{\sigma}({\Bf k};\varepsilon) = 
\lim_{\eta\downarrow 0} \wt{\Sigma}_{\sigma}({\Bf k};\varepsilon \pm 
i\eta)$, $\varepsilon\, {\Ieq<>}\, \mu$. It is trivially verified that
\begin{equation}
\label{e81}
Z_{{\bF k}_{\star};\sigma} =
\frac{1}{1 -\hbar \beta_{{\bF k}_{\star};\sigma} }.
\end{equation}
In our subsequent
considerations we assume ${\Bf\nabla}_{\bF k} \beta_{{\bF k};\sigma}$ 
to be bounded for ${\Bf k}$ in a neighbourhood of ${\Bf k}_{\star}$ 
(see Eq.~(\ref{e94}) below). For ${\sf n}_{\sigma}({\Bf k})$ 
discontinuous at ${\Bf k}={\Bf k}_{\star}$ and ${\Bf k}$ in a 
neighbourhood of ${\Bf k}_{\star}$, we employ the decomposition 
${\cal C} = {\cal C}_{\rm s} \cup {\cal C}_{\rm r}$ where 
${\cal C}_{\rm r} \equiv {\cal C}\backslash {\cal C}_{\rm s}$; the 
contour ${\cal C}_{\rm s}$ consists of the union of two semi-circles 
of infinitesimal radius centred at $\varepsilon_{{\bF k}_{\star};
\sigma}^-$ on the real axis of the complex $z$ plane (see 
Fig.~\ref{fi4}). Thus we write
\begin{equation}
\label{e82}
I_{m;\sigma}({\Bf k}) \equiv I_{m;\sigma}^{(\rm s)}({\Bf k})
+ I_{m;\sigma}^{(\rm r)}({\Bf k}),
\end{equation}
where $I_{m;\sigma}^{(\rm s)}({\Bf k})$ and  $I_{m;\sigma}^{(\rm r)}
({\Bf k})$ are the contributions due to contours ${\cal C}_{\rm s}$ 
and ${\cal C}_{\rm r}$ respectively. The decomposition of 
$I_{m;\sigma}({\Bf k})$ according to Eq.~(\ref{e82}) can be viewed 
as arising from the decomposition of $A_{\sigma}({\Bf k};\varepsilon)$ 
in terms of its regular and singular contributions, as presented in 
Eq.~(\ref{e8}) above. Following our considerations in \S\S~3.1 and 
3.2, for ${\Bf k}$ at which Eq.~(\ref{e3}) has no solution in the 
interval $(-\infty,\mu)$, $I_{m;\sigma}^{(\rm s)}({\Bf k})$ is 
identically vanishing. 

From Eq.~(\ref{e79}), making use of the general result in
Eq.~(\ref{e22}), we have
\footnote{\label{f24}
Below (as well as elsewhere in this paper) $\mp$ in 
${\Bf\nabla}_{\bF k} f({\Bf k})\vert_{{\bF k}={\bF k}_{\star}^{\mp}}$ 
denotes left/right gradients of $f({\Bf k})$, defined as the 
{\sl limit} of ${\Bf\nabla}_{\bF k} f({\Bf k})$ for ${\Bf k}$ 
approaching ${\Bf k}_{\star}$ from the left/right; here 
left/right is defined by $\hat{\Bf n}({\Bf k}_{\star}) \cdot 
({\Bf k}-{\Bf k}_{\star}) \,\Ieq{>}{<}\, 0$. For the definition 
of $\hat{\Bf n}({\Bf k}_{\star})$ see text preceding 
Eq.~(\protect\ref{e72}) above. } 
\begin{eqnarray}
\label{e83}
&&{\Bf\nabla}_{\bF k}
\varepsilon_{{\bF k};\sigma}^{<}\vert_{{\bF k}={\bF k}_{\star}^-} 
-{\Bf\nabla}_{\bF k}
\varepsilon_{{\bF k};\sigma}^{<}\vert_{{\bF k}={\bF k}_{\star}^+} 
\nonumber\\
&&\;\;\;\;
= \frac{1}{{\sf n}_{\sigma}({\Bf k}_{\star}^+) }
\Big\{\left.\Big( {\Bf\nabla}_{\bF k} I_{1;\sigma}({\Bf k})
- \varepsilon_{{\bF k}_{\star};\sigma}^-
{\Bf\nabla}_{\bF k} I_{0;\sigma}({\Bf k}) 
\Big)\right|_{{\bF k}={\bF k}_{\star}^-} \nonumber\\
&&\;\;\;\;\;\;\;\;\;\;\;\;\;\;\;\;\;\;\;\;
\left. -\Big( {\Bf\nabla}_{\bF k} I_{1;\sigma}({\Bf k})
- \varepsilon_{{\bF k}_{\star};\sigma}^-
{\Bf\nabla}_{\bF k} I_{0;\sigma}({\Bf k}) 
\Big)\right|_{{\bF k}={\bF k}_{\star}^+} \nonumber\\
&&\;\;\;\;\;\;\;\;\;\;\;\;\;\;\;\;\;\;\;\;
- Z_{{\bF k}_{\star};\sigma}
{\Bf\nabla}_{\bF k} 
\varepsilon_{{\bF k};\sigma}^{<}\vert_{{\bF k}=
{\bF k}_{\star}^-} \Big\}.
\end{eqnarray}
In arriving at Eq.~(\ref{e83}) we have made use of the assumed 
continuity of $\varepsilon_{{\bF k};\sigma}^{<}$ in a neighbourhood 
of ${\Bf k}={\Bf k}_{\star}$, implying, in conjunction with the 
assumption of discontinuity of ${\sf n}_{\sigma}({\Bf k})$ at 
${\Bf k}={\Bf k}_{\star}$, the equality of 
$\varepsilon_{{\bF k}_{\star};\sigma}^{<}$ with 
$\varepsilon_{{\bF k}_{\star};\sigma}^{-}$ (see Eq.~(\ref{e27})). 

The subsequent notation will be considerably simplified by 
introducing the following decomposition
\footnote{\label{f25}
Note that here
${\Bf\nabla}_{\bF k}^{(\rm x)}\varepsilon_{{\bF k};\sigma}^{<}$,
${\rm x=r,s}$, stands for an entire symbol, that is 
${\Bf\nabla}_{\bF k}^{(\rm x)}$ does not on its own denote 
an independent operation. }
\begin{equation}
\label{e84}
{\Bf\nabla}_{\bF k}
\varepsilon_{{\bF k};\sigma}^{<} \equiv
{\Bf\nabla}_{\bF k}^{(\rm r)} 
\varepsilon_{{\bF k};\sigma}^{<} +
{\Bf\nabla}_{\bF k}^{(\rm s)} \varepsilon_{{\bF k};\sigma}^{<},
\end{equation}
where 
\begin{equation}
\label{e85}
{\Bf\nabla}_{\bF k}^{(\rm r)} 
\varepsilon_{{\bF k};\sigma}^{<}\vert_{{\bF k}={\bF k}_{\star}^-}
= {\Bf\nabla}_{\bF k}^{(\rm r)} 
\varepsilon_{{\bF k};\sigma}^{<}\vert_{{\bF k}={\bF k}_{\star}^+},
\end{equation}
and 
\begin{equation}
\label{e86}
{\Bf\nabla}_{\bF k}^{(\rm s)} 
\varepsilon_{{\bF k};\sigma}^{<}
\vert_{{\bF k}={\bF k}_{\star}^+} = 0.
\end{equation}
In this way, making use of 
\begin{equation}
\label{e87}
1+\frac{Z_{{\bF k}_{\star};\sigma}}
{{\sf n}_{\sigma}({\Bf k}_{\star}^+)} \equiv 
\frac{{\sf n}_{\sigma}({\Bf k}_{\star}^-)}
{{\sf n}_{\sigma}({\Bf k}_{\star}^+)},
\end{equation}
Eq.~(\ref{e83}) can be written in the following equivalent form
\begin{eqnarray}
\label{e88}
&&{\Bf\nabla}_{\bF k}^{(\rm s)}
\varepsilon_{{\bF k};\sigma}^{<}\vert_{{\bF k}={\bF k}_{\star}^-} 
\nonumber\\
&&\;\;\;\;
= \frac{1}{{\sf n}_{\sigma}({\Bf k}_{\star}^-) }
\Big\{\left.
\Big( {\Bf\nabla}_{\bF k} I_{1;\sigma}({\Bf k})
- \varepsilon_{{\bF k}_{\star};\sigma}^-
{\Bf\nabla}_{\bF k} I_{0;\sigma}({\Bf k}) 
\Big)\right|_{{\bF k}={\bF k}_{\star}^-} \nonumber\\
&&\;\;\;\;\;\;\;\;\;\;\;\;\;\;\;\;\;\;\;\;
\left. -\Big( {\Bf\nabla}_{\bF k} I_{1;\sigma}({\Bf k})
- \varepsilon_{{\bF k}_{\star};\sigma}^-
{\Bf\nabla}_{\bF k} I_{0;\sigma}({\Bf k}) 
\Big)\right|_{{\bF k}={\bF k}_{\star}^+} \nonumber\\
&&\;\;\;\;\;\;\;\;\;\;\;\;\;\;\;\;\;\;\;\;
- Z_{{\bF k}_{\star};\sigma}
{\Bf\nabla}_{\bF k}^{(\rm r)} 
\varepsilon_{{\bF k};\sigma}^{<}\vert_{{\bF k}=
{\bF k}_{\star}^-} \Big\}.
\end{eqnarray}

We now proceed with the determination of the terms on the RHS of 
Eq.~(\ref{e88}). It is readily verified that (see footnote \ref{f24})
\begin{equation}
\label{e89}
{\Bf\nabla}_{\bF k} 
I_{0;\sigma}^{(\rm r)}({\Bf k})\vert_{{\bF k}
={\bF k}_{\star}^{\mp}} =
\left. {\Bf\nabla}_{\bF k} {\sf n}_{\sigma}({\Bf k})\right|_{{\bF k} 
={\bF k}_{\star}^{\mp}},
\end{equation}
\begin{eqnarray}
\label{e90}
\left. {\Bf\nabla}_{\bF k}
I_{1;\sigma}^{(\rm r)}({\Bf k})\right|_{{\bF k}={\bF k}_{\star}^{\mp}} 
&=&\big({\Bf\nabla}_{\bF k} {\sf n}_{\sigma}({\Bf k})
\vert_{{\bF k}={\bF k}_{\star}^{\mp}}\big) 
\varepsilon_{{\bF k}_{\star}^{\mp};\sigma}^{<} \nonumber\\
&+& \big({\Bf\nabla}_{\bF k}^{(\rm r)} 
\varepsilon_{{\bF k};\sigma}^{<} 
\vert_{{\bF k}={\bF k}_{\star}^{\mp}}\big)
{\sf n}_{\sigma}({\Bf k}_{\star}^{\mp}).
\end{eqnarray}
Following the Dyson equation, making use of Eqs.~(\ref{e80}) and 
(\ref{e81}), we have 
\begin{eqnarray}
\label{e91}
\left. {\Bf\nabla}_{\bF k} \wt{G}_{\sigma}({\Bf k};z)
\right|_{{\bF k}={\bF k}_{\star}^-}
&\sim& \frac{\hbar^2 Z_{{\bF k}_{\star};\sigma} }
{ (z-\varepsilon_{{\bF k}_{\star}^-;\sigma})^2 }\,
{\Bf v}_{{\bF k}_{\star}^-;\sigma}(z), \nonumber\\
& &\;\;\;\;\;\;\;\;\;\;\;\;\;\;\;\;\;\;\;
\mbox{\rm for}\;\;\;
z\to \varepsilon_{{\bF k}_{\star}^-;\sigma},
\end{eqnarray}
where (see Eq.~(\ref{e69}))
\begin{equation}
\label{e92}
{\Bf v}_{{\bF k}_{\star}^-;\sigma}(z) {:=}
Z_{{\bF k}_{\star};\sigma} 
\big( \frac{1}{\hbar} \left. {\Bf\nabla}_{\bF k}
\varepsilon_{\bF k} \right|_{{\bF k}={\bF k}_{\star}^-}
+ \left. {\Bf\nabla}_{\bF k}
\wt{\Sigma}_{\sigma}({\Bf k};z) 
\right|_{{\bF k}={\bF k}_{\star}^-}\big).
\end{equation}
Thus, on the basis of Eqs.~(\ref{e76}) and (\ref{e91}) and 
the Cauchy theorem, we have
\begin{eqnarray}
\label{e93}
\left. {\Bf\nabla}_{\bF k} 
I_{m;\sigma}^{(\rm s)}({\Bf k})\right|_{{\bF k}={\bF k}_{\star}^-}
&=& \hbar Z_{{\bF k}_{\star};\sigma} 
\int_{{\cal C}_{\rm s}} \frac{ {\rm d}z}{2\pi i}\;
\frac{ z^m\, {\Bf v}_{{\bF k}_{\star}^-;\sigma}(z) }
{ (z - \varepsilon_{{\bF k}_{\star}^-;\sigma} )^2 } \nonumber\\
&=& \hbar Z_{{\bF k}_{\star};\sigma}\,
\left. \frac{\partial}{\partial z}
\big( z^m {\Bf v}_{{\bF k}_{\star}^-;\sigma}(z) \big)
\right|_{z=\varepsilon_{{\bF k}_{\star}^-;\sigma}}. 
\nonumber\\
\end{eqnarray}
From Eqs.~(\ref{e92}) and (\ref{e80}) we obtain
\begin{equation}
\label{e94}
\left. \frac{\partial}{\partial z}\, 
{\Bf v}_{{\bF k}_{\star}^-;\sigma}(z)\right|_{z=
\varepsilon_{{\bF k}_{\star}^-;\sigma} }
\!\!\! = Z_{{\bF k}_{\star};\sigma} \left.
{\Bf\nabla}_{\bF k} \beta_{{\bF k};\sigma} 
\right|_{{\bF k}={\bF k}_{\star}^-},
\end{equation}
so that from Eq.~(\ref{e93}) it follows that
\begin{equation}
\label{e95}
{\Bf\nabla}_{\bF k} \left. 
I_{0;\sigma}^{(\rm s)}({\Bf k})\right|_{{\bF k}={\bF k}_{\star}^-}
= \hbar Z_{{\bF k}_{\star};\sigma}^2
\left. {\Bf\nabla}_{\bF k} \beta_{{\bF k};\sigma} 
\right|_{{\bF k}={\bF k}_{\star}^-},
\end{equation}
\begin{eqnarray}
\label{e96}
&&{\Bf\nabla}_{\bF k} \left. 
I_{1;\sigma}^{(\rm s)}({\Bf k})\right|_{{\bF k}={\bF k}_{\star}^-}
= \hbar Z_{{\bF k}_{\star};\sigma} \nonumber\\
&&\;\;\;\;\;\;\;
\times\Big( {\Bf v}_{{\bF k}_{\star}^-;\sigma}
(\varepsilon_{{\bF k}_{\star}^-;\sigma})
+Z_{{\bF k}_{\star};\sigma}
\varepsilon_{{\bF k}_{\star}^-;\sigma}
\left. {\Bf\nabla}_{\bF k} \beta_{{\bF k};\sigma}
\right|_{{\bF k}={\bF k}_{\star}^-} \Big).
\end{eqnarray}

Combining the results in Eqs.~(\ref{e89}) and (\ref{e90}) we have
\begin{eqnarray}
\label{e97}
&&\left.\Big( {\Bf\nabla}_{\bF k} I_{1;\sigma}^{(\rm r)}({\Bf k})
- \varepsilon_{{\bF k}_{\star};\sigma}^-
{\Bf\nabla}_{\bF k} I_{0;\sigma}^{(\rm r)}({\Bf k}) 
\Big)\right|_{{\bF k}={\bF k}_{\star}^{\mp}} \nonumber\\
&&\;\;\;\;\;\;\;\;\;\;\;\;\;\;\;\;\;\;\;\;\;\;\;\;\;\;
= \big({\Bf\nabla}_{\bF k}^{(\rm r)}
\varepsilon_{{\bF k};\sigma}^{<}\vert_{{\bF k}
={\bF k}_{\star}^{\mp}}\big)\, 
{\sf n}_{\sigma}({\Bf k}_{\star}^{\mp}),
\end{eqnarray}
while combining Eqs.~(\ref{e95}) and (\ref{e96}) we obtain
\begin{equation}
\label{e98}
\left.\Big( {\Bf\nabla}_{\bF k} I_{1;\sigma}^{(\rm s)}({\Bf k})
- \varepsilon_{{\bF k}_{\star};\sigma}^-
{\Bf\nabla}_{\bF k} I_{0;\sigma}^{(\rm s)}({\Bf k}) 
\Big)\right|_{{\bF k}={\bF k}_{\star}^{-}} 
= \hbar Z_{{\bF k}_{\star};\sigma}\,
{\Bf v}_{{\bF k}_{\star};\sigma}^{-},
\end{equation}
where we have made use of (see Eqs.~(\ref{e69}) and (\ref{e92}))
\begin{eqnarray}
{\Bf v}_{{\bF k}_{\star};\sigma}^-
\equiv {\Bf v}_{{\bF k}_{\star}^-;\sigma}(z=
\varepsilon_{{\bF k}_{\star}^-;\sigma}). \nonumber
\end{eqnarray}
Evidently (see Eq.~(\ref{e86})),
\begin{equation}
\label{e99}
\left.\Big( {\Bf\nabla}_{\bF k} I_{1;\sigma}^{(\rm s)}({\Bf k})
- \varepsilon_{{\bF k}_{\star};\sigma}^-
{\Bf\nabla}_{\bF k} I_{0;\sigma}^{(\rm s)}({\Bf k}) 
\Big)\right|_{{\bF k}={\bF k}_{\star}^{+}} = 0.
\end{equation}
From Eqs.~(\ref{e88}), (\ref{e97}), (\ref{e98}), (\ref{e99}), 
making use of Eqs.~(\ref{e82}) and (\ref{e85}) we arrive at
\begin{eqnarray}
\label{e100}
&&{\Bf\nabla}_{\bF k}^{(\rm s)}
\varepsilon_{{\bF k};\sigma}^{<}\vert_{{\bF k}={\bF k}_{\star}^-} 
= \frac{\hbar Z_{{\bF k}_{\star};\sigma} }
{{\sf n}_{\sigma}({\Bf k}_{\star}^-) }
\, {\Bf v}_{{\bF k}_{\star};\sigma}^{-},
\end{eqnarray}
which is simply the expression in Eq.~(\ref{e63}) in disguise; 
this is readily verified by employing the results in Eqs.~(\ref{e85}) 
and (\ref{e86}). We observe that a finite $Z_{{\bF k}_{\star};
\sigma}$ gives rise to a finite discontinuity in ${\Bf\nabla}_{\bF k}
\varepsilon_{{\bF k};\sigma}^{<}$ at ${\Bf k}={\Bf k}_{\star}$ (see 
Fig.~\ref{fi2}).

An important aspect of the result in Eq.~(\ref{e100}) is the
following. In \S~4.1 we have emphasized that, for metallic GSs of 
fermions interacting through potentials which are as long ranged 
as the Coulomb potential, ${\Bf\nabla}_{\bF k}
\varepsilon_{{\bF k};\sigma}^{<}$ logarithmically diverges for 
${\Bf k}$ approaching any region where ${\sf n}_{\sigma}({\Bf k})$ 
undergoes a discontinuous change \cite{BF03}. Since
\begin{equation}
\label{e101}
{\Bf\nabla}_{\bF k}^{(\rm s)} 
\varepsilon_{{\bF k};\sigma}^{<}\vert_{{\bF k}
={\bF k}_{\star}^-} \equiv
{\Bf\nabla}_{\bF k} 
\varepsilon_{{\bF k};\sigma}^{<}\vert_{{\bF k}
={\bF k}_{\star}^-} -
{\Bf\nabla}_{\bF k} 
\varepsilon_{{\bF k};\sigma}^{<}\vert_{{\bF k}
={\bF k}_{\star}^+}, 
\end{equation}
in follows that any logarithmically divergent contribution that 
may be contributing to ${\Bf\nabla}_{\bF k} \varepsilon_{{\bF k};
\sigma}^{<}$ for ${\Bf k} \to {\Bf k}_{\star}^{\mp}$, which by 
necessity has the same sign for ${\Bf k}\to {\Bf k}_{\star}^-$
and ${\Bf k}\to {\Bf k}_{\star}^+$,
\footnote{\label{f26}
See footnote \protect\ref{f16}. Further, for a simple model,
such as the electron-gas model, this aspect can be demonstrated 
by explicit calculation. }
cancel. Thus, even though for the above-mentioned systems 
${\Bf\nabla}_{\bF k} \varepsilon_{{\bF k};\sigma}^{<}$ is 
divergent for ${\Bf k} \to {\Bf k}_{\star}^{\mp}$, for these 
systems, following Eqs.~(\ref{e100}) and (\ref{e101}), 
${\Bf v}_{{\bF k}_{\star};\sigma}^{-}$ is bounded.

Finally, we point out that Eq.~(\ref{e100}) is {\sl not} applicable 
to cases in which  ${\Bf k}_{\star} \in {\cal S}_{{\sc f};\sigma}$. 
This aspect is rooted in the failure of Eq.~(\ref{e85}) for 
${\Bf k}_{\star} \in {\cal S}_{{\sc f};\sigma}$ which, in turn 
originates from the fact that, on all points of ${\cal S}_{{\sc f};
\sigma}$, $\varepsilon_{{\bF k};\sigma}^{<}$ acquires its maximum 
value (namely $\varepsilon_{\sc f}$) \cite{BF02a,BF03}. As can be 
explicitly shown, for ${\Bf k}_{\star} \in {\cal S}_{{\sc f};\sigma}$, 
the LHS of Eq.~(\ref{e85}) is equal to ${\Bf\nabla}_{\bF k}^{(\rm r)}
\varepsilon_{{\bF k};\sigma}^{>}\vert_{{\bF k}={\bF k}_{\star}^+}$.

\subsection*{\bf\S~5. \sc Summary and discussion}
\label{s5}

In this paper we have presented a set of complementary pieces of 
evidence indicating that the single-particle energy dispersions as 
deduced through the ARPES are in fact $\varepsilon_{{\bF k};\sigma}^{<}$
(Eq.~(\ref{e39})) rather than $\varepsilon_{{\bF k};\sigma}$ (more 
precisely, $\varepsilon_{{\bF k};\sigma}^-$) that formally is a 
solution of the quasi-particle equation (Eq.~(\ref{e3})). We have 
shown that, in addition to $\varepsilon_{{\bF k};\sigma}^{<}$ 
coinciding with $\varepsilon_{{\bF k};\sigma}^-$ on the Fermi surface 
${\cal S}_{{\sc f};\sigma}$ of metallic GSs, the two energy dispersions 
are equal at {\sl any} point inside the underlying Fermi sea FS$_{\sigma}$ 
where the GS momentum distribution function ${\sf n}_{\sigma}({\Bf k})$ 
undergoes a discontinuous change and $\varepsilon_{{\bF k};\sigma}^{<}$
is continuous; at the time of writing this paper the question remains 
open whether this property is retained when the last-mentioned 
discontinuity in ${\sf n}_{\sigma}({\Bf k})$ is replaced by a more 
general kind of singularity. One of the conclusions that we draw from 
the analysis of the experimental data obtained by Zhou {\sl et al.} 
\cite{XJZ03} as well as by other workers 
\cite{TV99,PVB00,PDJ01,AK01,AL01} is that (in particular), along the 
nodal directions of the 1BZs of the cuprate compounds investigated 
by the latter workers, ${\sf n}_{\sigma}({\Bf k})$ is discontinuous 
at a point ${\Bf k}_{\star}$ located in the close vicinity of the nodal 
Fermi point ${\Bf k}_{{\sc f};\sigma}$ ($\|{\Bf k}_{{\sc f};\sigma}-
{\Bf k}_{\star}\|$ typically amounts to some $5$\% of 
$\|{\Bf k}_{{\sc f};\sigma}\|$). From this and our aforementioned 
finding with regard to the equality of $\varepsilon_{{\bF k};\sigma}^-$ 
and $\varepsilon_{{\bF k};\sigma}^{<}$ at ${\Bf k}_{\star}$ and 
${\Bf k}_{{\sc f};\sigma}$ we have concluded that in the interval 
along the diagonal direction of the 1BZ between ${\Bf k}_{\star}$ 
and ${\Bf k}_{{\sc f};\sigma}$ the two energy dispersions 
$\varepsilon_{{\bF k};\sigma}^-$ and $\varepsilon_{{\bF k};
\sigma}^{<}$ should be in general experimentally indistinguishable 
(see Fig.~\ref{fi2}). For fermions interacting through the long-range 
Coulomb potential, $\varepsilon_{{\bF k};\sigma}^-$ and 
$\varepsilon_{{\bF k};\sigma}^{<}$ have distinct behaviours in the 
neighbourhoods of points where ${\sf n}_{\sigma}({\Bf k})$ is 
discontinuous (\S~4.1). Since, according to our analyses, the 
${\sf n}_{\sigma}({\Bf k})$ corresponding to systems studied by 
Zhou {\sl et al.} \cite{XJZ03} must be either continuous or only 
slightly discontinuous at ${\Bf k}={\Bf k}_{{\sc f};\sigma}$, we have 
arrived at the conclusion that the influence of the long-range of the 
Coulomb interaction potential should be strongest at ${\Bf k}=
{\Bf k}_{\star}$. The experimental data obtained by Zhou {\sl et al.} 
\cite{XJZ03} indeed show that for the underdoped sample 
(La$_{2-x}$Sr$_{x}$)CuO$_4$, corresponding to for instance $x=0.03$, 
the measured energy dispersion shows behaviour in the vicinity of 
${\Bf k}_{\star}$ that markedly differs from similar energy 
dispersions corresponding to optimally-doped and over-doped 
samples of (La$_{2-x}$Sr$_{x}$)CuO$_4$. This difference can be 
viewed as consisting of two separate contributions, namely a 
finite discontinuity and a logarithmic divergence in the
slope of the energy dispersion at ${\Bf k}={\Bf k}_{\star}$.
The former is consistent with the exact result obtained in \S~3.4 
according to which $\varepsilon_{{\bF k};\sigma}^{<}$ can undergo 
a discontinuous change at points inside FS$_{\sigma}$ where 
${\sf n}_{\sigma}({\Bf k})$ is discontinuous, and the latter
is feasible only if the two-body interaction potential is as 
long ranged as the Coulomb potential. We point out that, in cases 
where $\varepsilon_{{\bF k};\sigma}^{<}$ is discontinuous at the 
point of discontinuity of ${\sf n}_{\sigma}({\Bf k})$, that is
${\Bf k}={\Bf k}_{\star}$, neither $\varepsilon_{{\bF k}_{\star}^-;
\sigma}^{<}$ nor $\varepsilon_{{\bF k}_{\star}^+;\sigma}^{<}$ 
coincides with $\varepsilon_{{\bF k}_{\star};\sigma}^{-}$; here
$\|{\Bf k}_{\star}^-\| < \|{\Bf k}_{\star}^+\|$. However, in cases 
where ${\sf n}_{\sigma}({\Bf k}_{\star}^-)/Z_{{\bf k}_{\star};\sigma} 
\approx 1$, and thus ${\sf n}_{\sigma}({\Bf k}_{\star}^+) \approx 0$, 
it can be rigorously shown that $\varepsilon_{{\bF k}_{\star}^-;
\sigma}^{<} \approx \varepsilon_{{\bF k}_{\star};\sigma}^{-}$.

Our point of departure in this paper has been based on the consideration 
that single-particle energy dispersions as measured by the ARPES 
coincide with the average value of $\varepsilon$ with respect to the 
normalized energy distribution function $\hbar^{-1} A_{\sigma}({\Bf k};
\varepsilon)/{\sf n}_{\sigma}({\Bf k})$; this average value we have 
denoted by $\varepsilon_{{\bF k};\sigma}^{<}$. Throughout this paper 
we have presented a host of evidence to support this point of view. 
On the basis of this viewpoint we have proposed the standard variance 
$\Delta\varepsilon_{{\bF k};\sigma}^{<}$ of $\varepsilon$ with respect 
to the above-mentioned normalized energy distribution function as 
corresponding to the measured width of the peak in $A_{\sigma}
({\Bf k};\varepsilon)$ centred at $\varepsilon_{{\bF k};\sigma}^{<}$. 
From the expression for $\Delta\varepsilon_{{\bF k};\sigma}^{<}$ we 
have inferred that sharp variations in ${\sf n}_{\sigma}({\Bf k})$
are directly reflected in the behaviour of $\Delta\varepsilon_{{\bF k};
\sigma}^{<}$. We have shown for instance that the combination of
a discontinuous ${\sf n}_{\sigma}({\Bf k})$ and continuous
$\varepsilon_{{\bF k};\sigma}^{<}$ at ${\Bf k}={\Bf k}_{\star}$ 
(whereby $\varepsilon_{{\bF k}_{\star};\sigma}^{<} =
\varepsilon_{{\bF k}_{\star};\sigma}^{-}$) implies a discontinuity
in $\Delta\varepsilon_{{\bF k};\sigma}^{<}$ at ${\Bf k}={\Bf k}_{\star}$ 
with the change in $\Delta\varepsilon_{{\bF k};\sigma}^{<}$ being of 
the same sign as that in ${\sf n}_{\sigma}({\Bf k})$, that is
${\sf n}_{\sigma}({\Bf k}_{\star}^{-}) \, {\Ieq><}\, 
{\sf n}_{\sigma}({\Bf k}_{\star}^{+})\;$ $\iff\;$
$\Delta\varepsilon_{{\bF k}_{\star}^-;\sigma}^{<} \, {\Ieq><}\,
\Delta\varepsilon_{{\bF k}_{\star}^+;\sigma}^{<}$.
For (La$_{2-x}$Sr$_x$)CuO$_4$ at $x=0.063$, Zhou {\sl et al.} 
\cite{XJZ03} observed a {\sl decrease} in the electron scattering 
rate (i.e. the width of the MDC) at the same energy as a 
discontinuity in the gradient of the energy dispersion is observed 
on transposing ${\Bf k}$ from `below' to `above' ${\Bf k}_{\star}$, 
corresponding to $\|{\Bf k}\| < \|{\Bf k}_{\star}\|$ and
$\|{\Bf k}\| > \|{\Bf k}_{\star}\|$ respectively. This decrease 
in the scattering rate is in full conformity with the sign 
associated with $Z_{{\bF k}_{\star};\sigma} {:=} {\sf n}_{\sigma}
({\Bf k}_{\star}^-)-{\sf n}_{\sigma}({\Bf k}_{\star}^+)$ which has 
to be positive in order for the change in the energy dispersion as 
described by the expression deduced in this paper (Eqs.~(\ref{e63}) 
and (\ref{e100})) to conform with the experimentally observed 
change (see footnote \ref{f19}). A sudden change in the spectral width 
at the energy corresponding to the `break' in the single-particle 
energy dispersion has also been reported in \cite{PVB00,AK01,AL01}.

Since $\sum_{{\bF k},\sigma} \varepsilon_{\bF k} {\sf n}_{\sigma}
({\Bf k})$ amounts to the kinetic energy corresponding to the GS of 
the Hamiltonian $\wh{H}$ in Eq.~(\ref{e35}), one observes that a 
finite discontinuity in ${\sf n}_{\sigma}({\Bf k})$ in the interior 
of the underlying Fermi sea, of the type indicated above, necessarily 
results in a higher kinetic energy than the kinetic energy 
corresponding to a comparable ${\sf n}_{\sigma}({\Bf k})$ which, 
however, is free from the latter discontinuity; in this connection, 
recall that ${\sf n}_{\sigma}({\Bf k}) \ge 0$ and that $\sum_{{\bF k},
\sigma} {\sf n}_{\sigma}({\Bf k}) = N$, independent of the the 
strength of the interparticle interaction, so that a discontinuity in 
${\sf n}_{\sigma}({\Bf k})$ over a set of points $\{ {\Bf k}_{\star}\}$ 
in the interior of the Fermi sea, with ${\sf n}_{\sigma}
({\Bf k}_{\star}^-) > {\sf n}_{\sigma}({\Bf k}_{\star}^+)$, is 
necessarily accompanied by a `spillage' of ${\sf n}_{\sigma}({\Bf k})$ 
into the region outside the Fermi sea. We thus observe that the 
interpretation as put forward in the present paper of the experimental 
observations in \cite{XJZ03,TV99,PVB00,PDJ01,AK01,AL01} also provides 
a natural explanation concerning the experimental observations of 
the excess `kinetic' energy in the normal states of the doped cuprate 
compounds \cite{SS01,MPMKL02} in comparison with the `kinetic' energies 
of the GSs of conventional metals; in arriving at the latter statement 
we have relied on the supposition that the gapless excitations along 
the nodal directions of the 1BZ (below the pertinent superconducting 
transition temperatures) are akin to the gapless excitations of the 
normal states of the investigated cuprate compounds. Experimental
observations, at temperatures elevated above the pertinent 
superconducting transition temperatures, along various directions 
of the 1BZ \cite{PVB00} (concerning change in the spectral widths see 
\cite{AL01}), are in conformity with the above-mentioned supposition.

Although the considerations in this paper provide a consistent
scenario concerning the main aspects of the experimental data in 
\cite{XJZ03,TV99,PVB00,PDJ01,AK01,AL01}, a fundamental question 
remains to be answered, namely whether a single-band model which, 
moreover, exclusively takes account of the electronic degrees of 
freedom (Eq.~(\ref{e35})), allows for a uniform metallic GS whose 
corresponding single-particle spectral function involves a coherent 
contribution below $\varepsilon_{\sc f}$ characteristic of a 
well-defined quasi-particle excitation at ${\Bf k}={\Bf k}_{\star}$ 
located inside the underlying Fermi sea (i.e. at some finite 
distance from ${\cal S}_{{\sc f};\sigma}$). Note that, although the 
existence of a well-defined quasiparticle excitation at ${\Bf k}
={\Bf k}_{\star}$ is sufficient for rendering ${\sf n}_{\sigma}
({\Bf k})$ discontinuous at ${\Bf k}={\Bf k}_{\star}$, this 
discontinuity does not necessarily imply that the pertinent excitation 
energy should lie by a finite amount below the Fermi energy; 
conversely, in cases where the excitation energy at issue is less 
than the Fermi energy by a non-vanishing amount, it is necessary (in 
fact by definition) that ${\Bf k}_{\star}$ be located away from 
${\cal S}_{{\sc f};\sigma}$ (see paragraph following Eq.~(\ref{e12}) 
above). Recent Monte Carlo results corresponding to projected 
variational wave functions \cite{AP01,AP03} concerning the conventional 
single-band Hubbard Hamiltonian (involving only an on-site interaction 
between fermions of opposite spin) on a square lattice, are {\sl not} 
supportive of the possibility of a noticeable discontinuity in 
${\sf n}_{\sigma}({\Bf k})$ in the interior of the underlying Fermi sea. 
\footnote{\label{f27}
It is interesting to note that the ${\sf n}_{\sigma}({\Bf k})$ 
as reported in \protect\cite{AP01,AP03} (along the nodal direction 
of the 1BZ of a square lattice and corresponding to a projected 
variational wave function for $U/t = 12$ and $t'=t/4$) unequivocally 
violates the result $[{\sf n}_{\sigma}({\Bf k}_{{\sc f};\sigma}^-) + 
{\sf n}_{\sigma}({\Bf k}_{{\sc f};\sigma}^+)]/2 = 1/2$. Following 
the analysis in \cite{BF02a} and in full conformity with the 
conclusion arrived at herein, we observe that the underlying metallic 
GS is {\sl not} a Fermi liquid. }
Furthermore, although for the hole-doping fraction $x$ approaching 
zero, the discontinuity of ${\sf n}_{\sigma}({\Bf k})$ at the nodal
Fermi point ${\Bf k}_{{\sc f};\sigma}$ diminishes (according to the 
authors of \cite{AP01,AP03}, $Z_{{\bF k}_{{\sc f};\sigma}}$ scales 
like $x$ for sufficiently small $x$), for $x$ corresponding to optimal 
doping (here $x_{\rm opt}=0.18$) it is relatively too large for the 
following two reasons. Firstly, since $0 \le {\sf n}_{\sigma}({\Bf k}) 
\le 1$, the larger the value of $Z_{{\bF k}_{{\sc f};\sigma}}$, the 
smaller is the maximal value that $Z_{{\bF k}_{\star};
\sigma}/{\sf n}_{\sigma}({\Bf k}_{\star}^-)$, or $\lambda_{\star}$
(see Eq.~(\ref{e68})), can possibly attain; following Eq.~(\ref{e63}), 
a small value for the latter quantity corresponds to a small `kink' 
in the single-particle energy dispersion 
$\varepsilon_{{\bF k};\sigma}^{<}$ at ${\Bf k}={\Bf k}_{\star}$. 
Secondly, in principle a non-vanishing (and in practice a relatively 
large) $Z_{{\bF k}_{{\sc f};\sigma}}$ corresponding to a nodal 
Fermi point at optimal doping does not conform with the experimental 
observations by Valla {\sl et al.} \cite{TV99} according to which 
the self-energy in the neighbourhood of a nodal Fermi wave vector
along a nodal direction of the 1BZ satisfies the scaling behaviour 
characteristic of the self-energies of marginal Fermi liquids. 
Assuming that, insofar as our present considerations are concerned, 
the variational results in \cite{AP01,AP03} are sufficiently accurate, 
we have no alternative but to pronounce the conventional Hubbard 
Hamiltonian (as opposed to `extended' Hubbard Hamiltonians and as 
opposed to Hamiltonians, such as the Hubbard-Holstein Hamiltonian 
(see, e.g., \cite{BKT00}), in which the fermionic field is coupled 
to an {\it external} bosonic field) as being inadequate for describing 
the phenomena observed in \cite{XJZ03} and 
\cite{TV99,PVB00,PDJ01,AK01,AL01}. This conclusion contradicts the 
finding by Randeria {\sl et al.} \cite{MR03} whose analysis is based 
on the very same computational results (i.e. those reported in 
\cite{AP01,AP03}) as relied upon by us above. 

Concerning the interpretation of the experimental observation
in \cite{XJZ03,TV99,PVB00,PDJ01,AK01,AL01} as presented in 
\cite{MR03}, denoting the Fermi velocity at high binding energies by 
$v_{\sc f}^{\rm high}$ (to be compared with what we have in this 
paper denoted by ${\Bf v}_{{\bF k}_{\star}^-;\sigma}^{<}$) and that 
at low binding energies by $v_{\sc f}^{\rm low}$ (to be compared 
with ${\Bf v}_{{\bF k}_{\star}^+;\sigma}^{<}$), Randeria {\sl et al.}
\cite{MR03} deduce that $v_{\sc f}^{\rm high} = v_{\sc f}^{\rm low}/Z$ 
(compare with Eq.~(\ref{e63}) above) where $Z$ stands for what we have 
in this paper denoted by $Z_{{\bF k}_{{\sc f};\sigma}}$, to be strictly 
distinguished from $Z_{{\bF k}_{\star};\sigma}$. The expression as 
deduced in \cite{MR03} for $v_{\sc f}^{\rm low}$ is the standard 
expression for the Fermi velocity in (isotropic) Fermi-liquids (see 
Eq.~(7) in \cite{MR03}). Through stating that ``at intermediate to 
high energies, $\vert\partial\Sigma'/\partial\omega\vert \ll 1$'', 
where $\Sigma'$ is the real part of self-energy as evaluated at 
``${\Bf k}={\Bf k}_{\sc f}$'', Randeria {\sl et al.} \cite{MR03} 
deduced an expression for $v_{\sc f}^{\rm high}$ which up to the 
numerical factor $Z$ (by the latter statement this is identified with 
unity) coincides with that for $v_{\sc f}^{\rm low}$ and hence the 
above-quoted expression relating the two velocities. It is evident 
that the crucial, and in our opinion manifestly unjustified, step 
taken by Randeria {\sl et al.} \cite{MR03} in deducing the expression 
$v_{\sc f}^{\rm high} = v_{\sc f}^{\rm low}/Z$ is that of effecting 
the condition ``$\vert\partial\Sigma'/\partial\omega\vert \ll 1$'' in 
the expression concerning $v_{\sc f}^{\rm high}$, for the essential 
aspect of the experimental observations being interpreted concerns 
a change in energy dispersions (or discontinuity in Fermi velocities) 
over a relatively very narrow range of energies (centred around, for 
instance, approximately $\varepsilon_{\star}= \varepsilon_{\sc f} - 
70$ meV insofar as the experiments reported in \cite{XJZ03} are 
concerned) whereby an appeal to the relationship 
``$\vert\partial\Sigma'/\partial\omega\vert \ll 1$'' is seen to 
be entirely unwarranted. The only possibility that would justify 
an appeal to the latter relationship would be the discontinuity 
of ``$\partial\Sigma'/\partial\omega$'' at $\omega_{\star} = 
\varepsilon_{\star}/\hbar$, with $\Sigma'$ reducing to a 
nearly constant function
\footnote{\label{f28}
For the uniform GSs of the single-band Hubbard Hamiltonian,
which are dealt with in \protect\cite{MR03}, 
$\lim_{\vert\varepsilon\vert\to\infty} \Sigma_{\sigma}
({\Bf k};\varepsilon) = \hbar^{-1} U n_{\bar\sigma}$, where
$n_{\bar\sigma}$ is the number of fermions with spin index 
$\bar\sigma$ (the index complementary to $\sigma$) per site and $U$ 
is the on-site interaction energy. The above-mentioned nearly-constant 
function must therefore be close to the latter limiting value. }
of $\omega$ for {\sl all} $\omega < \omega_{\star}$. By the 
Kramers-Kr\"onig relation the imaginary part $\Sigma''$ of the 
self-energy would be almost vanishing for {\sl all} $\omega < 
\omega_{\star}$, which is neither realistic nor enjoined by the 
experimental observations under consideration. We therefore conclude 
that the considerations by Randeria {\sl et al.} \cite{MR03} do not 
shed light on the observations as reported in 
\cite{XJZ03,TV99,PVB00,PDJ01,AK01,AL01}. 

It is important to point out that according to a widely, but not 
universally, held view (for a review see \cite{ZXS02}; see also 
\cite{PBA01}), both the `kink' in the single-particle energy 
dispersion and the decrease in the spectral width at ${\Bf k}=
{\Bf k}_{\star}$ are to be attributed to electron-phonon 
interaction \cite{AL01,ZG01}.
\footnote{\label{f29}
Concerning the role envisioned for the electron-phonon interaction 
in the context of high-temperature superconductivity in the cuprate 
compounds see \protect\cite{AM94} and for the interplay between 
electron-electron and electron-phonon interactions see 
\protect\cite{MLK00}. }
Counterarguments, disfavouring phonons and attributing the latter 
phenomena to antiferromagnetic spin fluctuations have been put forward 
\footnote{\label{f30}
For completeness we mention that we fail to comprehend the 
methodology with the aid of which the authors of 
\protect\cite{MEB03} deduced from their computational results 
concerning the single-particle spectral function (denoted in 
\protect\cite{MEB03} by $N({\Bf k},\omega)$) a kink in the 
single-particle energy dispersion. In our judgement, connecting 
the maxima of a series of $N({\Bf k},\omega)$ corresponding to 
various ${\Bf k}$ (considered as parameter), as the authors of 
\protect\cite{MEB03} did (see Figs.~4 -- 6 in \protect\cite{MEB03}, 
noting that in these $\omega$ is measured in milielectron volts 
and ${\Bf k}$ in units of the inverse lattice constant), cannot 
provide unequivocal information with regard to the dispersion of 
the underlying single-particle excitation energies. }
\cite{EN00,MEB01,MEB03} (see also the discussions in \cite{PDJ01,AK01} 
which concern experiments as well as those in \cite{PVB00}, examining 
experimental observations in the light of a number of different 
theoretical possibilities). We do {\sl not} consider electron-phonon 
interaction as being the primary cause for the `kink' in the 
experimentally measured energy dispersions. This becomes evident by 
linearly extrapolating the data in Fig.~1a of \cite{XJZ03} at high 
binding energies (corresponding to energies between approximately $0.2$ 
and $0.1$ eV below the Fermi energy $\varepsilon_{\sc f}$) towards the 
nodal Fermi point ${\Bf k}={\Bf k}_{{\sc f};\sigma}$. One observes 
that at ${\Bf k}={\Bf k}_{{\sc f};\sigma}$ the extrapolated energy 
dispersions corresponding to {\sl all} doping concentrations (in total 
$10$ energy dispersions corresponding to $x \in [0.03, 0.3]$) intersect 
the energy axis at the same point (this with remarkable accuracy) 
{\sl above} the Fermi energy, namely at approximately $0.075$ eV 
{\sl above} $\varepsilon_{\sc f}$ (instead of at, or nearly at, 
$\varepsilon_{\sc f}$ if electron-phonon interaction were directly at 
work; see later); this excess energy is almost equal to the binding 
energy (i.e. $\varepsilon_{\sc f}-\varepsilon_{\star}$) at which `kink' 
in the energy dispersions has been observed. The conventional 
electron-phonon theory \cite{GDM90} (see in particular \S~6.4 herein), 
according to which $\|{\Bf v}_{{\bF k}_{\star}^+;\sigma}\|/\|
{\Bf v}_{{\bF k}_{\star}^-;\sigma}\| = 1/(1+\lambda)$, where 
$\lambda$ is the coupling constant of the electron-phonon 
interaction, or the electron-phonon contribution to the mass 
enhancement (for a comprehensive overview see \cite{AC69}), 
requires that the change in the energy dispersion correspond 
to a clockwise `winding' around $\varepsilon_{\sc f}$ (inside 
an energy window of width $\hbar\omega_{\rm ph}$ centred at
$\varepsilon_{\sc f}$,
\footnote{\label{f31}
See, for example, Fig.~26.1 in \protect\cite{AM81}; see also 
Fig.~6.16 in \protect\cite{GDM90} and note that the electron-phonon
self-energy is vanishing at $\varepsilon=\varepsilon_{\sc f}$, or 
using the notation adopted in \cite{GDM90}, at $u=0$. } where
$\omega_{\rm ph}$ the the characteristic phonon frequency) of the 
electronic dispersion with respect to that in the absence of 
phonons; the `kink' in the single-particle energy dispersions in 
systems of electrons coupled to phonons is thus seen to correspond 
to a sharp crossover of the phonon-dressed dispersion into that of 
electrons in the absence of electron-phonon interaction (from which 
the phonon-dressed energy dispersion becomes rapidly indistinguishable) 
at some short distance outside the above-mentioned energy window;
linear extrapolation of electron-phonon dressed energy dispersions 
from {\sl outside} the above-mentioned crossover region (and
corresponding to $\|{\Bf k}\| < \|{\Bf k}_{\star}\|$) must 
therefore at ${\Bf k}={\Bf k}_{{\sc f};\sigma}$ very nearly 
intersect the energy axis at $\varepsilon_{\sc f}$.
Experimentally, this aspect is evident from the high-resolution 
photoemission data concerning the $(0001)$ surface state of beryllium
(also referred to as the $\ol{\Gamma}$-surface state of Be$(0001)$) as 
presented in Fig.~3 of the work by Hengsberger {\sl et al.} 
\cite{HPSGB99} (see also \cite{LSJB00}). A comparison of the data 
in Fig.~3 of \cite{VFJH99} with those in Fig.~4 of \cite{LSJB00} 
leads one to conclude that the above-mentioned clockwise `winding' 
around $\varepsilon_{\sc f}$ of the phonon-dressed single-particle 
energy dispersions with respect to their undressed counterparts is 
also present in the energy dispersion of the $(110)$ surface-state 
of molybdenum \cite{VFJH99}. With reference to our above statement with 
regard to the extrapolation of the energy dispersions in Fig.~1a of 
\cite{XJZ03} corresponding to $\|{\Bf k}\| < \|{\Bf k}_{\star}\|$ 
towards ${\Bf k}= {\Bf k}_{{\sc f};\sigma}$, it is evident that the 
`kinks' observed in the single-particle energy dispersions in the 
hole-doped cuprate compounds 
\cite{XJZ03,TV99,PVB00,PDJ01,AK01,AL01} {\sl cannot} be a direct 
consequence of the electron-phonon interaction.  

In order to examine the possibility of an indirect role played by 
phonons in bringing about `kink' in the energy dispersions as observed 
in \cite{XJZ03,TV99,PVB00,PDJ01,AK01,AL01}, it is instructive to
recall a model proposed by Fr\"ohlich \cite{HF50,HF51} (see also 
\S\S~1.8 and 3.16 in \cite{GR65}). Fr\"ohlich \cite{HF50,HF51} 
showed that, for sufficiently large value of coupling constant, 
the interaction of independent electrons with phonons gives rise 
to an exotic electronic GS, which Fr\"ohlich showed to be a 
superconducting state. 
\footnote{\label{f32}
Schafroth \protect\cite{MRS51} has, however, shown that, as the 
theory of Fr\"ohlich is based on a finite-order perturbation theory 
concerning electron-phonon interaction (explicitly, a second-order 
theory), it fails to describe the Meissner effect. For an 
exposition of the work by Schafroth see \protect\cite{GR65} 
(appendix A1.8 herein). }
In this state, the Fermi sea of electrons consists of two 
concentric regions, separated by a narrow region of unoccupied 
single-particle states (see Figs.~1 and 2 in \cite{HF50}); Fr\"ohlich 
explicitly showed that this GS, corresponding to a `shell'
distribution of electrons in the ${\Bf k}$ space, is characterized 
by the equality of the widths of the intervening unoccupied region 
and the outer occupied region of the ${\Bf k}$ space. For such a 
state, on transposing ${\Bf k}$ from the interior of the inner 
Fermi sea to the exterior of the outer Fermi sea, ${\sf n}_{\sigma}
({\Bf k})$ undergoes three discontinuities, from unity to zero, 
from zero to unity and again from unity to zero respectively. Each 
of these discontinuities would in turn give rise to discontinuities 
in $\varepsilon_{{\bF k};\sigma}^{<}$ (in principle; \S~3.4) and 
${\Bf\nabla}_{\bF k} \varepsilon_{{\bF k};\sigma}^{<}$, in the manner 
described in this paper (\S\S~3.4, 4.2 and 4.3), were it not for the 
fact that these discontinuities all correspond to single-particle 
excitations whose energies coincide with $\varepsilon_{\sc f}$. It 
follows that the Fr\"ohlich state is {\sl not} the appropriate state 
insofar as the observations in \cite{XJZ03,TV99,PVB00,PDJ01,AK01,AL01} 
are concerned. In spite of this, the crucial role that phonons play 
in bringing about the Fr\"ohlich state (at least in principle), with 
the above-indicated discontinuities in the associated 
${\sf n}_{\sigma}({\Bf k})$, makes evident that phonons can in 
principle be vital in regard to the observations reported in 
\cite{XJZ03,TV99,PVB00,PDJ01,AK01,AL01}; 
in this connection it should be noted that the binding energies 
$\varepsilon_{\sc f}-\varepsilon_{\star}$ corresponding to `kinks' 
in the measured single-particle energy dispersions of the investigated 
cuprate compounds very nearly coincide with the energies of the 
longitudinal optic phonons in these \cite{AL01} (for a detailed 
discussion of this subject see \cite{ZXS02}).

We point out that the prediction by Fr\"ohlich \cite{HF50} of the 
above-mentioned `shell' state relies on a second-order perturbation 
expansion of the GS energy of the coupled electron-phonon system. Soon 
after its predication, this state was asserted by Wentzel \cite{GW51} 
and Kohn and Vachaspati \cite{KV51} as not feasible through its 
interception by a structural lattice instability at an electron-phonon 
coupling constant smaller than that required for the formation of the 
Fr\"ohlich state; Wentzel \cite{GW51} conceded, however, that interaction 
of electrons, not taken account of in Fr\"ohlich's considerations, 
would render the Fr\"ohlich state feasible. However, according to 
Bardeen \cite{JB51} (see the closing paragraph of section III herein), 
who has established a modified stability condition for the lattice, 
the Fr\"ohlich state is viable.
\footnote{\label{f33}
Anderson \protect\cite{PWA69} presented the contents of the 
Fr\"ohlich model in a modern setting, indicating the diagrammatic 
representations of the underlying exchange processes. Subsequently, 
with reference to a Migdal theorem \protect\cite{ABM58}, Anderson 
\protect\cite{PWA69} stated (p.~1346): ``{\sl But in fact 
Migdal resolved the problem even more completely by showing that 
$\Sigma$ is a much more sensitive function of $\omega$ than of 
$k$ --- i.e., it is local in space, retarded in time, thus depends 
on $\omega$ much more sharply --- so that the large correction 
to $(E_k)_{\rm eff}$ comes from $\partial\Sigma/\partial\omega$: ... 
and $Z=1/[1 - (\partial\Sigma/\partial\omega)]$ is always positive: 
the Fr\"ohlich-Bardeen instability simply does not occur. The
complete treatment shows that phonon instability and the first
singularity of $Z$ actually occur precisely at the same coupling 
strength.}'' Two comments are in order. Firstly, the electronic 
state on which the Migdal theorem at issue is based is that of 
the non-interacting free-fermion model (this is readily verified 
through inspecting \protect\cite{ABM58} in conjunction with 
\protect\cite{HF50}); in other words, it is assumed that the scale 
of the electronic excitation energies is determined by a single 
electronic mass parameter. For specifically strongly-correlated 
electron systems there is no {\sl a priori} reason to believe that 
the GS of this model would at all be an appropriate starting point 
for dealing with the problem of electron-phonon interaction (this 
is exactly the same problem which in fact hinders a rigorous 
formulation of a theory concerning the superconducting states of 
the cuprate compounds whose low-lying single-particle excitations 
in the normal state are not quasi-particle-like; see however 
\protect\cite{CSAS93}), so that the above-mentioned Fr\"ohlich-Bardeen 
instability cannot on its own rule out the existence of the 
Fr\"ohlich state. Secondly, it is generally, even though perhaps 
not universally, accepted that non-adiabatic processes are significant 
in the copper-oxide-based high-temperature superconductors, leading 
to the `failure' of the Migdal theorem referred to above. See 
\protect\cite{PSG95}. For completeness, we have carried out 
calculations \protect\cite{BF99c} on a uniform two-dimensional model 
in which the GS is the Fr\"ohlich state. From the first-order 
self-energy in terms of the dynamically screened interaction 
function we have deduced quasi-particle lifetimes which are 
generically by one order of magnitude shorter than the lifetimes of 
quasi-particles in the conventional two-dimensional model at similar 
densities. We have further studied \protect\cite{BF99c} a variety of 
the properties of this model in the superconducting state within the 
Bardeen-Cooper-Schrieffer (BCS) framework, despite the fact that the 
aforementioned short quasi-particle lifetimes undermines the use of 
the BCS formalism. For instance, in the weak-coupling limit we obtain 
that $2\Delta(T=0)/[k_{\sc b} T_{\rm c}]$ is relatively close to $4$ (to 
be compared with the conventional BCS value of $3.52$). We point out 
that a recent \protect\cite{BCGKS03} non-selfconsistent first-order 
calculation (in which the effective interaction has been deduced 
from diffusion Monte Carlo results concerning the static charge and 
spin structure factors and random-phase approximation Ans\"atze 
concerning the dynamical density-density and spin-spin susceptibilities) 
concerning two-dimensional liquid $^3$He indicates that on increasing 
the aerial density of $^3$He atoms the effective mass diverges; a 
subsequent increase of the density suggests what Boronat {\sl et al.}
\protect\cite{BCGKS03} refer to as being indicative of a transition 
to ``{\sl anomalous occupation numbers}'' and to which we would refer 
as the Fr\"ohlich `shell' state. Interestingly, according to Boronat
{\sl et al.} \protect\cite{BCGKS03} ``{\sl Both spin and density 
fluctuations have profound effects.}'' }

Concerning the `universality' (within an experimental uncertainty 
of approximately $20$\%) of the Fermi velocities at low binding 
energies as reported by Zhou {\sl et al.} \cite{XJZ03}, following 
the analysis in \S~3.5, this `universality' reveals two aspects; 
firstly, the `universal' energy $\varepsilon_{\sc f} 
-\varepsilon_{\star}$, where $\varepsilon_{\star}\equiv
\varepsilon_{{\bF k}_{\star};\sigma}^{-}\equiv 
\varepsilon_{{\bF k}_{\star};\sigma}^{<}$, is associated with 
some bosonic excitations that are external to the electronic
degrees of freedom, supporting the possibility that longitudinal 
optical phonons, though {\sl not} directly (see above), can 
indirectly be the sought-after bosonic excitations; secondly, a 
rigid, that is doping-independent, tight-binding energy dispersion 
$\varepsilon_{\bF k}$ underlying the many-body Hamiltonian in 
Eq.~(\ref{e35}), is {\sl not} capable of reliably describing the 
cuprate compounds, specifically those investigated by Zhou {\sl et al.}
\cite{XJZ03}; the bare mass corresponding to a rigid tight-binding 
energy dispersion is relatively strongly dependent on the level of 
doping (see caption of Fig.~\ref{fi3}), and this is most likely 
inherited by the corresponding dressed mass and consequently the 
interacting Fermi velocity, thus contradicting the above-mentioned 
experimental observations.

Although in this paper we have mainly focused on the ARPES data
along the nodal directions of the 1BZs of the cuprate compounds
as reported in \cite{XJZ03,TV99,PVB00,PDJ01,AK01,AL01}, our approach 
is applicable along any direction of the 1BZ. In the light of the 
observations as reported in \cite{PVB00} (see specifically 
Fig.~3 in \cite{PVB00}) we conclude that a discontinuity in 
${\sf n}_{\sigma}({\Bf k})$ in the interior of the Fermi sea may 
be a generic property of the normal states of all cuprate 
superconductors. With reference to our observation in this paper 
that along the nodal directions of the 1BZs of the cuprate compounds 
investigated in \cite{XJZ03,TV99,PVB00,PDJ01,AK01,AL01}, the 
quasi-particle equation must have at least (see footnote \ref{f9}) 
two solutions $\varepsilon_{{\bF k}_{\star};\sigma}^- < \mu$ 
and $\varepsilon_{{\bF k}_{\star};\sigma}^+ > \mu$ 
corresponding to true Landau quasi-particles (in the sense of 
$Z_{{\bF k}_{\star};\sigma} > 0$), it is interesting to recall 
an earlier observation presented in \cite{BF02a}; here it has 
been rigorously shown that, within the framework of the conventional 
single-band Hubbard Hamiltonian, for ${\Bf k}$ located in the 
pseudogap regions of the reciprocal space, the quasi-particle 
equation must have (at least) two solutions, one strictly below 
$\mu$ and one strictly above $\mu$, and that these solutions do 
not correspond to quasi-particle excitations of the Landau type, 
but to resonances. It is of interest to investigate whether, and 
if so, how these two distinct energies are related to 
$\varepsilon_{{\bF k}_{\star};\sigma}^-$ and 
$\varepsilon_{{\bF k}_{\star};\sigma}^+$ which correspond to Landau
quasiparticles. We should emphasize that, whereas the above-mentioned 
resonances correspond to wave vectors in the pseudogap regions of the 
underlying putative Fermi surfaces (in \cite{BF02a} it has been 
shown that 
\footnote{\label{f34}
Here ${\cal S}_{{\sc f};\sigma}^{(0)}$ stands for the Fermi 
surface corresponding to $\varepsilon_{\bF k}+\epsilon_{\sigma}$,
in which $\epsilon_{\sigma}$ has been chosen such that 
the interior of ${\cal S}_{{\sc f};\sigma}^{(0)}$ contains the 
same number of ${\Bf k}$ points as contained in the interior of 
${\cal S}_{{\sc f};\sigma}$, namely $N_{\sigma}$, the total
number of fermions in the GS with spin index $\sigma$. }
${\cal S}_{{\sc f};\sigma}^{(0)}\backslash 
{\cal S}_{{\sc f};\sigma}$, which may or may not be empty,
constitutes the pseudogap region of the wave-vector space), the 
wave vector ${\Bf k}_{\star}$ is located in the {\sl interior} of 
Fermi seas, i.e. at some finite distance from 
${\cal S}_{{\sc f};\sigma}$.  

We close this section by reiterating that, although we have been 
able to provide a consistent scenario for the experimental
observations in \cite{XJZ03,TV99,PVB00,PDJ01,AK01,AL01}, a number of 
outstanding questions remain. These are \\

\indent
(i) whether a single-band model is capable of possessing a uniform 
metallic GS whose momentum distribution function is discontinuous 
in the interior of the underlying Fermi sea (specifically along the 
nodal directions of the 1BZ) {\sl and} at the pertinent wave vector, 
denoted in this paper by ${\Bf k}_{\star}$, the single-particle 
spectral function involves a singular contribution at $\varepsilon
=\varepsilon_{\star}$ where $\varepsilon_{\star}$ lies some $50$ 
to $70$ meV below the Fermi energy $\varepsilon_{\sc f}$, \\

\indent
(ii) whether a uniform GS is sufficient, or inhomogeneity, such as 
`stripes' (whether static or dynamic) \cite{JZ99,CEKO02,NCMS03} are 
essential for the single-particle spectral function $A_{\sigma}
({\Bf k};\varepsilon)$ to possess the properties indicated above
and\\ 

\indent
(iii) whether the above-mentioned properties expected of 
$A_{\sigma}({\Bf k};\varepsilon)$ can arise through collective 
excitations {\sl internal} to the electronic system (charge and spin 
excitations) \cite{CPS02} or coupling of electrons to {\sl external} 
bosonic modes (phonons) are essential, or a combination of both.\\ 

\noindent
As for the role that phonons can play in bringing about 
the above-mentioned properties concerning $A_{\sigma}({\Bf k};
\varepsilon)$, we believe that the experimental data considered 
by us in this paper \cite{XJZ03,TV99,PVB00,PDJ01,AK01,AL01} rule 
out a primary role; however, a secondary (even though possibly vital) 
role is not excluded. As an example of the interesting possibilities 
that can become available through the mediation of the electron-phonon 
interaction, we have briefly discussed the Fr\"ohlich `shell' state. 
We have, however, shown that this state {\sl cannot} be the 
appropriate state to which the experimental observations
as reported in \cite{XJZ03,TV99,PVB00,PDJ01,AK01,AL01} can be
ascribed.

%________________________
\vspace{3mm}
\subsection*{\bf\sc Acknowledgements}
\label{ack}
With pleasure I thank Professor Zhi-xun Shen for kindly 
communicating to me the experimental observations by Zhou 
{\sl et al.} \cite{XJZ03} prior to publication, Professor 
Philip Stamp for bringing Refs.~\cite{EMS91,EO94,EOMS94} to my
attention and Professor Fei Zhou (Vancouver) for discussion. 
With appreciation I acknowledge hospitality and support by 
Spinoza Institute.

%________________________
\vspace{0.6cm}
\centerline{\bf ---------------------}
\vspace{-4.42mm}
\centerline{\bf ------------------------------}
\vspace{-4.42mm}
\centerline{\bf ---------------------}

\begin{appendix}

\section{An approximate $\varepsilon_{\lowercase{\bF k};\sigma}^{<}$ }
\label{app}

In this appendix we deduce an approximate expression for 
$\varepsilon_{{\bF k};\sigma}^{<}$ by assuming ${\sf n}_{\sigma}
({\Bf k}) \in \{0,1\}$, subject to the condition $\sum_{\bF k} 
{\sf n}_{\sigma}({\Bf k}) = N_{\sigma}$, $\forall\sigma$, where 
$N_{\sigma}$ stands for the number of particles with spin index 
$\sigma$ in the GS. The ${\sf n}_{\sigma}({\Bf k})$ pertaining 
to the $(N_{\sigma}+N_{\bar\sigma})$-particle GS of $\wh{H}_0$ 
in Eq.~(\ref{e37}) is only a specific example of the type of 
${\sf n}_{\sigma}({\Bf k})$ considered in this appendix. Many 
aspects in this appendix coincide with those in \S~A.2 of appendix 
A in \cite{BF02b}; however, the approach adopted here is more direct 
and the details are more closely related to the subject matters of 
the present paper.

We consider the following form for the many-body interacting 
Hamiltonian $\wh{H}$:
\begin{eqnarray}
\label{ea1}
&&\wh{H} =
\sum_{\sigma} \int {\rm d}^dr\;
\hat\psi_{\sigma}^{\dag}({\Bf r}) h_{0}({\Bf r})
\hat\psi_{\sigma}({\Bf r}) \nonumber\\
&&\;\;\;
+\frac{1}{2}\sum_{\sigma,\sigma'}
\int {\rm d}^dr {\rm d}^dr'\;
\hat\psi_{\sigma}^{\dag}({\Bf r})
\hat\psi_{\sigma'}^{\dag}({\Bf r}')\,
v({\Bf r}-{\Bf r}')\,
\hat\psi_{\sigma'}({\Bf r}')
\hat\psi_{\sigma}({\Bf r}), \nonumber\\
\end{eqnarray}
where $h_{0}({\Bf r})$ is the sum of the single-particle 
kinetic-energy operator and one-body external potential (e.g., 
ionic potential) and $v({\Bf r}-{\Bf r}')$ is the two-particle 
interaction potential which we leave unspecified at this stage. 
In \cite{BF02b} for the single-particle spectral function 
$A_{\sigma}({\Bf r},{\Bf r}';\varepsilon)$ pertaining to the GS 
of the Hamiltonian in Eq.~(\ref{ea1}) we have obtained the 
following exact expression
\footnote{\label{fa1}
The RHS of Eq.~(\protect\ref{ea2}) is simply ${\cal D}_{\sigma}
({\Bf r},{\Bf r}')$ defined in appendix E of \protect\cite{BF02b}
(see in particular Eqs.~(E5), (E12) and (E13) therein) of which the 
second contribution is equal to $-{\cal B}_{\sigma}({\Bf r},{\Bf r}')$
defined in appendix B of \protect\cite{BF02b} (see Eq.~(B29) therein). }
\begin{eqnarray}
\label{ea2}
&&\frac{1}{\hbar}\int_{-\infty}^{\mu} {\rm d}\varepsilon\;
\varepsilon A_{\sigma}({\Bf r},{\Bf r}';\varepsilon) =
h_0({\Bf r})\, \Gamma^{(1)}({\Bf r}'\sigma,{\Bf r}\sigma) \nonumber\\
&&\;\;\;\;
+\int {\rm d}^dr''\; v({\Bf r}-{\Bf r}'')\, \sum_{\sigma'}
\Gamma^{(2)}({\Bf r}'\sigma,{\Bf r}''\sigma';
{\Bf r}\sigma,{\Bf r}''\sigma'),
\end{eqnarray}
where $\Gamma^{(m)}$, $m=1,2$, is the $m$-particle ground-state 
density matrix, for which we have (for details and various properties 
of this function we refer the reader to appendix B in \cite{BF02b})
\begin{eqnarray}
\label{ea3}
&&\Gamma^{(m)}({\Bf r}_1\sigma_1,\dots, {\Bf r}_m\sigma_m;
{\Bf r}_1'\sigma_1',\dots,{\Bf r}_m'\sigma_m') \nonumber\\
&&{:=} \langle\Psi_{N;0}\vert
\hat\psi_{\sigma_1}^{\dag}({\Bf r}_1)\dots
\hat\psi_{\sigma_m}^{\dag}({\Bf r}_m)
\hat\psi_{\sigma_m'}({\Bf r}_m')\dots
\hat\psi_{\sigma_1'}({\Bf r}_1') \vert\Psi_{N;0}\rangle;
\nonumber\\
\end{eqnarray}
the function $\Gamma^{(1)}({\Bf r}'\sigma;{\Bf r}\sigma)$ is more
commonly denoted by $\varrho_{\sigma}({\Bf r}',{\Bf r})$. 

Now we specialize to cases in which one has the following spectral 
representation for the single-particle density matrix:
\begin{equation}
\label{ea4}
\varrho_{\sigma}({\Bf r}',{\Bf r}) =
\sum_{\bF k} {\sf n}_{\sigma}({\Bf k})\,
\psi_{\bF k}({\Bf r}) 
\psi_{\bF k}^*({\Bf r}'), 
\end{equation}
where, for uniform GSs defined on the continuum, one has
\begin{equation}
\label{ea5}
\psi_{\bF k}({\Bf r}) = 
\frac{1}{\Omega^{1/2}}\, {\rm e}^{i {\bF k}\cdot {\bF r} },
\end{equation}
and, for uniform GSs defined on a regular lattice 
consisting of $N_{\sc l}$ sites (subject to periodic
boundary condition)
\begin{equation}
\label{ea6}
\psi_{\bF k}({\Bf r}) =
\frac{1}{N_{\sc l}^{1/2}}\,
\sum_{j} {\rm e}^{i {\bF k}\cdot {\bF R}_j }\,
\phi({\Bf r}-{\Bf R}_j),
\end{equation}
where $\{ {\Bf R}_j\, \|\, j=1,\dots,N_{\sc l} \}$ defines the 
above-mentioned lattice and $\phi_{j}({\Bf r})$ is a normalized 
`atomic' orbital localized around ${\Bf r}={\Bf 0}$. Since 
$\psi_{\bF k}({\Bf r})$ in Eq.~(\ref{ea6}) is a Bloch function, 
it can be written as follows:
\begin{equation}
\label{ea7}
\psi_{\bF k}({\Bf r}) = \frac{1}{\Omega^{1/2}}\,
{\rm e}^{ i {\bF k}\cdot {\bF r} }\,
u_{\bF k}({\Bf r}),
\end{equation}
where $u_{\bF k}({\Bf r})$ stands for a normalized periodic function 
of ${\Bf r}$ satisfying $u_{\bF k}({\Bf r}) = u_{\bF k}({\Bf r}+
{\Bf R}_j)$, $\forall j$. In what follows we explicitly deal with 
the Bloch function in Eq.~(\ref{ea7}) and deduce the results 
corresponding to the Bloch function in Eq.~(\ref{ea5}) through 
identifying $u_{\bF k}({\Bf r})$ with unity. 

With
\begin{equation}
\label{ea8}
h_{0}({\Bf r}) \psi_{\bF k}({\Bf r}) = 
\varepsilon_{\bF k} \psi_{\bF k}({\Bf r}), 
\end{equation}
it follows that
\footnote{\label{fa2}
For time-reversal-symmetric GSs, for which $\varepsilon_{-\bF k}\, 
{\sf n}_{\sigma}(-{\Bf k}) \equiv \varepsilon_{\bF k}\, 
{\sf n}_{\sigma}({\Bf k})$ holds, one readily verifies that the 
RHS of Eq.~(\protect\ref{ea9}) is invariant under the exchange 
of ${\Bf r}$ and ${\Bf r}'$ and consequently so is also the LHS. }
\begin{equation}
\label{ea9}
h_0({\Bf r})\, \varrho_{\sigma}({\Bf r}',{\Bf r})
= \sum_{\bF k} \varepsilon_{\bF k}\, 
{\sf n}_{\sigma}({\Bf k})\, \psi_{\bF k}({\Bf r})\, 
\psi_{\bF k}^*({\Bf r}').
\end{equation}
For uniform GSs we have
\begin{equation}
\label{ea10}
A_{\sigma}({\Bf r},{\Bf r}';\varepsilon)
= \sum_{\bF k} A_{\sigma}({\Bf k};\varepsilon)\,
\psi_{\bF k}({\Bf r}) \psi_{\bF k}^*({\Bf r}'),
\end{equation}
so that through the orthogonality relation
\begin{equation}
\label{ea11}
\int {\rm d}^dr\; \psi_{\bF k}^*({\Bf r})
\psi_{{\bF k}'}({\Bf r}) = \delta_{{\bF k},{\bF k}'},
\end{equation}
from Eq.~(\ref{ea10}) we obtain
\begin{equation}
\label{ea12}
A_{\sigma}({\Bf k};\varepsilon) =
\int {\rm d}^dr {\rm d}^dr'\;
\psi_{\bF k}^*({\Bf r}) A_{\sigma}({\Bf r},{\Bf r}';\varepsilon)
\psi_{\bF k}({\Bf r}').
\end{equation}
Thus from Eqs.~(\ref{ea2}), (\ref{ea9}), (\ref{ea11}) and (\ref{ea12})
we deduce that
\begin{equation}
\label{ea13}
\frac{1}{\hbar}
\int_{-\infty}^{\mu} {\rm d}\varepsilon\;
\varepsilon\, A_{\sigma}({\Bf k};\varepsilon)
= \varepsilon_{\bF k}\, {\sf n}_{\sigma}({\Bf k})
+ {\sf g}\, \beta_{{\bF k};\sigma}^{<},
\end{equation}
where with reference to Eqs.~(\ref{e39}) and (\ref{e40}) we have
\begin{eqnarray}
\label{ea14}
&&\beta_{{\bF k};\sigma}^{<} \equiv
\int {\rm d}^dr {\rm d}^dr'\; 
\psi_{\bF k}^*({\Bf r}) \psi_{\bF k}({\Bf r}') \nonumber\\
&&\;\;
\times \int {\rm d}^dr''\; w({\Bf r}-{\Bf r}'')\, \sum_{\sigma'}
\Gamma^{(2)}({\Bf r}'\sigma,{\Bf r}''\sigma';
{\Bf r}\sigma,{\Bf r}''\sigma'),
\end{eqnarray}
in which as in Eq.~(\ref{e37}) we have introduced the dimensionless 
two-body interaction potential $w \equiv {\sf g}^{-1}\, v$.

From the representation in Eq.~(\ref{ea4}) and the orthogonality
relation in Eq.~(\ref{ea11}) we have
\begin{equation}
\label{ea15}
\int {\rm d}^dr''\; 
\varrho_{\sigma}({\Bf r}',{\Bf r}'') 
\varrho_{\sigma}({\Bf r}'',{\Bf r}) 
= \sum_{\bF k} \big({\sf n}_{\sigma}({\Bf k})\big)^2\,
\psi_{\bF k}({\Bf r}) \psi_{\bF k}^*({\Bf r}'),
\end{equation}
so that, in cases where ${\sf n}_{\sigma}({\Bf k})$ takes the 
values $0$ and $1$ over the entire range of ${\Bf k}$, it follows 
that $\varrho_{\sigma}({\Bf r}',{\Bf r})$ is {\sl idempotent} (in 
the representation-free notation, idempotency is defined through 
$\varrho_{\sigma} \varrho_{\sigma} = \varrho_{\sigma}$). 
\footnote{\label{fa3}
The fact that the single-particle density matrix pertaining to 
a non-interacting GS is idempotent is associated with the very 
specific condition where ${\sf n}_{\sigma}({\Bf k})$ coincides 
with the characteristic function of the non-interacting Fermi 
sea (equal to unity inside and equal to zero outside the 
mentioned Fermi sea). The expressions that are presented in 
Eqs.~(\protect\ref{ea20}) and (\protect\ref{ea21}) below are 
therefore of wider applicability than solely to the $\Gamma^{(2)}$ 
pertaining to non-interacting GSs. } 
In such 
cases (and only in such case) the following result is exact:
\begin{eqnarray}
\label{ea16}
\Gamma^{(2)}(x_1,x_2;x_1',x_2') = 
\left|
\begin{array}{cc}
\Gamma^{(1)}(x_1;x_1') & \Gamma^{(1)}(x_1;x_2') \\ \\
\Gamma^{(1)}(x_2;x_1') & \Gamma^{(1)}(x_2;x_2') 
\end{array} \right|,
\end{eqnarray}
in which $x_i \equiv {\Bf r}_i\sigma_i$. By the conservation of 
spin we have \cite{BF02b}
\begin{equation}
\label{ea17}
\Gamma^{(1)}({\Bf r}\sigma,{\Bf r}'\sigma') = 
\delta_{\sigma,\sigma'} \, \Gamma^{(1)}({\Bf r}\sigma,{\Bf r}'\sigma)
\equiv \delta_{\sigma,\sigma'}\,\varrho_{\sigma}({\Bf r},{\Bf r}').
\end{equation}
Under the assumption that ${\sf n}_{\sigma}({\Bf k}) \in \{0, 1\}$,
we thus have
\begin{eqnarray}
\label{ea18}
&&\int {\rm d}^dr''\; w({\Bf r}-{\Bf r}'')\, \sum_{\sigma'}
\Gamma^{(2)}({\Bf r}'\sigma,{\Bf r}''\sigma';
{\Bf r}\sigma,{\Bf r}''\sigma') \nonumber\\
&&\;\;\;
= \frac{1}{\sf g}\, v_{\sc h}({\Bf r};[n]) 
\varrho_{\sigma}({\Bf r}',{\Bf r})\nonumber\\
&&\;\;\;
- \int {\rm d}^dr''\; w({\Bf r}-{\Bf r}'')\,
\varrho_{\sigma}({\Bf r}',{\Bf r}'') 
\varrho_{\sigma}({\Bf r}'',{\Bf r}), 
\end{eqnarray}
where $v_{\sc h}({\Bf r};[n])$ is the Hartree potential 
corresponding to the total number density $n({\Bf r}) \equiv 
\sum_{\sigma} n_{\sigma}({\Bf r})$, where $n_{\sigma}({\Bf r})
\equiv \varrho_{\sigma}({\Bf r},{\Bf r})$, that is
\begin{equation}
\label{ea19}
v_{\sc h}({\Bf r};[n]) \equiv \int {\rm d}^dr'\;
v({\Bf r}-{\Bf r}')\, n({\Bf r}').
\end{equation}
For two-body potentials for which $\int {\rm d}^dr'\;
v({\Bf r}-{\Bf r}') \equiv \Omega\, \bar{v}({\Bf q}={\Bf 0})$ is 
unbounded, it is necessary to effect a regularization procedure 
which takes account of the attractive interaction of the particles 
with some uniformly distributed background charge through which 
$v_{\sc h}({\Bf r};[n])$ is replaced by $v_{\sc h}({\Bf r};[n-n_0])$ 
where $n_0 {:=} N/\Omega$. For uniform GSs defined on the continuum, 
this procedure entirely removes $v_{\sc h}({\Bf r};[n])$ from the
formalism. For uniform GSs defined on a lattice, $v_{\sc h}({\Bf r};
[n])$ is equal to a constant, independent of ${\Bf r}$. Since in this 
appendix we deal with uniform GSs, in what follows we denote the 
Hartree potential by $v_{\sc h}$ which represents a constant which 
may or may not be vanishing.

From the above observations, some straightforward algebra, and 
Eqs.~(\ref{ea14}) and (\ref{ea18}) we obtain
\begin{eqnarray}
\label{ea20}
\beta_{{\bF k};\sigma}^{<}
&=& \Big\{\frac{1}{\sf g}\, v_{\sc h}
-\int \frac{{\rm d}^d k'}{(2\pi)^d}\; 
\bar{w}(\|{\Bf k}-{\Bf k}'\|)\, 
{\sf n}_{\sigma}({\Bf k}')\Big\}\,
{\sf n}_{\sigma}({\Bf k}) \nonumber\\
&\equiv& \frac{\hbar}{\sf g}\, 
\Sigma_{\sigma}^{\sc hf}({\Bf k})\,
{\sf n}_{\sigma}({\Bf k}),
\end{eqnarray}
where $\Sigma_{\sigma}^{\sc hf}({\Bf k})$ stands for the 
Hartree-Fock part of the self-energy. For the uniform GSs of the
single-band Hubbard Hamiltonian, $\Sigma_{\sigma}^{\sc hf}({\Bf k})= 
\hbar^{-1} U n_{\bar\sigma}$ (independent of ${\Bf k}$) where 
$n_{\bar\sigma} {:=} N_{\bar\sigma}/N_{\sc l}$; from this and 
Eq.~(\ref{ea20}) it follows that $\beta_{{\bF k};\sigma}^{<} = 
n_{\bar\sigma} {\sf n}_{\sigma}({\Bf k})$ and, through Eq.~(\ref{e40}), 
$\xi_{{\bF k};\sigma}=n_{\bar\sigma}$. In arriving at Eq.~(\ref{ea20}) 
we have considered the thermodynamic limit and replaced $\Omega^{-1} 
\sum_{{\bF k}'}$ by $(2\pi)^{-d} \int {\rm d}^d k'$. This integral 
covers the entire available wave-vector space which for systems defined 
on a lattice consists of the 1BZ corresponding to the underlying 
Bravais lattice; for these systems, the vector ${\Bf k}-{\Bf k}'$ on 
the RHS of Eq.~(\ref{ea20}) is to be understood as representing 
${\Bf k}-{\Bf k}'+ {\Bf K}_0$ where the reciprocal lattice vector 
${\Bf K}_0$ (corresponding to an Umklapp process) is a function of 
${\Bf k}$ and ${\Bf k}'$ and ensures that ${\Bf k}-{\Bf k}'+ {\Bf K}_0 
\in {\rm 1BZ}$. 

From Eq.~(\ref{ea20}) it is observed that within the framework of 
the approximation where ${\sf n}_{\sigma}({\Bf k}) \in \{0,1\}$, 
$\beta_{{\bF k};\sigma}^{<}$ is proportional to ${\sf n}_{\sigma}
({\Bf k})$ so that for $\xi_{{\bF k};\sigma}$ as defined in 
Eq.~(\ref{e40}) we have
\begin{equation}
\label{ea21}
\xi_{{\bF k};\sigma} = \frac{\hbar}{\sf g}\,
\Sigma_{\sigma}^{\sc hf}({\Bf k}).
\end{equation}
Thus, following Eq.~(\ref{e39}), according to the present 
approximation we have
\begin{equation}
\label{ea22}
\varepsilon_{{\bF k};\sigma}^{<}
= \varepsilon_{\bF k} + \hbar \Sigma_{\sigma}^{\sc hf}({\Bf k}).
\end{equation}
The unsound (see, however, next paragraph) nature of this expression in 
its functional form becomes evident by realizing the fact that by the 
assumed stability of the GS of the system it is required that 
$\varepsilon_{{\bF k};\sigma}^{<} < \mu$, $\forall {\Bf k}$, while 
the expression on the RHS of Eq.~(\ref{ea22}) in general does not 
satisfy this requirement. For instance, by considering the uniform 
electron-gas system for which one has the unbounded energy dispersion 
$\varepsilon_{\bF k}=\hbar^2 \|{\Bf k}\|^2/[2 m_e]$, one immediately 
observes that for sufficiently large $\|{\Bf k}\|$ the RHS of 
Eq.~(\ref{ea22}) can be made to exceed any constant value. This 
aspect is made the more explicit by considering the energy dispersion 
$\varepsilon_{{\bF k};\sigma}^{>}$ in Eq.~(\ref{e48}) (see also
Eq.~(\ref{e23})). Under the conditions for which Eq.~(\ref{ea20}) 
applies, Eq.~(\ref{e48}) yields
\begin{equation}
\label{ea23}
\varepsilon_{{\bF k};\sigma}^{>} = 
\varepsilon_{\bF k} + \hbar\Sigma_{\sigma}^{\sc hf}({\Bf k}),
\end{equation}
which identically coincides with the (approximate) result in 
Eq.~(\ref{ea22}). This result, similar to that in Eq.~(\ref{ea22}), 
clearly exposes the  shortcoming of the approximation according to 
which ${\sf n}_{\sigma}({\Bf k}) \in \{0,1\}$; in \cite{BF02a,BF03} 
is has been shown that the exact $\varepsilon_{{\bF k};\sigma}^{>}$ 
satisfies $\varepsilon_{{\bF k};\sigma}^{>} > \mu$, $\forall {\Bf k}$.

The peculiarities of the expressions in Eqs.~(\ref{ea22}) and
(\ref{ea23}) are understood by considering the fictitious 
single-particle spectral function ${\cal A}_{\sigma}({\Bf k};
\varepsilon)$ introduced in Eq.~(\ref{e32}) and to which the 
energies $\varepsilon_{{\bF k};\sigma}^{<}$ and 
$\varepsilon_{{\bF k};\sigma}^{>}$ correspond. It is observed that 
${\cal A}_{\sigma}({\Bf k};\varepsilon)$ yields the same result 
by substituting ${\cal A}_{\sigma}({\Bf k};\varepsilon)$ for the 
exact $A_{\sigma}({\Bf k};\varepsilon)$ in Eq.~(\ref{e39}), that is
\begin{equation}
\label{ea24}
\frac{ \int_{-\infty}^{\mu} {\rm d}\varepsilon\;
\varepsilon\, {\cal A}_{\sigma}({\Bf k};\varepsilon) }
{ \int_{-\infty}^{\mu} {\rm d}\varepsilon\;
{\cal A}_{\sigma}({\Bf k};\varepsilon) }
= \varepsilon_{{\bF k};\sigma}^{<},\;\;\; \forall {\Bf k}.
\end{equation}
The same applies to $\varepsilon_{{\bF k};\sigma}^{>}$ in 
Eq.~(\ref{e23}). As is evident, the validity of Eq.~(\ref{ea24}) 
and of its counterpart concerning $\varepsilon_{{\bF k};\sigma}^{>}$ 
crucially depends on the properties $\varepsilon_{{\bF k};\sigma}^{<} 
< \mu$ and $\varepsilon_{{\bF k};\sigma}^{>} > \mu$, $\forall 
{\Bf k}$. In order to gain insight into the consequences of the 
approximate framework considered in this appendix, leading to 
identical energy dispersions for $\varepsilon_{{\bF k};\sigma}^{<}$ 
and $\varepsilon_{{\bF k};\sigma}^{>}$, for all ${\Bf k}$, it is 
appropriate to substitute the expressions in Eqs.~(\ref{ea22}) 
and (\ref{ea23}) into Eq.~(\ref{e32}) from which it follows that
\begin{equation}
\label{ea25}
{\cal A}_{\sigma}({\Bf k};\varepsilon) = \hbar\, 
\delta(\varepsilon-\varepsilon_{{\bF k};\sigma}^{\sc hf}),
\end{equation}
where 
\begin{equation}
\label{ea26}
\varepsilon_{{\bF k};\sigma}^{\sc hf} {:=}
\varepsilon_{\bF k} + \hbar\Sigma_{\sigma}^{\sc hf}({\Bf k}). 
\end{equation}
One observes that, within the approximate framework in which 
${\sf n}_{\sigma}({\Bf k}) \in \{0,1\}$, the single-particle 
energy dispersion in fact consists of a single branch, namely
$\varepsilon_{{\bF k};\sigma}^{\sc hf}$, and not of two identical 
branches. Thus the identity of $\varepsilon_{{\bF k};\sigma}^{<}$ 
and $\varepsilon_{{\bF k};\sigma}^{>}$ in our approximate framework 
amounts to no fundamental defect, rather to a manifestation of 
the restricted space of the single-particle excitations available 
to states (or GSs) for which the associated ${\sf n}_{\sigma}
({\Bf k})$ satisfies ${\sf n}_{\sigma}({\Bf k}) \in \{0,1\}$, 
$\forall {\Bf k}$. With reference to specifically our considerations 
in \S~4, resulting in the conclusion that for cases in which 
$v({\Bf r}-{\Bf r}')$ is the long-range Coulomb potential (or one 
as long-ranged as this), ${\Bf\nabla}_{\bF k}\varepsilon_{{\bF k};
\sigma}^{\Ieq><}$ is logarithmically divergent for ${\Bf k}$ 
approaching regions (e.g. ${\cal S}_{{\sc f};\sigma}$) in which 
${\sf n}_{\sigma}({\Bf k})$ is discontinuous, it is observed that 
this aspect is appropriately preserved by ${\Bf\nabla}_{\bF k}
\varepsilon_{{\bF k};\sigma}^{\sc hf}$ in which 
$\varepsilon_{{\bF k};\sigma}^{\sc hf}$ is purported to approximate 
$\varepsilon_{{\bF k};\sigma}^{\Ieq><}$.
\hfill $\Box$

\end{appendix}

%________________________
\pagebreak

%________________________

% 1.
\pagebreak
%\widetext
%\clearpage

% 2.
\pagebreak
%\widetext
%\clearpage

% 3.
\pagebreak
%\widetext
%\clearpage

% 4.
\pagebreak
%\widetext
%\clearpage


\vspace{1.5cm}
\centerline{\bf References}
\begin{references}
\label{ref}
\vspace{-1.3cm}

% 1.
\bibitem{DHS03}
A. Damascelli, Z. Hussain, and Z.-X. Shen,
{\it Rev. Mod. Phys.} {\bf 75}, 473 (2003).

% 2.
\bibitem{SD95}
Z.-X. Shen, and D.~S. Dessau,
{\it Physica Reports}, {\bf 253}, 1-162 (1995).

% 3.
\bibitem{XJZ03}
X.~J. Zhou, T. Yoshida, A. Lanzara, P.~V. Bogdanov,
S.~A. Kellar, K.~M. Shen, W.~L. Yang, F. Ronning,
T. Sasagawa, T. Kakeshita, T. Noda, H. Eisaki, S. Uchida,
C.~T. Lin, F. Zhou, J.~W. Xiong, W.~X. Ti, Z.~X. Zhao,
A. Fujimori, Z. Hussain, and Z.-X. Shen,
{\it Nature}, {\bf 423}, 398 (2003).

% 4.
\bibitem{TV99}
T. Valla, A.~V. Fedorov, P.~D. Johnson, B.~O. Wells,
S.~L. Hulbert, Q. Li, G.~D. Gu, and N. Koshizuka,
{\it Science}, {\bf 285}, 2110 (1999).

% 5.
\bibitem{PVB00}
P.~V. Bogdanov, A. Lanzara, S.~A. Kellar, X.~J. Zhou, 
E.~D. Lu, W.~J. Zheng, G. Gu, J.-I. Shimoyama,
K. Kishio, H. Ikeda, R. Yoshizaki, Z. Hussain, and Z.-X. Shen,
{\it Phys. Rev. Lett.} {\bf 85}, 2581 (2000).

% 6.
\bibitem{PDJ01}
P.~D. Johnson, T. Valla, A.~V. Fedorov, Z. Yusof, B.~O. Wells,
Q. Li, A.~R. Moodenbaugh, G.~D. Gu, N. Koshizuka, C. Kendziora,
S. Jian, and D.~G. Hinks,
{\it Phys. Rev. Lett.} {\bf 87}, 177007 (2001).

% 7.
\bibitem{AK01}
A. Kaminski, M. Randeria, J.~C. Campuzano, M.~R. Norman, 
H. Fretwell, J. Mesot, T. Sato, T. Takahashi, and 
K. Kadowaki,
{\it Phys. Rev. Lett.} {\bf 86}, 1070 (2001).

% 8.
\bibitem{AL01}
A. Lanzara, P.~V. Bogdanov, X.~J. Zhou, S.~A. Kellar, 
D.~L. Feng, E.~D. Lu, T. Yoshida, H. Eisaki, A. Fujimori,
K. Kishio, J.-I. Shimoyama, T. Noda, S. Uchida, Z. Hussain,
and Z.-X. Shen,
{\it Nature} {\bf 412}, 510 (2001).

% 9.
\bibitem{BF02a}
B. Farid,
{\it Phil. Mag.} {\bf 83}, 2829 (2003). 
{\tt cond-mat/0211244}

% 10.
\bibitem{BF03}
B. Farid,
{\it Phil. Mag.}, {\bf 84}, 109 (2004). 
{\tt cond-mat/0304350}

% 11.
\bibitem{SS01}
A.~F. Santander-Syro, R.~P.~S.~M. Lobo, N. Bontemps,
Z. Konstantinovic, Z.~Z. Li, and H. Raffy,
{\it Europhys. Lett.} {\bf 62}, 568 (2003).
{\tt cond-mat/0111539}

% 12.
\bibitem{MPMKL02}
H.~J.~A. Molegraaf, C. Presura, D. van der Marel,
P.~H. Kes, and M. Li,
{\it Science} {\bf 295}, 2239 (2002).

% 13.
\bibitem{PWA97}
P.~W. Anderson,
{\sl THE Theory of Superconductivity in the High-$T_c$
Cuprates} (Princeton University Press, 1997).

% 14.
\bibitem{PN66}
D. Pines, and P. Nozi\`eres,
{\sl The Theory of Quantum Liquids}, Vol.~I,
{\sl Normal Fermi Liquids} (Benjamin, New York, 1966).

% 15.
\bibitem{REA95}
M. Randeria, H. Ding, J.-C. Campuzano, A. Bellman,
G. Jennings, T. Yokoya, T. Takahashi,
H. Katayama-Yoshida, T. Mochiku, and K. Kadowaki,
{\it Phys. Rev. Lett.} {\bf 74}, 4951 (1995).

% 16.
\bibitem{CDNR96} 
J.~C. Campuzano, H. Ding, M.~R. Norman, M. Randeria,
A.~F. Bellman, T. Yokoya, T. Takahashi, 
H. Katayama-Yoshida, T. Mochiku, and K. Kadowaki,
{\it Phys. Rev.} B~{\bf 53}, R14737 (1996).

% 17.
\bibitem{PLS98}
W.~O. Putikka, M.~U. Luchini, and R.~R.~P. Singh,
{\it Phys. Rev. Lett.} {\bf 81}, 2966 (1998).

% 18.
\bibitem{BF99a}
B. Farid,
{\it Phil. Mag.} B~{\bf 79}, 1097 (1999).

% 19.
\bibitem{BF02b}
B. Farid,
{\it Phil. Mag.} B, {\bf 82}, 1413 (2002);
(E) {\it ibid.}, {\bf 82}, 1817 (2002).

% 20.
\bibitem{BF99b}
B. Farid,
{\sl Electron Correlation in the Solid State},
edited by N.~H. March (Imperial College Press, 
London, 1999), chapter 3.

% 21.
\bibitem{ABM57}
A.~B. Migdal, 
{\it Soviet Phys. JETP}, {\bf 5}, 333 (1957).

% 22.
\bibitem{JML60}
J.~M. Luttinger,
{\it Phys. Rev.} {\bf 119}, 1153 (1960).

% 23.
\bibitem{GDM90}
G.~D. Mahan, {\sl Many-Particle Physics}, second edition
(Plenum, New York, 1990).

% 24.
\bibitem{JEH01}
J.~E. Hirsch,
{\it Phys. Rev. Lett.} {\bf 87}, 206402 (2001).

% 25.
\bibitem{JEH03}
J.~E. Hirsch,
{\it Phys. Rev.} B~{\bf 67}, 035103 (2003).

% 26.
\bibitem{HL67}
A.~B. Harris, and R.~V. Lange,
{\it Phys. Rev.} {\bf 157}, 295 (1967).

% 27.
\bibitem{EMS91}
H. Eskes, M.~B.~J. Meinders, and G.~A. Sawatzky,
{\it Phys. Rev. Lett.} {\bf 67}, 1035 (1991). 

% 28.
\bibitem{EO94}
H. Eskes, and A.~M. Ole\'s,
{\it Phys. Rev. Lett.} {\bf 73}, 1279 (1994). 

% 29.
\bibitem{EOMS94}
H. Eskes, A.~M. Ole\'s, M.~B.~J. Meinders, and W. Stephan,
{\it Phys. Rev.} B~{\bf 50}, 17980 (1994).

% 30.
\bibitem{VAB61}
V.~A. Belyakov,
{\it Soviet Phys. JETP}, {\bf 13}, 850 (1961).

% 31.
\bibitem{SM80}
R. Sartor, and C. Mahaux, 
{\it Phys. Rev.} C~{\bf 21}, 1546 (1980).

% 32.
\bibitem{VLSRAR89}
C.~M. Varma, P.~B. Littlewood, S. Schmitt-Rink,
E. Abrahams, and A.~E. Ruckenstein,
{\it Phys. Rev. Lett.} {\bf 63} 1996 (1989);
{\it ibid.} {\bf 64}, 497 (1990).

% 33.
\bibitem{ZXS02}
Z.-X. Shen, A. Lanzara, S. Ishihara, and N. Nagaosa,
{\it Phil. Mag.} B~{\bf 82}, 1349 (2002). 
{\tt cond-mat/0108381}

% 34.
\bibitem{AP01}
A. Paramekanti, M. Randeria, and N. Trivedi,
{\it Phys. Rev. Lett.} {\bf 87}, 217002 (2001).

% 35.
\bibitem{AP03}
A. Paramekanti, M. Randeria, and N. Trivedi, \\
{\tt cond-mat/0305611}.

% 36.
\bibitem{BKT00}
J. Bon\v{c}a, T. Katra\v{s}nik, and S.~A. Trugman,
{\it Phys. Rev. Lett.} {\bf 84}, 3153 (2000).

% 37.
\bibitem{MR03}
M. Randeria, A. Paramekanti, and N. Trivedi, \\
{\tt cond-mat/0307217}.

% 38.
\bibitem{PBA01}
P.~B. Allen,
{\it Nature}, {\bf 412}, 494 (2001).

% 39.
\bibitem{ZG01}
R. Zeyher, and A. Greco,
{\it Phys. Rev.} B~{\bf 64}, 140510 (2001).

% 40.
\bibitem{AM94}
A.~S. Alexandrov, and N.~F. Mott,
{\it Rep. Prog. Phys.} {\bf 57}, 1197 (1994).

% 41.
\bibitem{MLK00}
M.~L. Kuli\'c,
{\it Physics Reports} {\bf 338}, 1 (2000). 

% 42.
\bibitem{MEB03}
D. Manske, I. Eremin, and K.~H. Bennemann,
{\it Phys. Rev.} B~{\bf 67}, 134520 (2003). 

% 43.
\bibitem{EN00}
M. Eschrig, and M.~R. Norman,
{\it Phys. Rev. Lett.} {\bf 85}, 3261 (2000).

% 44.
\bibitem{MEB01}
D. Manske, I. Eremin, and K.~B. Bennemann,
{\it Phys. Rev. Lett.} {\bf 87}, 177005 (2001).

% 45.
\bibitem{AC69}
P.~B. Allen, and M.~L. Cohen,
{\it Phys. Rev.} {\bf 187}, 525 (1969). 

% 46.
\bibitem{AM81}
N.~W. Ashcroft, and N.~D. Mermin,
{\sl Solid State Physics} (Holt-Saunders, Philadelphia, 1981).

% 47.
\bibitem{HPSGB99}
M. Hengsberger, D. Purdie, P. Segovia, M. Arnierm and Y. Baer,
{\it Phys. Rev. Lett.} {\bf 83}, 592 (1999).

% 48.
\bibitem{LSJB00}
S. LaShell, E. Jensen, and T. Balasubramanian,
{\it Phys. Rev.} B~{\bf 61}, 2371 (2000).

% 49.
\bibitem{VFJH99}
T. Valla, A.~V. Fedorov, P.~D. Johnson, and S.~L. Hulbert,
{\it Phys. Rev. Lett.} {\bf 83}, 2085 (1999).

% 50.
\bibitem{HF50}
H. Fr\"ohlich,
{\it Phys. Rev.} {\bf 79}, 845 (1950).

% 51.
\bibitem{HF51}
H. Fr\"ohlich,
{\it Proc. Phys. Soc.} A~{\bf 64}, 129-134 (1951).

% 52.
\bibitem{GR65}
G. Rickayzen,
{\sl Theory of Superconductivity}
(Interscience Publishers, New York, 1965).

% 53.
\bibitem{MRS51}
M.~R. Schafroth, 
{\it Hel. Phys. Acta}, {\bf 24}, 645 (1951).

% 54.
\bibitem{GW51}
G. Wentzel,
{\it Phys. Rev.} {\bf 83}, 168 (1951).

% 55.
\bibitem{KV51}
W. Kohn, and [no initial(s)] Vachaspati, 
{\it Phys. Rev.} {\bf 83}, 462 (1951).

% 56.
\bibitem{JB51}
J. Bardeen,
{\it Rev. Mod. Phys.} {\bf 23}, 261 (1951).

% 57.
\bibitem{PWA69}
P.~W. Anderson, 
{\sl Superconductivity}, Vol. 2, edited by R.~D. Parks 
(Marcel Dekker, New York, 1969), chapter 23, pp. 1343-1358.

% 58.
\bibitem{ABM58}
A.~B. Migdal, 
{\it Soviet Phys., JETP}, {\bf 7}, 996 (1958).

% 59.
\bibitem{CSAS93}
S. Chakravarty, A. Sudb\o, P.~W. Anderson, and S. Strong,
{\it Science} {\bf 261}, 337 (1993).

% 60.
\bibitem{PSG95}
L. Pietronero, S. Str\"assler, C. Grimaldi,
{\it Phys. Rev.} B~{\bf 52}, 10516 (1995).

% 61.
\bibitem{BF99c}
B. Farid, 
unpublished (1999).

% 62.
\bibitem{BCGKS03}
J. Boronat, J. Casulleras, V. Grau, E. Krotscheck, and
J. Springer,
{\it Phys. Rev. Lett.} {\bf 91}, 085302 (2003).
{\tt cond-mat/0307493}

% 63.
\bibitem{JZ99}
J. Zaanen,
{\it Physica} C~{\bf 317}-{\bf 318}, 217 (1999).

% 64.
\bibitem{CEKO02}
E.~W. Carlson, V.~J. Emery, S.~A. Kivelson, and D. Orgad,
{\sl The Physics of Superconductors}, Vol. 2, edited by 
K.-H. Bennemann and J.~B. Ketterson (Springer Verlag,
Berlin, 2004), to be published.
{\tt cond-mat/0206217}

% 65.
\bibitem{NCMS03}
A.~H. Castro Neto, and C. Morais Smith,
{\sl Physics and Chemistry of Materials with Low Dimensional
Structure}, edited by D. Baeriswyl and L. DeGiorgi
(Kluwer, Dordrecht, 2003), to be published. 
{\tt cond-mat/0304094}

% 66.
\bibitem{CPS02}
A.~V. Chubukov, D. Pines, and J. Schmalian,
{\sl The Physics of Superconductors}, Vol. 1, edited by 
K.-H. Bennemann and J.~B. Ketterson (Springer Verlag,
Berlin, 2003), 495.
{\tt cond-mat/0201140}



\end{references}
\end{document}